%% file: paper.tex
\documentclass{aa}  

\usepackage{graphicx}
\usepackage{lscape}
\usepackage{natbib} 
\usepackage{siunitx}
\usepackage{url}

\usepackage[varg]{txfonts}

\bibpunct{(}{)}{;}{a}{}{,} 

\begin{document}

\title{Wolf-Rayet stars in the Small Magellanic Cloud
}

   \subtitle{I. Analysis of the single WN stars}

   \author{R. Hainich\inst{1}
          \and D. Pasemann\inst{1,\,}\inst{2}
          \and H. Todt\inst{1}
          \and T. Shenar\inst{1}
          \and A. Sander\inst{1}
          \and W.-R. Hamann\inst{1}
          }

   \institute{Institut f\"ur Physik und Astronomie,
              Universit\"at Potsdam,
              Karl-Liebknecht-Str. 24/25, D-14476 Potsdam, Germany \\
              \email{rhainich@astro.physik.uni-potsdam.de}
              \and 
              Charité,
              Humboldt-Universit\"at zu Berlin,
              Charitéplatz 1,
              D-10117 Berlin, Germany
              }
   \date{Received <date> / Accepted <date>}


\abstract
%
%
{
Wolf-Rayet (WR) stars have a severe impact on their environments owing to their strong ionizing radiation fields and powerful stellar winds. Since these winds are considered to be driven by radiation pressure, it is theoretically expected that the degree of the wind mass-loss depends on the initial metallicity of WR stars.
}
%
%
{
Following our comprehensive studies of WR stars in the Milky Way, M\,31, and the LMC, we derive stellar parameters and mass-loss rates for all seven putatively single WN stars known in the SMC. Based on these data, we discuss the impact of a low-metallicity environment on the mass loss and evolution of WR stars.
}
%
%
{
The quantitative analysis of the WN stars is performed with the Potsdam Wolf-Rayet (PoWR) model atmosphere code. The physical properties of our program stars are obtained from fitting synthetic spectra to multi-band observations.
}
%
%
{
In all SMC WN stars, a considerable surface hydrogen abundance is detectable. The majority of these objects have stellar temperatures 
exceeding $75\,\mathrm{kK}$, while their luminosities range from $10^{5.5}$ to $10^{6.1}\,L_\odot$. The WN stars in the SMC exhibit on average lower mass-loss rates and weaker winds than their counterparts in the Milky Way, M\,31, and the LMC.
}
%
%
{
By comparing the mass-loss rates derived for WN stars in different Local Group galaxies, we conclude that a clear dependence of the wind mass-loss 
on the initial metallicity is evident, supporting the current paradigm that WR winds are driven by radiation. A metallicity effect on the evolution of massive stars is obvious from the HRD positions of the SMC WN stars at high temperatures and high luminosities. 
Standard evolution tracks are not able to reproduce these parameters and the observed surface hydrogen abundances.
Homogeneous evolution might provide a better explanation for their evolutionary past.
}


\maketitle

\section{Introduction}
\label{sect:intro}

The Small Magellanic Cloud (SMC) is a unique laboratory for studying massive stars in a low-metallicity environment. With a distance of 60.2\,kpc, corresponding to a distance modulus of $18.9\,\mathrm{mag}$ \citep{Westerlund1997,Harries2003,Hilditch2005,Keller2006}, it is close enough to allow for detailed spectroscopy of its massive star population. Moreover, the interstellar extinction in the direction and within the SMC is on average quite low \citep[e.g., ][]{Grieve1986,Schwering1991,Bessell1991,Schlegel1998,Larsen2000,Haschke2011}.
The metallicity of SMC stars is in general subsolar, but the individual values are strongly age dependent \citep{Piatti2011,Piatti2012}. 
Even in comparison to the metal deficient LMC, the metal content of SMC stars is still on average a factor of 2.5 lower at all stellar ages \citep{Piatti2013}.

Stellar winds of hot stars are assumed to be driven by the transfer of momentum from photons to the matter of the stellar atmosphere by scattering and absorption in spectral lines \citep[for a review see, e.g.,][]{Puls2008}. For this reason, the strength of the winds and the associated mass-loss is a function of the opacity and should thus depend on the metal content of the stellar atmosphere \citep[e.g., ][]{Kudritzki1987,Puls2000,Vink2000,Vink2001}. For Wolf-Rayet (WR) stars, the dependence of the mass-loss rate on the metallicity was theoretically predicted by \citet{Graefener2008} and \citet{Vink2005}. Since the SMC has a subsolar metallicity, considerably weaker radiation-driven stellar winds are expected in the SMC than in the Galaxy.

It is well known that in the SMC, the WR stars of the nitrogen sequence (WN stars) exhibit emission line spectra with relatively weak and narrow emission lines \citep[e.g., ][]{Conti1989}. \citet{Crowther2006b} found that the line luminosities of the WN stars in the SMC are a factor of four to five lower than their LMC counterparts. Moreover, distinct {\em absorption} lines can be recognized in eight of the twelve WR stars known in the SMC. These characteristics were either interpreted as a sign of binarity \citep{Massey2003} or attributed to photospheric lines that are inherent to the SMC WN stars because of their relatively weak stellar winds \citep{Conti1989,Massey2003}. \citet{Foellmi2003a} and \citet{Foellmi2004} showed that these absorption lines are blue-shifted but that they exhibit a constant radial velocity, concluding that these lines must form in the wind of these objects. Later on, reduced mass-loss rates of the SMC WN stars (compared to their LMC and MW counterparts) were supported by the findings of \citet{Crowther2006}, \citet{Nugis2007}, and \citet{Bonanos2010}. 

The WR population of the SMC is small but exceptional. It comprises only twelve objects. In contrast more than 140 WR stars are known in the LMC, and more than 630 in the Galaxy. In the SMC, almost all WR stars belong to the WN sequence. The only exception is the binary system SMC\,AB\,8 whose primary belongs to the rare class of WR stars showing prominent oxygen lines in their spectra (WO type). Wolf-Rayet stars of the carbon sequence (WC stars) are completely absent in the SMC. This is in strong contrast to the Galactic WR population where the number of WC and WN stars is approximately equal. For comparison, the WN/WC ratio in the LMC is roughly four. These differences in the specific WR populations are generally attributed to the different metallicities of these galaxies. The same trend of an increasing WN/WC ratio with decreasing metallicity is also found in other Local Group Galaxies \citep{Lopez-Sanchez2010}.

In addition to the WN/WC ratio also the distribution of the subtypes within each sequence varies between galaxies of different metallicities.
Using stellar atmosphere models, \citet{Crowther2006b} showed that a down scaling of the mass-loss rate at lower metallicity has a direct influence on the spectral appearance of WR stars, resulting in weaker emission line spectra and earlier WN subtypes for significantly subsolar metallicity objects. Indeed, almost all SMC WN stars exhibit an early spectral subtype (WNE: WN2--WN5), the only exceptions being SMC\,AB\,4 and the secondary in the SMC\,AB\,5 binary system. In contrast, the ratio of early- to late-type WN stars (WNL: WN6--WN11) in the MW is almost unity, while the majority of the LMC WN stars belong to the WNE category. 

Despite these characteristic differences between WR populations of individual galaxies, the impact of the metallicity on the strength of WR winds is not yet well established by means of spectral analyses. \citet{Crowther2006} and \citet{Nugis2007} empirically studied the wind mass-loss as a function of the metallicity based on a subset of the WR stars known in the Galaxy, LMC, and SMC. \citet{Vink2005} derived a theoretical prediction for the dependence of the mass-loss rate of WNL stars on the metallicity by means of Monte-Carlo simulations. 

Empirical and theoretical studies indicate a power-law dependence of the mass-loss rate $\dot{M}$ on the metallicity $Z$ with an exponent in the range of 0.8--1.0. However, a comprehensive evaluation of the metallicity effect on WR  winds, considering a large number of objects from the Local Group Galaxies is still missing but imperative. The present study addresses this important problem, building on our recent analyses of WN stars in the Galaxy \citep{Hamann2006}, M31 \citep{Sander2014}, and the LMC \citep{Hainich2014}.

The paper is structured as follows: In Sect.\,\ref{sect:sample} we briefly characterize the sample investigated in this work before presenting the data our analysis is based on in Sect.\,\ref{sect:data}. We then describe the Potsdam Wolf-Rayet (PoWR) atmosphere models and the procedure of the analysis in Sect.\,\ref{sect:method}. The results are compiled in Sect.\,\ref{sect:results}. In Sect.\,\ref{sect:evolution} we discuss the evolutionary status of the WN stars in the SMC. Section\,\ref{sect:conclusions} gives a summary and our general conclusions. The appendices encompass additional tables (Appendix\,\ref{sec:addtables}), comments on the individual objects (Appendix\,\ref{sect:comments}), and spectral fits of all stars in our sample (Appendix\,\ref{sect:specfits}). 

\section{The sample}
\label{sect:sample}

The number of known WR stars in the SMC has increased since the first comprehensive compilation by \citet{Azzopardi1979} from eight to twelve \citep{Morgan1991,Massey2001,Massey2003}. We identify each star by a running number, following the name scheme SMC\,AB\,\# which is based on the work conducted by \citet{Azzopardi1979} and which is one of the suggested names by the SIMBAD database. Alias names of the objects are included in the comments on individual objects (Appendix\,\ref{sect:comments}).

\begin{figure*}
\centering
\includegraphics[width=\hsize]{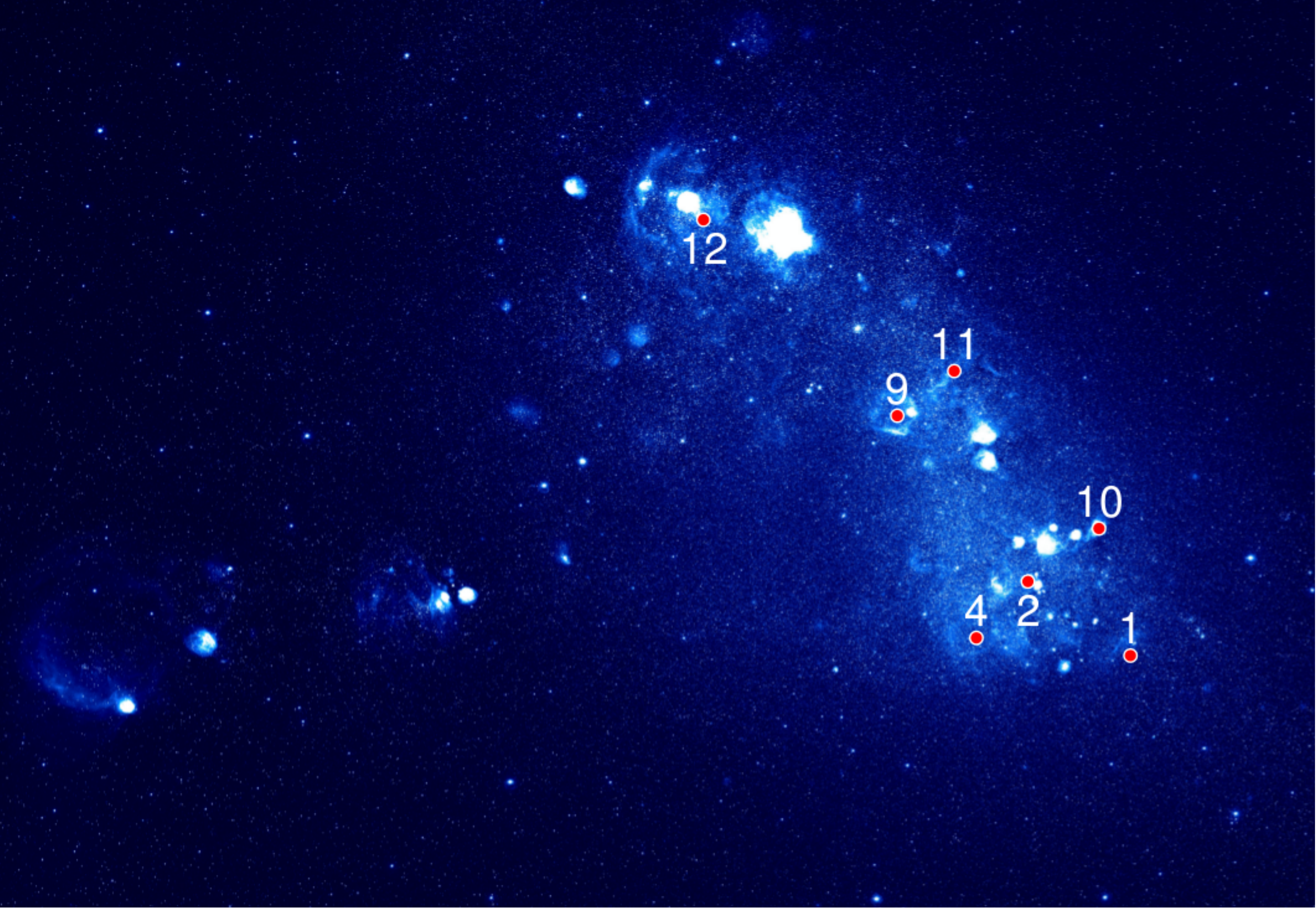}
\caption{The position of the putatively single WN stars in the SMC. The stars are identified by their number, following the scheme SMC\,AB\,\# (see text for details). The background image shows the \ion{O}{iii} nebular emission in the SMC. The image was retrieved from the Magellanic Cloud Emission-Line Survey \citep[MCELS, ][]{Smith2005}.} 
\label{fig:smc_sample}
\end{figure*}

In the present study, we concentrate on the putatively single WN stars in the SMC. \citet{Foellmi2003a} and \citet{Foellmi2004} carefully studied the binary status of the SMC WN stars by means of radial-velocity shifts, line profile variations, photometric  measurements, and X-ray data. According to these authors, SMC\,AB\,1, 2, 4, 9, 10, 11, and 12 are probably single stars. We kept SMC\,AB\,9 in our sample although \citet{Foellmi2003a} found weak indications that this object might actually be a binary. Thus, the sample investigated in this work comprises seven objects. The spatial positions of these stars in the SMC are shown in Fig.\,\ref{fig:smc_sample}. A detailed analysis of the known binary WR systems in the SMC will be the subject of a forthcoming paper (Shenar et al., in prep).

\section{Observational data}
\label{sect:data}

For most of our targets, high-resolution spectra taken with the Ultraviolet and Visual Echelle Spectrograph (UVES) mounted on Unit Telescope number 2 (UT2) of the Very Large Telescope (VLT) are available. The fully reduced UVES data in a wavelength range of 3960--4960\,\AA\ and 5040--6010\,\AA\ were retrieved from the ESO archive for advanced data products. For details on the data reduction we refer the reader to the ESO phase 3 data release website\footnote{\url{https://www.eso.org/observing/dfo/quality/PHOENIX/UVES_ECH/processing.html}}. We rectified the spectra by means of the spectral continua which were identified by a careful adjustment of a low-order polynomial to the observed data. 

For those objects without available UVES spectra, we used low-resolution spectra (6.7--2.8\,\AA) obtained by \citet{Foellmi2003a} in the spectral range of 3700--6830\,\AA. 
Although these datasets were primarily intended for radial velocity studies, they are well suited for our analysis owing to their high signal-to-noise ratio (S/N). 
Detailed information on the instrumentation used and the data reduction can be found in \citet{Foellmi2003a}. These spectra were normalized and co-added by the former authors. 
For three objects in our sample, we also used optical spectra observed by \citet{Torres-Dodgen1988}. In contrast to the spectra described above, these datasets are flux-calibrated. While the S/N of these low-resolution spectra is relatively modest, they are only used to construct the observed spectral energy distribution (SED). 

Additionally, archival spectra taken with the Far Ultraviolet Spectroscopic Explorer (FUSE) and the International Ultraviolet Explorer (IUE) were retrieved from the MAST archive. High-resolution FUSE spectroscopy in the FUV (960--1190\,\AA) is available for four objects in our sample. To improve the S/N ratio, these spectra were rebinned to 0.1\,\AA, resulting in a spectral resolution which is still sufficient to resolve the interstellar absorption lines. Low-dispersion IUE spectra (resolution of $\sim 6$\,\AA) in the range of 1200--3300\,\AA\ were downloaded for three stars. Since the IUE and FUSE data are flux-calibrated, these spectra are rectified by the reddened model continuum. A compilation of the spectra used in the spectral analysis of each sample star can be found in Table\,\ref{table:spectra}.

In addition to flux calibrated spectra, we used narrow-band magnitudes given by \citet{Crowther2006b} and \citet{Torres-Dodgen1988} to construct the observed SED in the optical range. Photometric measurements in the $I$ band are taken from \citet{Monet2003} and \citet{Bonanos2010}. In the near infrared we used photometry ($J, H, K_\mathrm{S}$) from the Two Micron All Sky Survey \citep[2MASS,\ ][]{Cutri2012b}. Furthermore, we applied Spitzer-IRAC magnitudes published by \citet{Bonanos2010} and the SAGE LMC and SMC IRAC Source Catalog (IPAC 2009). For SMC\,AB\,2, 4, \& 11 we also incorporated photometry from the WISE all sky survey \citep{Cutri2012a}. Additional photometry was used for SMC\,AB\,11 as described in Appendix\,\ref{sect:comments}.

\section{The analysis}
\label{sect:method}
 
For the spectral analysis we used synthetic spectra calculated with the PoWR model atmosphere code.
By fitting the model spectra to the observational data, we derive the fundamental parameters of the stars and their winds. 

\subsection{The models}
\label{sect:models}

PoWR is a state-of-the-art code for expanding stellar atmospheres. Its basic assumption is a spherically symmetric, stationary mass-loss. The models account for deviations from local thermodynamic equilibrium (non-LTE), wind inhomogeneities, and iron line blanketing. The radiative transfer equation is solved in the co-moving frame consistently with the rate equations for statistical equilibrium, while the energy conservation is also ensured. Some details on the numerical methods and the assumptions applied in the code can be found in \citet{Graefener2002} and \citet{Hamann2003,Hamann2004}. 

The inner boundary of the model atmospheres is located at a Rosseland continuum optical depth of $\tau_\mathrm{ross} = 20$, which per definition corresponds to the stellar radius $R_*$. The stellar temperature $T_*$ is related to this radius by the Stefan-Boltzmann law 
\begin{equation}
\label{eq:sblaw}
L = 4 \pi \sigma R_*^2 T_*^4~.
\end{equation} 
The stellar temperature thus represents an effective temperature at $\tau_\mathrm{ross} = 20$. 

In the subsonic part of the stellar atmosphere, the velocity field is calculated self-consistently to fulfill the hydrodynamic equation \citep{Sander2015}.
The velocity field in the supersonic part of the stellar wind is prescribed by a so-called $\beta$-law \citep{Castor1979,Pauldrach1986}. For most of the models calculated in the context of this work, the exponent $\beta$ is set to unity. Only in the case of SMC\,AB\,10 is a double-$\beta$ law \citep{Hillier1999,Graefener2005} in the form described by \citet{Todt2015} used since it gives a slightly better fit (for details see Appendix\,\ref{sect:comments}).

Wind inhomogeneities are accounted for by means of the so-called mircoclumping approach, which assumes optically thin clumps \citep{Hillier1991,Hamann1998} that fill a volume fraction $f_\mathrm{V}$, while the interclump medium is void. The density contrast between the clumps and an homogeneous model with the same mass-loss rate is given by the clumping factor $D = {f_\mathrm{V}}^{-1}$. The models used here account for a depth dependency of $D(r)$, where the clumping starts at the sonic point, increases outward, and reaches the full value of $D$ at 30\,\% of the terminal velocity.

The main wind parameters of a stellar atmosphere model are combined in the so-called transformed radius $R_\mathrm{t}$ \citep{Hamann1998} 
\begin{equation}
\label{eq:rt}
R_{\mathrm{t}} = 
R_* \left(\frac{\varv_\infty}{2500 \, \mathrm{km}\,\mathrm{s^{-1}}} 
\left/
\frac{\dot M \sqrt{D}}{10^{-4} \, M_\odot \, \mathrm{yr^{-1}}}\right)^{2/3}
\right.~,
\end{equation}
relating the mass-loss rate $\dot{M}$ with the terminal wind velocity $\varv_\infty$, the stellar radius $R_*$, and the clumping factor $D$. \citet{Schmutz1989} found that the equivalent width possess by emission lines in synthetic spectra is approximately independent from the specific value of the luminosity and the individual values of the wind parameters which enters $R_\mathrm{t}$ as long as the transformed radius, the stellar temperature, and the chemical composition remain unaltered. Therefore, models of WR stars can be easily scaled to different luminosities. According to Eq.\,(\ref{eq:rt}), the mass-loss rate has to be scaled proportional to $L^{3/4}$. Under the additional constraint of a constant $\varv_\infty$ also the line profiles are conserved. The scaling invariance for models of the same transformed radius can be understood from the fact that $R_\mathrm{t}^3$ is proportional to the ratio between the emission measure of the wind and the stellar surface area. 

The model calculations rely on complex model atoms of \element{H}, \element{He}, \element{N}, \element{C}, and in some cases also \element{O} and \element{P} (see Table\,\ref{table:model_atoms} for details). The iron group elements, comprising \element{Fe}, \element{Sc}, \element{Ti}, \element{Cr}, \element{Mn}, \element{Co}, and \element{Ni}, are treated in the ``superlevel approach'' \citep[see\ ][]{Graefener2002}.

\subsection{Abundances}
\label{sect:abundances}

The atmospheres of WN stars exhibit burning products of the CNO cycle because of envelope stripping due to the stellar winds and  internal mixing processes such as rotationally induced instabilities. Hence, the surface abundances of these objects are altered in comparison to that observed in unevolved stars such as B-type main sequence stars. Since the proton capture of \element[][14]{N} is the slowest reaction in the CNO cycle, most of the initial carbon and oxygen is converted into nitrogen. According to \citet{Schaerer1993}, the remaining carbon and oxygen is only $1/60$ of the nitrogen mass fraction.

\begin{table*}
\caption{Chemical composition (mass fractions in percent)} 
\label{table:abundances}      
\centering
\begin{tabular}{lSSSSSSS}
\hline\hline 
 & \multicolumn{1}{r}{Sun\tablefootmark{a}} 
 & \multicolumn{1}{c}{MW/M31} 
 & \multicolumn{1}{c}{LMC}
 & \multicolumn{4}{c}{\rule[1ex]{28mm}{0.5pt} SMC \rule[1ex]{28mm}{0.3pt} }
 \rule[0mm]{0mm}{2.2ex}\\
&& \multicolumn{1}{c}{WN\tablefootmark{b}} 
 & \multicolumn{1}{c}{WN\tablefootmark{c}} 
 & \multicolumn{1}{r}{B\,stars\tablefootmark{d}} 
 & \multicolumn{1}{r}{B\,stars\tablefootmark{e}} 
 & \multicolumn{1}{r}{\ion{H}{ii}\,\tablefootmark{f}}
 & \multicolumn{1}{r}{WN\tablefootmark{g}} \\
\hline 
C                   & 0.237  & 0.01                           & 0.0067                        & 0.02                          & 0.022 & 0.014   & 0.0025\rule[0mm]{0mm}{2.2ex} \\
N                   & 0.069  & 1.5                             & 0.40                           & 0.003                         & 0.018 & 0.0037 & 0.15 \\
O                   & 0.573  & \multicolumn{1}{c}{-} & \multicolumn{1}{c}{-} & 0.13                           & 0.17  & 0.12     & \multicolumn{1}{c}{-} \\
$\Sigma$CNO & 0.88    & 1.51                           & 0.41                           & 0.1323                       & 0.21   & 0.1377 & 0.1525\\
Fe                 & 0.129   & 0.14\tablefootmark{h} & 0.07\tablefootmark{h} & 0.035\tablefootmark{i}&0.027 &  \multicolumn{1}{c}{-}  & 0.03\tablefootmark{h} \\ 
\hline 
\end{tabular}
\tablefoot{
\tablefoottext{a}{\citet{Asplund2009}}
\tablefoottext{b}{as used in \citet{Hamann2006} and \citet{Sander2014}}
\tablefoottext{c}{as used in \citet{Hainich2014}}
\tablefoottext{d}{\citet{Hunter2007}}
\tablefoottext{e}{\citet{Korn2000}}
\tablefoottext{f}{\ion{H}{ii} regions \citep{Kurt1998}}
\tablefoottext{g}{as adopted in this work}
\tablefoottext{h}{including iron group elements}
\tablefoottext{i}{mean value from \citet{Trundle2007}}
}
\end{table*}

By default, we thus assume a nitrogen abundance which is roughly the sum of the observed CNO abundances for B-type stars and \ion{H}{ii} regions. 
For the initial composition of the SMC WN stars, we refer to stellar atmosphere analyses of B-type stars carried out by \citet{Korn2000}, \citet{Trundle2007} and \citet{Hunter2007}. Additionally, we take the chemical composition of \ion{H}{ii} regions in the SMC derived by \citet{Kurt1998} into account. As described above, we calculate the chemical composition of our grid models from the averaged abundances obtained by these authors. 
Accordingly, we used a nitrogen abundance of $X_{\element{N}} = 1.5 \cdot 10^{-3}$ (mass fraction) and a carbon abundance of $X_{\element{C}} = 2.5 \cdot 10^{-5}$. The iron abundances is taken from studies conducted by \citet{Korn2000} and \citet{Trundle2007}, resulting in an averaged value of $X_{\element{Fe}} = 3 \cdot 10^{-4}$. These basic model abundances, as well as the observed values obtained by the former authors, are listed in Table\,\ref{table:abundances}. 

For comparison, Table\,\ref{table:abundances} also includes the abundances used by \citet{Hamann2006} and \citet{Hainich2014} for the analyses of the MW and LMC WN populations, respectively. The CNO abundance of the SMC grid models is almost a tenth of the solar value, while the iron abundance is approximately a fifth. Even compared to the LMC, the SMC CNO and iron abundances are lower by more than a factor of two. 
We note that the abundances for the SMC WN stars reflect the basic composition of the grid models. The nitrogen abundance for all analyzed stars was individually adjusted during the fitting procedure. For those stars which show oxygen lines in their spectra oxygen was also included in the model calculation. The individual abundances derived in our analyses can be found in Tables\,\ref{table:parameters} and \ref{table:derived_abundances}. We note that the values derived for the \element{CNO} and \element{P} abundances are uncertain by up to a factor of two. 
The error in the derived hydrogen abundances is on the order of 5\,\%.

\subsection{Spectral fitting}
\label{subsec:spefit}

\begin{figure*}
\centering
\includegraphics[width=\hsize]{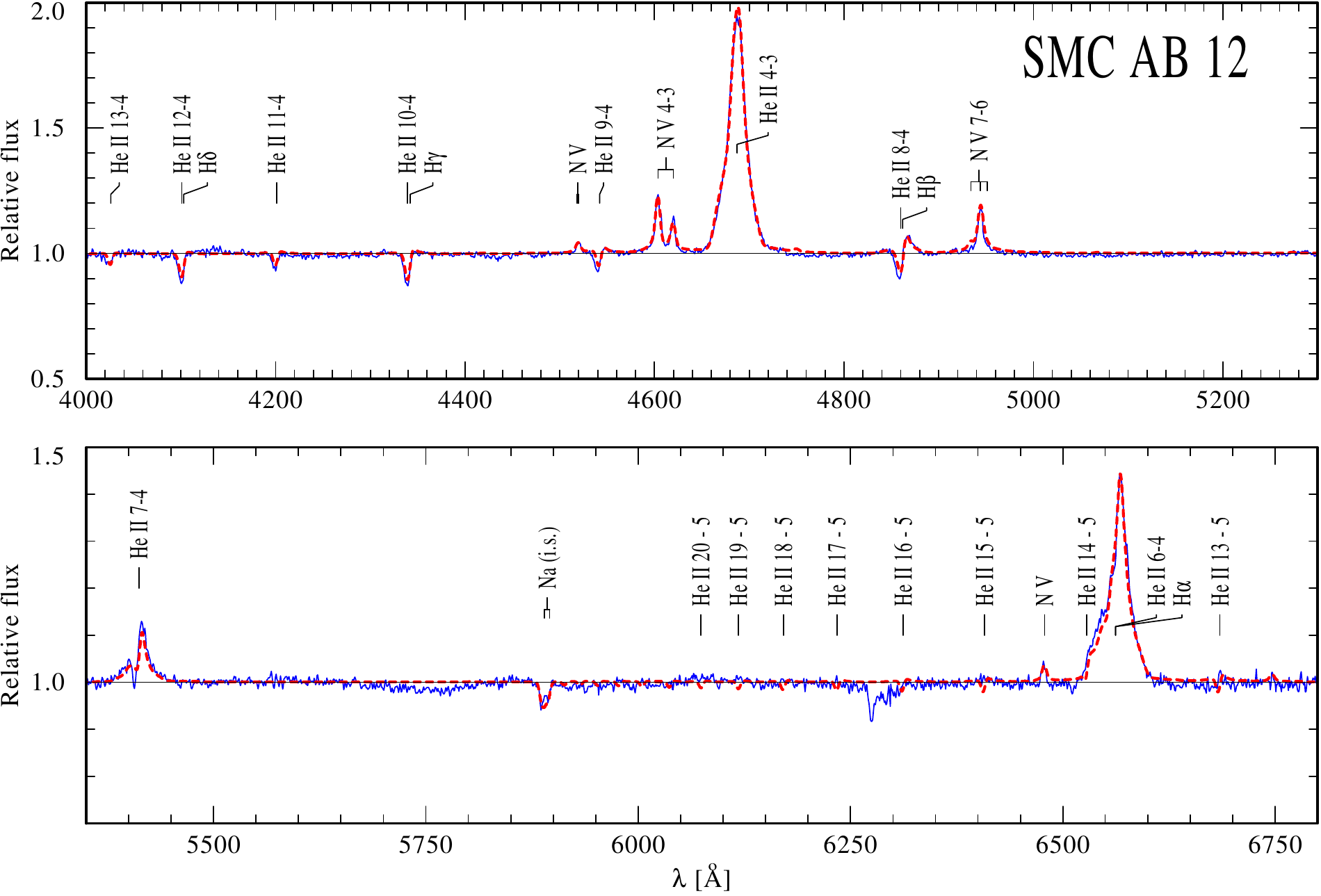}
\caption{Normalized line spectrum of SMC\,AB\,12 in comparison to the a PoWR model. The observation is shown by a thin (blue) solid line, whereas the best fitting model spectrum is overplotted by a thick (red) dashed line} 
\label{fig:linefit}
\end{figure*}

\begin{figure*}
\centering
\includegraphics[width=\hsize]{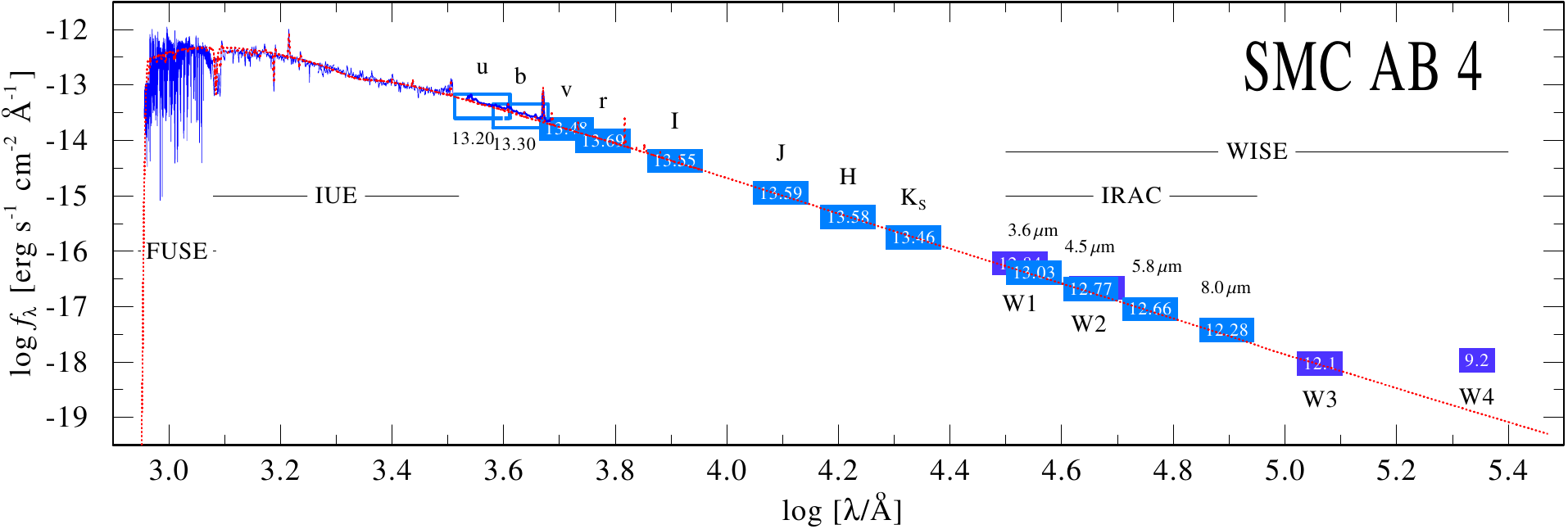}
\caption{The observed spectral energy distribution of SMC\,AB\,4, composed of flux calibrated spectra (FUSE, IUE, CTIO; blue thin lines) and photometry (blue boxes, labeled with magnitudes), compared to the model continuum (red dashed line) of the best fitting model. The model flux is corrected for interstellar extinction and accounts for the geometrical dilution due to the distance of the SMC.} 
\label{fig:sed}
\end{figure*}
 
The spectral analysis is based on three grids of stellar atmosphere models. The free parameters of these grids are the stellar temperature $T_*$ and the transformed radius $R_\mathrm{t}$, while the grids are distinguished by hydrogen abundances of 0.2, 0.4 and 0.6 (mass fraction). For further details on the grid spacing we refer the reader to \citet{Hamann2004}. The model grids are publicly available on the PoWR homepage\footnote{\url{www.astro.physik.uni-potsdam.de/PoWR.html}}. Suitable models are preselected by means of contour plots \citep{Sander2012} or a $\chi_{\nu}^2$-fitting technique \citep{Todt2013}. Subsequently, individual models with refined stellar parameters and abundances are calculated for each object in our sample. 

A typical example for a normalized line fit is shown in Fig.\,\ref{fig:linefit}. As mentioned in Sect.\,\ref{sect:intro}, distinct absorption lines are visible in the spectra of most SMC WN stars. These features are characterized by asymmetric line profiles and a strong blue-shift \citep{Foellmi2003a} that does not vary in time, indicating the severe influence of the stellar wind on these lines. With the exception of a few lines ($\ion{He}{ii}\,\lambda\lambda4200,\,4542$\AA\ in the spectra of SMC\,AB\,1 and 12), the absorption line component as well as the relatively weak emission-line component of the SMC WN-star spectra can be simultaneously reproduced by the models, illustrating that weak absorption lines do not necessarily imply binarity \citep[see also][]{Hamann2006}. 

The stellar temperatures of the model atmospheres are adjusted until the observed \ion{He}{I}\,/\,\ion{He}{II} and \ion{N}{III}\,/\,\ion{N}{IV}\,/\,\ion{N}{V} line ratios are adequately reproduced. For SMC\,AB\,2 and 4, this procedure leaves a temperature uncertainty of approximately $\pm 3\,\mathrm{kK}$. For the other stars in our sample, the error in the temperature might be higher because the available spectra for these objects only possess lines of one ionization stage per element. The individual error for each object will be discussed in Appendix\,\ref{sect:comments}. 

The transformed radius is adjusted such that the observed strengths of the emission lines are reproduced by the model. The error margin of $R_\mathrm{t}$ is on the order of $\pm 0.05\,\mathrm{dex}$, with the exception of SMC\,AB\,4, where it is twice as large. This translates to an uncertainty of the mass-loss rate of $\pm 0.075\,\mathrm{dex}$ ($\pm 0.15\,\mathrm{dex}$ for AB\,4). According to Eqs.\,(\ref{eq:sblaw}) and (\ref{eq:rt}), the uncertainty in the stellar temperature corresponds to an additional uncertainty in the mass-loss rate of $\pm0.025\,\mathrm{dex}$ for the cooler stars in our sample, while it is twice as large for the hotter objects. 
The terminal velocity and the degree of clumping are derived from the width of the emission lines and the strength of the electron scattering wings \citep{Hillier1991,Hamann1998}, respectively.
The accuracies in the determination of the terminal velocity and the clumping factor are on the order of 10\,\% and 50\,\%, respectively. 
These uncertainties contribute to the error margin of the mass-loss rate ($\dot{M} \propto v_\infty D^{-1/2}$), which is at most 0.2\,dex.

Simultaneously with the line fit, the simulated spectral energy distribution (SED) is fitted to flux-calibrated spectra and photometry in order to determine the luminosity and the color excess $E_{b-v}$ (see Fig.\,\ref{fig:sed} for an example). For this purpose the model flux is geometrically diluted according to a distance modulus of $18.9\,\mathrm{mag}$ \citep{Westerlund1997,Harries2003,Hilditch2005,Keller2006}. The reddening encompassed contributions from an internal SMC component and the Galactic foreground. For the latter we used the reddening law published by \citet{Seaton1979} and a color excess of $E_{b-v} = 0.05\,\mathrm{mag}$, which is the mean reddening derived by \citet{Grieve1986}, \citet{Bessell1991}, \citet{Schwering1991}, and \citet{Schlegel1998}. For the SMC component, the color excess is a free parameter that is individually adjusted, while the reddening law is adopted from the work of \citet{Gordon2003}.

The error margin of the luminosity has a couple of contributions: the uncertainty in the color excess, which is relatively small for the SMC stars; the uncertainty in the temperature determination; the approximated treatment of the ISM extinction; and the error inherent to the photometry and the flux calibrated spectra. Since the flux of WN stars scales approximately linear with the stellar temperature in the optical and IR (Rayleigh-Jeans domain), 
the temperature uncertainty contributes $0.03\,\mathrm{dex}$ to the luminosity error of the WN\,5--6 stars and about $0.1\,\mathrm{dex}$ to the error in the luminosity of the WN\,3--4 stars.
The uncertainty in the SED fit results in a luminosity error of approximately $0.05\,\mathrm{dex}$ for those objects where both flux-calibrated spectra and photometry is available, while it is twice as high for SED fits that are based on photometry alone. Therefore, the total error margin in the luminosity amounts to $0.1\,\mathrm{dex}$ for the cooler objects and $0.2\,\mathrm{dex}$ for the hotter objects without flux calibrated spectra.

\section{Results}
\label{sect:results}

\subsection{Stellar parameters}
\label{sect:parameters}

The basic stellar parameters derived for the SMC WN stars are compiled in Table\,\ref{table:parameters}. Furthermore, the number of hydrogen and helium  ionizing photons emitted by these objects and the associated Zanstra temperatures are given in Table\,\ref{table:zansT}. 

\begin{figure}
\centering
\includegraphics[angle=-90,width=\hsize]{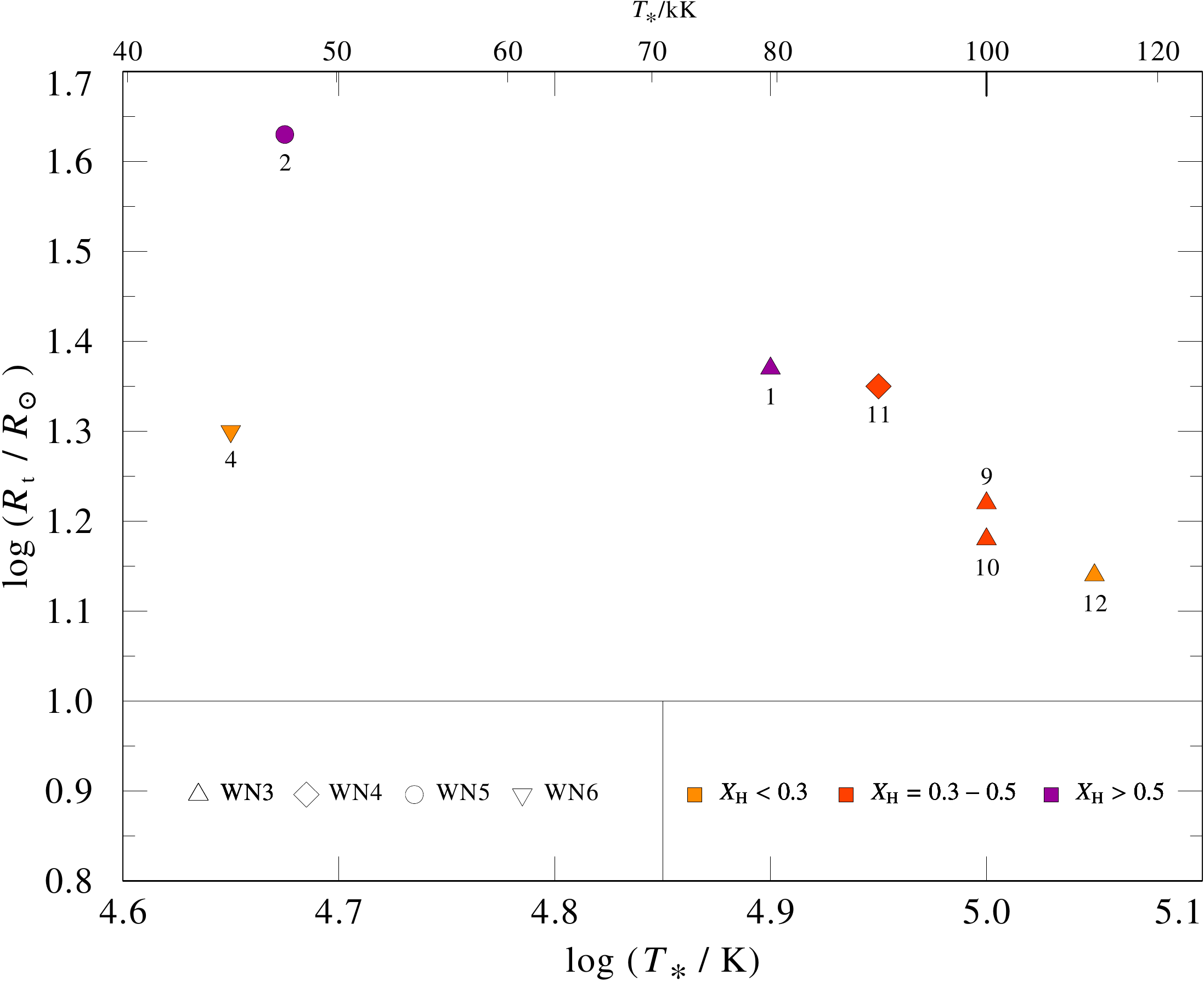}
\caption{The position of the analyzed SMC WN stars in the $\log T_*$-$\log R_\mathrm{t}$-plane. The symbols reflect the WN subtype (see inlet). The hydrogen abundances is indicated in three steps by means of the filling color (purple: $X_{\element{H}} \ge 50\,\%$, red: $X_{\element{H}} = 30 - 50\,\%$, orange: $X_{\element{H}} \le 30\,\%$).} 
\label{fig:wnstats}
\end{figure}

In Fig.\,\ref{fig:wnstats}, we show the location of our program stars in the $\log T_*$-$\log R_\mathrm{t}$-plane. Although the number of objects is rather small, a clear one-dimensional sequence is recognizable with SMC\,AB\,4 being the only outlier. In relation to the other SMC stars, the position of the latter object in this diagram implies a significantly denser wind and higher mass-loss rate. Again with the exception of SMC\,AB\,4, the surface hydrogen abundance of the SMC WN stars gradually decreases with increasing stellar temperature. Compared to their Galactic and LMC counterparts, the WN stars in the SMC have on average higher stellar temperatures, which in turn reflects their subtype distribution. 
 
In Table\,\ref{table:parameters}, we also give the wind efficiency, which is defined as the wind momentum divided by the momentum of the radiation field:
\begin{equation}
\label{eq:eta}
\eta = \frac{\dot{M} v_\infty c}{L}~.
\end{equation}
With their comparably weak winds, all WN stars in the SMC have wind efficiencies below unity. They thus clearly fall below the single scattering limit. In strong contrast, the early-subtype WN stars (WNE) have averaged wind efficiencies of 2.1 in the LMC \citep{Hainich2014} and 5.9 in the WM \citep{Hamann2006}. Hence they clearly exceed the so-called single scattering limit and enter the multi scattering domain, where photons need to be scattered $\eta$ times on average to drive the wind. 

Except for the weakness of their winds (see\ Sect.\,\ref{subsect:mdot}), the most intriguing property of all WN stars in the SMC is the occurrence of hydrogen in their atmospheres. For SMC\,AB\,1,\,2, and 4 this fact was already noticed by \citet{Conti1989}. Later, \citet{Foellmi2003a} and \citet{Foellmi2004} showed that all putatively single WN stars in the SMC possess hydrogen at their surface. Our spectral analysis quantifies these findings, yielding hydrogen abundances in the range of $X_{\element{H}} = 0.2-0.55$ (by mass). 
These values can be well understood in terms of weak stellar winds that were not able to completely remove the hydrogen rich layers of the stellar atmospheres of these stars. 

Early type WN stars that exhibit a high stellar temperature, hydrogen at the stellar surface, and an inherent absorption line component in their spectra can also be found in the LMC, although they are rare. Two examples are BAT99\,25 and BAT99\,74. Especially the latter exhibits striking similarities with SMC\,AB\,1, which is, however, more luminous by a factor of two. For comparison, no early type WN stars with absorption lines and a significant amount of hydrogen are known in the MW, indicating that metallicity is decisive for the evolution of massive stars (see Sect.\,\ref{subsubsect:mdot-z} and \ref{sect:evolution}). In this context, BAT99\,25 and BAT99\,74 might represent objects with initial metallicities significantly below the standard LMC metallicity, explaining their spectral appearance and their unusually low mass-loss rates. This concept is corroborated by SMC\,AB\,4. In contrast to the other SMC WN stars, AB\,4 exhibits spectral characteristics and a mass-loss rate that reflect rather typical LMC values, suggesting that the progenitor was more metal rich. 

Alternatively to the scenario stated above, interaction with a close stellar companion might explain the abundance pattern seen in some of these stars. Mass transfer via Roche-lobe overflow is able to efficiently remove parts of of the hydrogen rich atmosphere of the donor star, resulting in a WR star with a rejuvenated companion. This could explain, e.g., the relatively low hydrogen abundance derived for SMC\,AB\,4 with respect to its late spectral type. However, the study conducted by \citet{Foellmi2003a} revealed that the WN stars discussed in this paper do most likely not have a close companion, rendering it unlikely that these objects have been the donor stars in a binary system. Of course, more complex binary interactions might have happened. For example, some of the SMC WN stars could have been mass gainers in a binary interaction, while the system was disrupted by the supernova (SN) of a previous primary. In this case, however, it remains to be proven that the rejuvenated object can evolve into a WR star, which might be questionable at SMC metallicities. 

\begin{figure}
\centering
\includegraphics[width=\hsize]{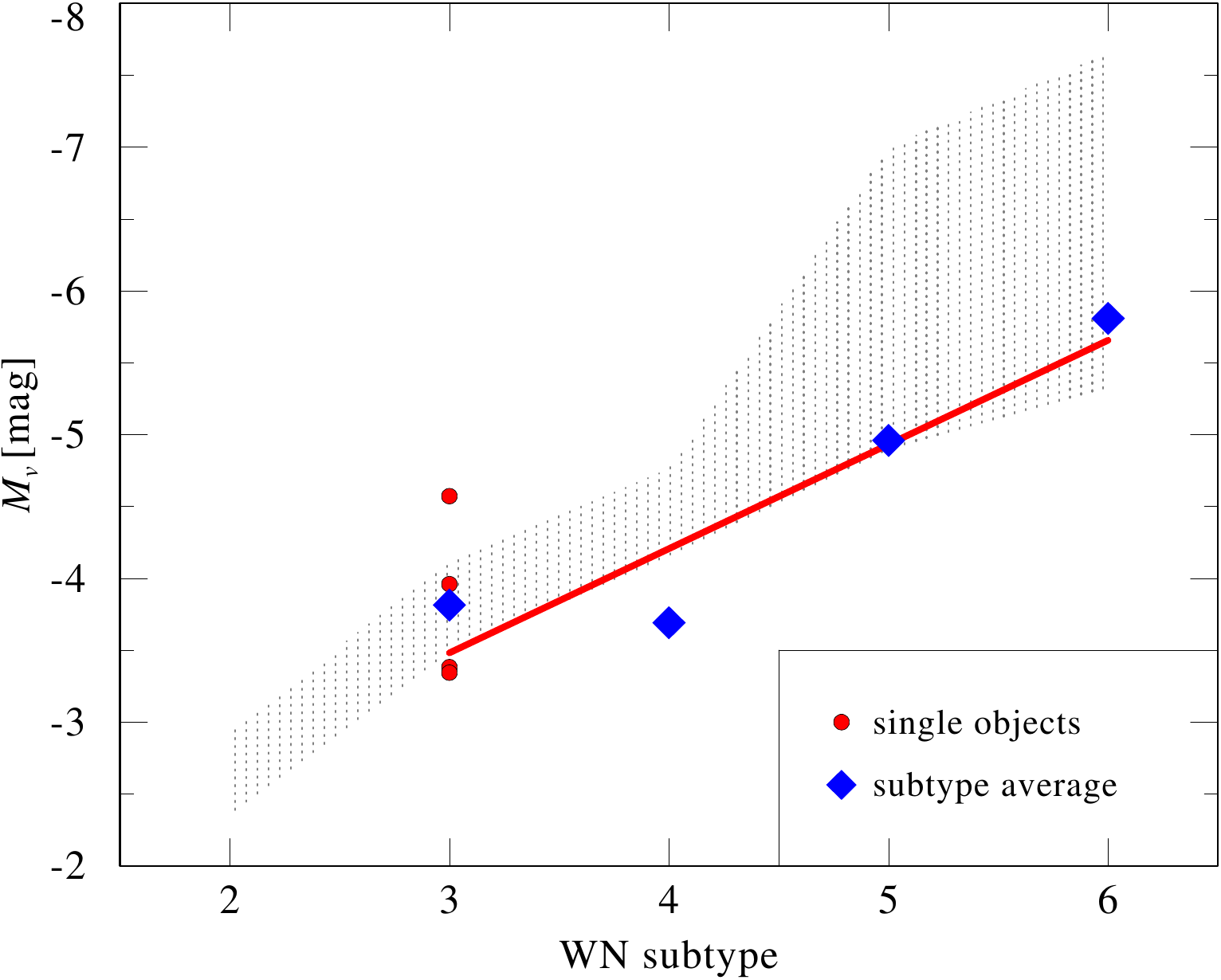}
\caption{Absolute monochromatic $v$-magnitudes \citep{Smith1968} of the SMC WN stars plotted over subtype number. Mean magnitudes for each subtype are represented by thick blue symbols, while red dots refer to individual objects. The straight red line is the linear regression to the SMC stars. The hashed area represents the scatter in the LMC correlation published by \citet{Hainich2014}.} 
\label{fig:mv_sptype_lmc}
\end{figure}

\begin{figure}
\centering
\includegraphics[width=\hsize]{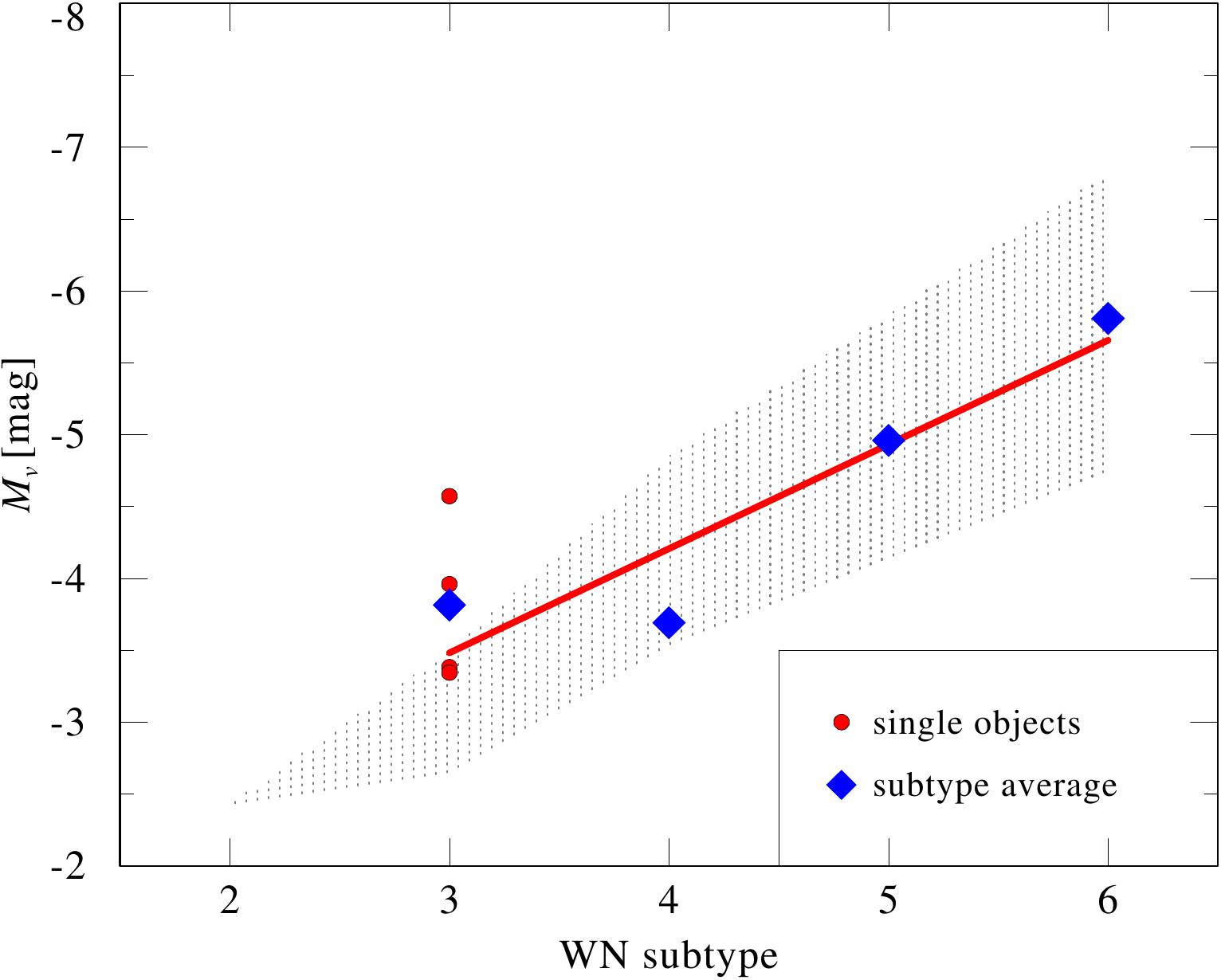}
\caption{Same as Fig.\,\ref{fig:mv_sptype_lmc}, but the hashed area refers to the scatter in the Galactic relation.} 
\label{fig:mv_sptype_gal}
\end{figure}

In Figs.\,\ref{fig:mv_sptype_lmc} and \ref{fig:mv_sptype_gal}, we plot the absolute visual magnitudes \citep[monochromatic colors as defined by][]{Smith1968} of our sample stars versus their spectral subtypes (WN\,2--WN\,6). The hashed areas in these figures refer to the corresponding relations for the LMC \citep{Hainich2014} and MW \citep[accounting only for the stars with known distances,][]{Hamann2006}, respectively. Although the SMC stars appear to be slightly brighter in the visual than the LMC stars, there is a fairly good agreement between the results from the three galaxies, indicating a universal correlation between $M_{v}$ and spectral type. This is quite astonishing, given that the WNE stars from individual galaxies appear to be very different, e.g., most WN\,2 to WN\,4 stars in the LMC and MW are hydrogen-free, while all their SMC counterparts exhibit hydrogen at the stellar surface. 

For three stars in our sample (SMC\,AB\,1, 2, 4), spectral analyses by means of stellar atmosphere models were conducted by \citet{Crowther2000} and \citet{Martins2009}. In general, the basic stellar and wind parameters deduced in our analysis agree well with the results published by these authors. The only exception is SMC\,AB\,1. For this object, we obtain a significantly higher stellar temperature and consequently also a higher luminosity (see Appendix\,\ref{sect:comments} for details) than the values derived by \citet{Martins2009}. 
\citet{Nugis2007} computed the luminosity for six stars in our sample by means of an adopted bolometric luminosity of $4.74\,\mathrm{mag}$ and bolometric corrections derived by \citet{Nugis2000}. In Table\,\ref{table:comp_mdot}, we compare these estimates with the luminosity derived from our elaborated SED fits. Except for SMC\,AB\,2, the luminosities obtained by \citet{Nugis2007} are significantly lower.

\begin{table*}
\small
\caption{Parameters of SMC WN stars }
\label{table:parameters}
\centering  
\begin{tabular}{clcScccScccccll}
\hline \hline \rule[0mm]{0mm}{3.0mm}   
    SMC\,AB &
    Subtype\tablefootmark{a} &
    $T_*$ &
    $\mathrm{log}\, R_\mathrm{t}$ &
    $\varv_{\infty}$ &
    $E_{b-v}$ &
    $M_{v}$ &
    $R_*$ &
    $\log \dot M$ &
    $D$ &
    $\log L$ &
    $\eta$ &
    $M$\tablefootmark{b} &
    $X_{\element{H}}$ &
    Comments \\
    &
    &
    $[\mathrm{kK}]$ &
    $[R_{\odot}]$ &
    $[\mathrm{km/s}]$ &
    $[\mathrm{mag}]$ &
    $[\mathrm{mag}]$ &
    $[R_{\odot}]$ &
    $[M_{\odot}/\mathrm{yr}]$ &
    &
    $[L_{\odot}]$ &
    &
    $[M_{\odot}]$ &
    & \\
\hline  \rule[0mm]{0mm}{4.0mm}  
\input{table-parameters}
\hline
\end{tabular}
\tablefoot{
\tablefoottext{a}{The subtype classifications follow \citet{Foellmi2003a} and \citet{Foellmi2004}}
\tablefoottext{b}{Masses calculated from the luminosity, using the mass–luminosity relation derived by \citet{Graefener2011}}
}
\end{table*}

\subsection{The Hertzsprung-Russell diagram}
\label{subsect:hrd}

\begin{figure}
\centering
\includegraphics[width=\hsize]{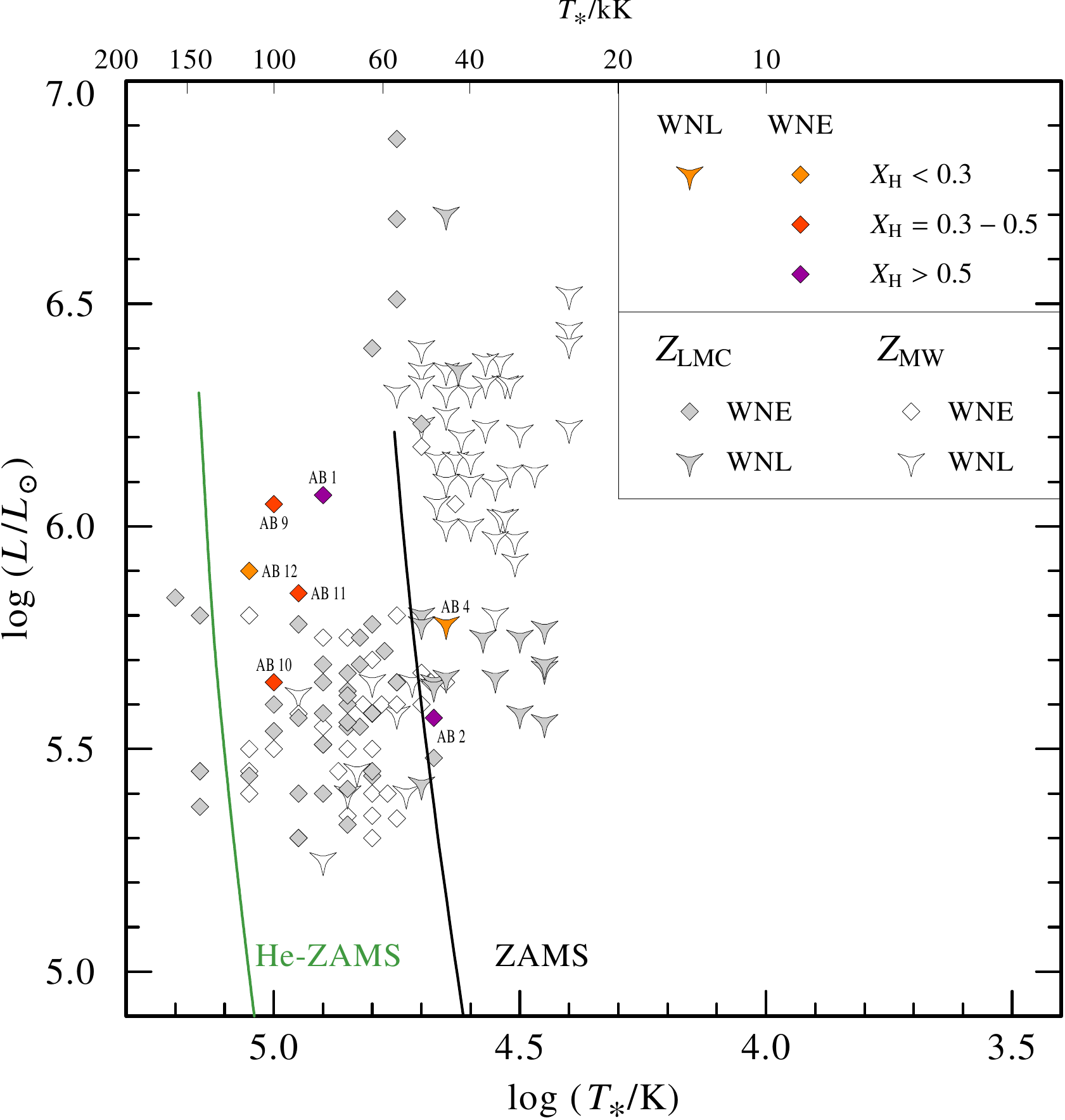}
\caption{The HRD of the single WN stars. The color-filled symbols represent the WN stars in the SMC analyzed in the present paper. The filling color of these symbols reflects the derived atmospheric hydrogen abundance (see inlet). The gray filled symbols and the open symbols refer to the LMC WN stars and the Galactic WN stars, respectively.} 
\label{fig:hrd_compare}
\end{figure}

The Hertzsprung-Russell diagram (HRD) of the single WR stars in the SMC is shown in Fig.\,\ref{fig:hrd_compare}. For comparison, the HRD positions obtained by \citet{Hainich2014} for the LMC WN stars and by \citet{Hamann2006}, \citet{Martins2008}, \citet{Liermann2010}, and \citet{Barniske2008} for the Galactic WN stars are also reproduced by gray and open symbols, respectively. For the SMC stars, the hydrogen abundance, as derived in this analysis, is color-coded. 

With the exception of SMC\,AB\,2, all SMC WNE stars are located on the hot side of the zero-age main sequence (ZAMS). These objects, while containing a considerable amount of hydrogen in their stellar atmosphere, are located in a temperature range ($\geq 79\,\mathrm{kK}$), where almost none of their LMC and Galactic counterparts show a measurable amount of hydrogen. In addition, these hot stars (except of SMC\,AB\,10) exhibit luminosities of $\log\,(L/L_\odot) \geq 5.85$ and thus occupy a parameter space where neither MW nor LMC WR stars can be found. Therefore, despite the low number statistics, one might conclude that the WNE stars in the SMC decisively differ from the Galactic and LMC population. In contrast, the HRD position and hydrogen abundance of SMC\,AB\,4, the only WNL star in the SMC, roughly corresponds to the WNL population in the MW and the LMC. 

None of the putatively single WN stars in the SMC has a luminosity of $\log\,(L/L_\odot) \leq 5.6$, in sharp contrast to the WN population in the MW and the LMC. This finding promotes a significantly higher minimum initial mass for the WR stars in the SMC, in line with theoretical expectations for its low metallicity (see Sect.\,\ref{sect:evolution}). Interestingly, this effect is not seen between the MW and LMC \citep{Hainich2014}, although the difference in their metallicity content is also not negligible.

\subsection{Stellar wind mass-loss}
\label{subsect:mdot}

As described above, the emission lines in SMC WR stars are systematically weaker than their Galactic and LMC counterparts \citep{Conti1989,Crowther2006b}. \citet{Crowther2006} and \citet{Nugis2007} studied a subset of the WN stars in the MW, LMC, and SMC, showing that the weak emission lines of the SMC WN stars correspond to a reduced mass-loss rate. The present study confirms these findings. All our sample stars have mass-loss rates of about $\log\,(\dot M / (\mathrm{ M_\odot}/\mathrm{yr})) \approx -5.6$, with the exception of SMC\,AB\,4, which displays a mass-loss rate roughly a factor of three higher than any other WN star in the SMC. 

In Table\,\ref{table:comp_mdot}, we compare the mass-loss rates obtained from our detailed spectral line fits with the results presented by \citet{Nugis2007}. These authors derived the mass-loss rates for six stars in our sample using the equivalent width of the \ion{He}{ii}\,$\lambda\,4686$ line and an empirically scaled formula. This comparison reveals a reasonable agreement, exhibiting the largest deviations for SMC\,AB\, 1 and 2, where the values presented by \citet{Nugis2007} are larger by about 50\,\%. 

\begin{figure}
\centering
\includegraphics[width=\hsize]{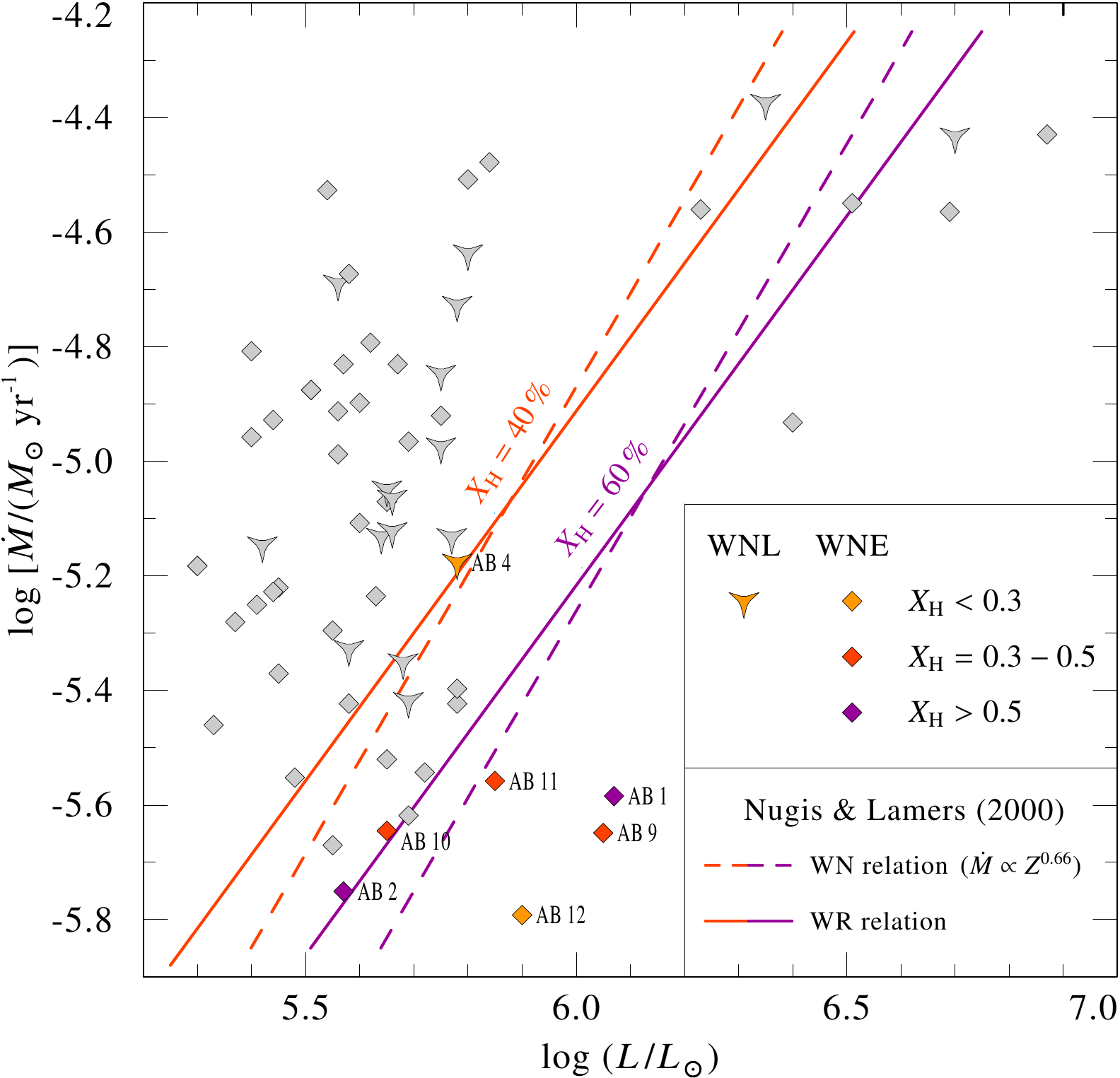}
\caption{The mass-loss rates as a function of the luminosity of the WN stars in the SMC as well as the LMC for the sake of comparison. The meaning of the symbols is the same as in Fig\,\ref{fig:hrd_compare}. 
The dashed lines refer to the mass-loss relation published by \citet{Nugis2000} for Galactic WN stars, which is scaled according to a metallicity dependence of $\dot{M} \propto Z^{0.66}$ \citep{Vink2005}. The straight lines are the general mass-loss prescription for WR stars given by \citet{Nugis2000}. Both relations are plotted for two surface hydrogen abundances. 
} 
\label{fig:mdot_comp}
\end{figure}

The mass-loss rates of our SMC WN sample and of the LMC WN stars \citep{Hainich2014} are compared in Fig.\,\ref{fig:mdot_comp}. 
While the SMC stars are on average slightly more luminous than the bulk of the LMC stars, their mass-loss rates are, on average, lower by more than a factor of four.
In Fig.\,\ref{fig:mdot_comp} we also plot the mass-loss relations given by \citet{Nugis2000} for WN stars (their Eq.\,20) and for WR stars in general (their Eq.\,22). These relations are based on mass-loss rates determined from radio emissions and emission line equivalent widths, accounting for the effect of clumping in both cases. While the universal WR calibration published by \citet{Nugis2000} intrinsically accounts for a metallicity dependence, the WN relation does not. We therefore scaled their WN relation according to a metallicity dependence of $\dot{M} \propto Z^{0.66}$ predicted by \citet{Vink2005}. Both prescriptions are shown for surface hydrogen abundances of $40\,\%$ and $60\,\%$ (by mass) in Fig.\,\ref{fig:mdot_comp}. 
The individual values predicted for our program stars can be found in Table\,\ref{table:comp_nugis}. The third and fifth column of this Table give the factor between the mass-loss rates predicted and the results obtained from our detailed model fit. In general, the latter values are considerably lower than the prediction of \citet{Nugis2000}, with SMC\,AB\,2 being the only exception. We note that the disagreement tends to be larger for lower surface hydrogen abundance.
Such discrepancies might imply that the relation suggested by \citet{Nugis2000} is oversimplified.

\begin{table}
\caption{Mass-loss rates predicted by the relations from \citet{Nugis2000}}
\label{table:comp_nugis}
\setlength{\tabcolsep}{4.6pt}
\centering  
\begin{tabular}{cSSSS}
\hline\hline  \rule[0mm]{0mm}{4.0mm} 
   &
   \multicolumn{2}{c}{WN relation\tablefootmark{a}} &
   \multicolumn{2}{c}{WR relation} 
   \\
   SMC\,AB &
   \multicolumn{1}{c}{$\log \dot{M}_\mathrm{N\,\&\,L}$} & 
   \multicolumn{1}{c}{$\dot{M}_\mathrm{N\,\&\,L}/\dot{M}$\tablefootmark{b}} & 
   \multicolumn{1}{c}{$\log \dot{M}_\mathrm{N\,\&\,L}$} & 
   \multicolumn{1}{c}{$\dot{M}_\mathrm{N\,\&\,L}/\dot{M}$\tablefootmark{b}}
   \\
    &
    $[M_{\odot}/\mathrm{yr}]$ &
    &
    $[M_{\odot}/\mathrm{yr}]$ &
   \\
\hline \rule[0mm]{0mm}{4.0mm} 
   1   &  -4.93     &  4.5     &  -4.96     &  4.2          \\
   2   &  -5.85     &  0.8     &  -5.68     &  1.2          \\
   4   &  -5.01     &  1.5     &  -5.03     &  1.4          \\
   9   &  -4.71     &  8.7     &  -4.79     &  7.2          \\
   10  & -5.36     &  1.9     &  -5.3       &  2.2          \\
   11  & -5.11     &  2.8      &  -5.11     & 2.8          \\
   12  & -4.56     &  17.0    &  -4.83     &  9.1           \\
\hline 
\end{tabular}
\tablefoot{
\tablefoottext{a}{scaled according to a metallicity dependence of $Z^{0.66}$ \citep{Vink2005}}
\tablefoottext{b}{ratio between the mass-loss rates predicted by \citet{Nugis2000} and the values derived in our analysis}
}
\end{table}

\subsubsection{Wind-momentum luminosity relationship}

The ``modified wind momentum'' is a measure for the strength of the stellar wind \citep{Kudritzki1995,Puls1996,Kudritzki1999} that is defined as 
\begin{equation}
\label{eq:dmom}
D_\mathrm{mom} = \dot{M} v_\infty R_*^{1/2}~.
\end{equation}
The theory of radiation driven winds predicts a tight relation of the form ${D_\mathrm{mom}}^\gamma \propto L$, the so-called wind-momentum luminosity relationship \citep[WLR,][]{Kudritzki1999}.

\begin{figure}
\centering
\includegraphics[width=\hsize]{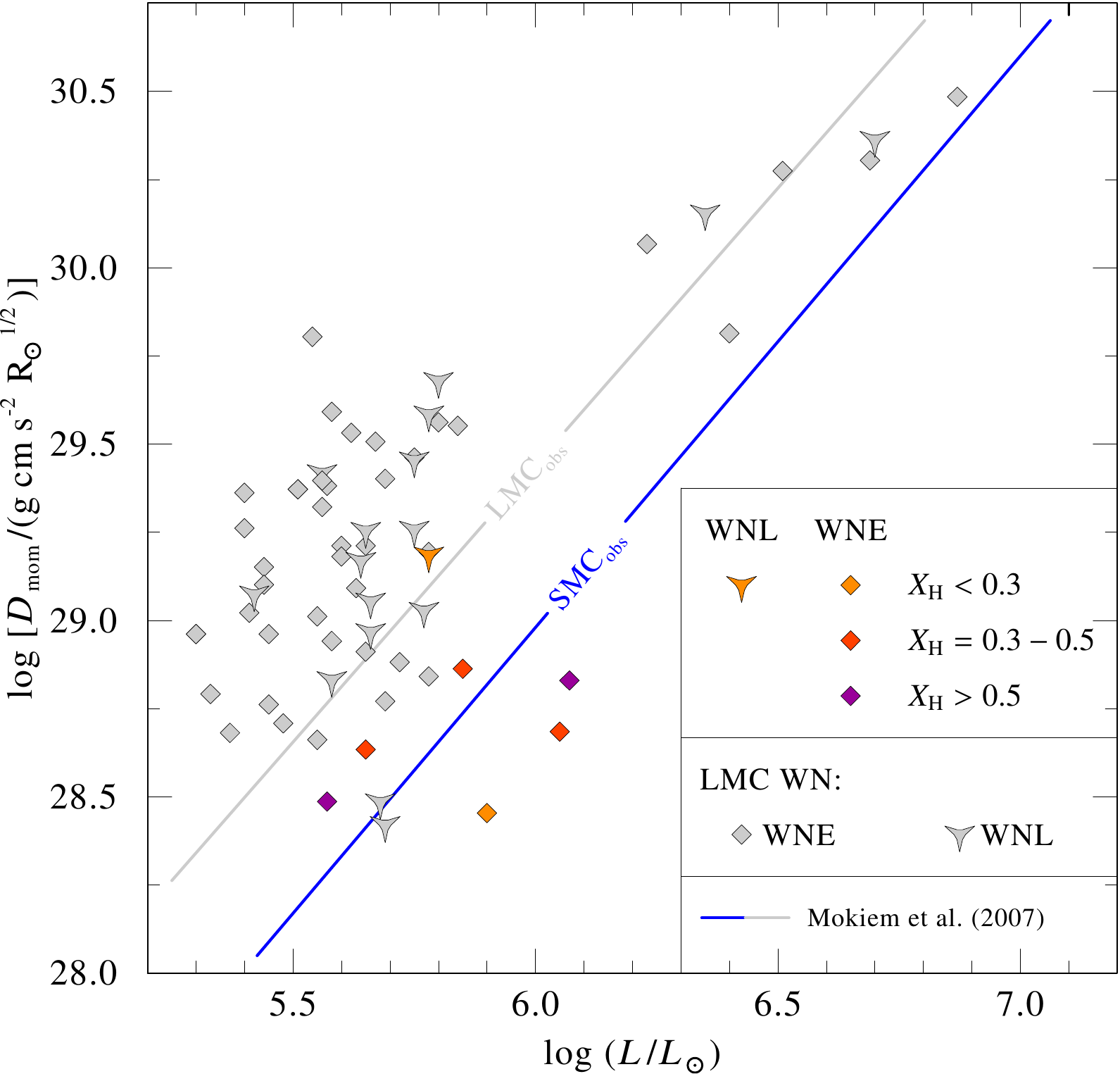}
\caption{The modified wind momentum $D_\mathrm{mom}$ as a function of the luminosity for the putatively single WN stars in the SMC (same symbols as in Fig.\,\ref{fig:hrd_compare}). For comparison the LMC WN stars (gray filled symbols) analyzed by \citet{Hainich2014} also shown. The blue and gray straight line is the clumping corrected wind luminosity relation for OB-type stars in the SMC and LMC, respectively, derived by \citet{Mokiem2007}.} 
\label{fig:dmom-l}
\end{figure}

In Fig.\,\ref{fig:dmom-l}, we plot the modified wind momenta versus the luminosities of the WN stars in the SMC and its LMC counterparts \citep{Hainich2014}. Furthermore, we show the empirical WLRs for OB-type stars derived by \citet{Mokiem2007}, using clumping corrected $D_\mathrm{mom}$ values for stars with H$\alpha$ emission. As expected from the theory of line driven winds, the WN stars in the SMC exhibit on average a lower wind momentum than their higher metallicity counterparts. This difference amounts to 0.45\,dex for those objects with similar luminosities.
Interestingly, the WN stars in the SMC posses modified wind momenta that are in the range of what is expected for equally luminous O-type stars.
In contrast to the SMC stars, most of the LMC WN stars exhibit winds which are significantly stronger than the wind of equally luminous O-type stars (on average almost a factor of six). Thus, the modified wind momentum supports the picture that the wind of the WN stars in the SMC are considerably weaker than their counterparts in the LMC. 

\subsubsection{Metallicity dependence}
\label{subsubsect:mdot-z} 

As highlighted in Sect.\,\ref{sect:intro}, it is expected that the mass-loss rate of massive stars varies as a function of the metallicity $Z$. By applying a modified CAK approach \citep{Castor1975,Abbott1982,Pauldrach1986} on O-type stars with Magellanic-Cloud metallicities, \citet{Kudritzki1987} derived a scaling relation of the form 
\begin{equation}
\label{eq:mdotz}
\dot{M} \sim Z^{\alpha}
\end{equation}
with $\alpha = 0.5$. \citet{Vink2000,Vink2001} obtained an exponent of $\alpha = 0.69$ by means of a Monte-Carlo approach to the wind dynamics. This theoretical prediction is supported by empirical studies \citep[e.g.,][]{Bouret2003,deKoter2006,Mokiem2007}. For WR-stars, the situation is less clear since only a limited number of observational constraints \citep{Crowther2006,Nugis2007,Hainich2014} and even fewer theoretical predictions \citep{Vink2005,Graefener2008} are available. 

\begin{figure}
\centering
\includegraphics[width=\hsize]{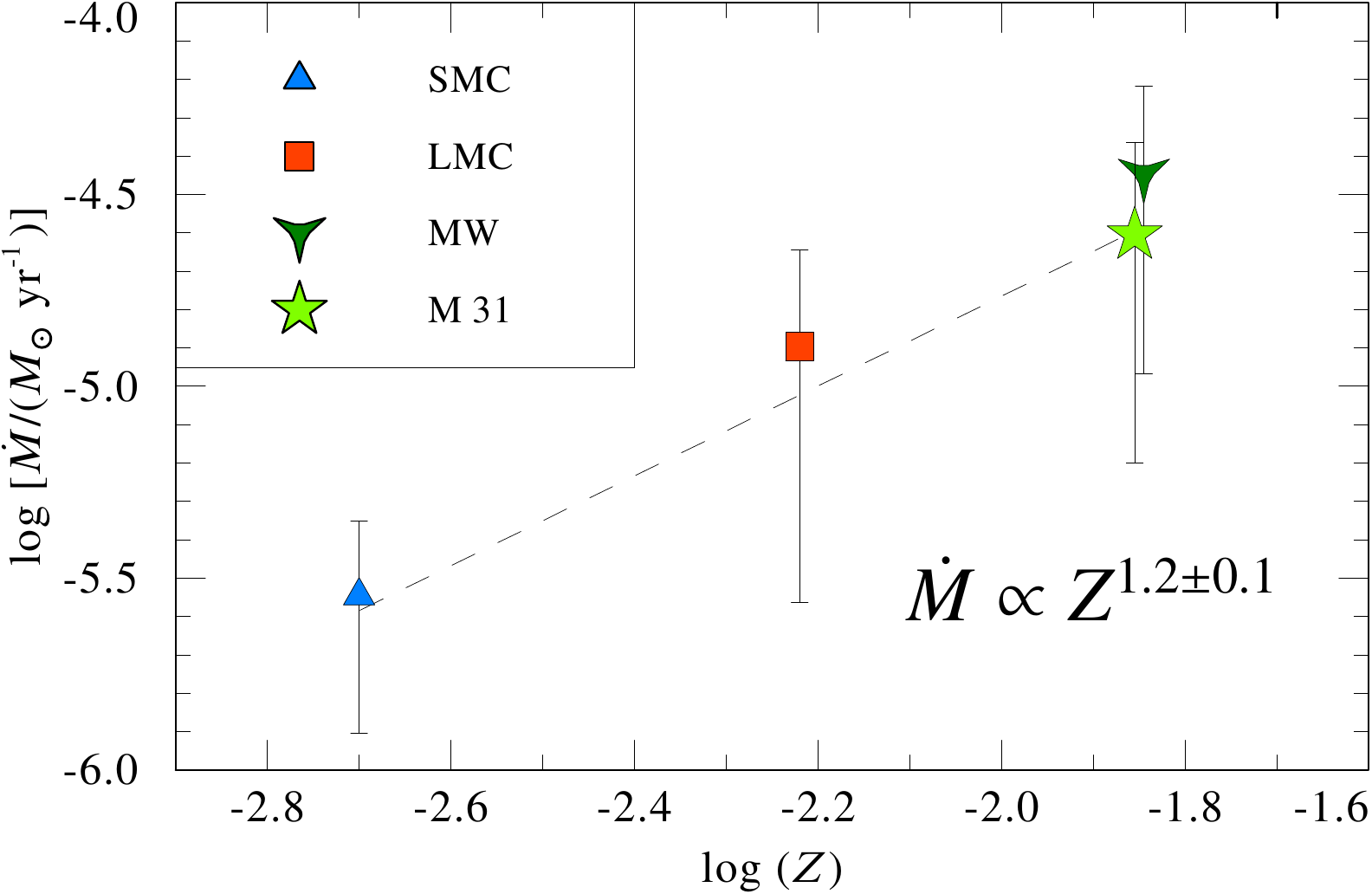}
\caption{The averaged mass-loss rates of the WN stars in the SMC (blue), the LMC (red), the MW (dark green), and the late type WN stars in M\,31 (light green) are plotted versus the logarithmic metallcity of their host galaxies. The error bars refer to the standard deviation in each sample. 
The dashed line is a linear regression.
} 
\label{fig:mdot-z}
\end{figure}

In recent years, a large number of WN stars were spectroscopically analyzed in the MW \citep{Hamann2006,Martins2008,Liermann2010,Oskinova2013}; in M\,31 \citep{Sander2014}; and in the LMC \citep{Hainich2014}, facilitating a comprehensive investigation of the relation between their wind mass-loss and metallicity. 
The averaged mass-loss rates of the WN stars in the MW, the late type WN stars in M\,31, the WN stars in the LMC, and the SMC WN stars analyzed in this work are plotted in Fig.\,\ref{fig:mdot-z} as a function of the metallicity of their host galaxy.
For the MW, we assume that a metallicity of $Z_\odot = 0.014$, derived for the solar neighborhood by \citet{Przybilla2008}, is valid for the whole Galaxy, neglecting the observed metallicity gradient \citep[e.g.,][]{Mayor1976,Nordstroem2004,Balser2011,Bovy2014,Hayden2014}. Metallicities of $Z_\mathrm{SMC} = 0.14\,Z_\odot$ and $Z_\mathrm{LMC} = 0.43\,Z_\odot$ are assumed for the SMC and LMC stars, respectively \citep{Meynet2005,Eldridge2006,Brott2011,Georgy2013}.

In Fig.\,\ref{fig:mdot-z}, we also show a linear regression to the data for the full set of analyzed stars,  resulting in a power law with an exponent of $1.2 \pm 0.1$. 
This exponent, however, strongly depends on the assumed metallicities. If we alternatively assume, e.g., a metallicity of $0.57\,Z_\odot$ for the LMC and $0.29\,Z_\odot$ for the SMC \citep{Meynet2005} but keep the Galaxy at $Z_\odot = 0.014$, we would obtain a considerably steeper relation with an exponent of 1.8. An exponent of only 0.9 is achieved for a Galactic and M\,31 metallicity of 0.02, while keeping the LMC and SMC metallicity as in Fig.\,\ref{fig:mdot-z}. 

For comparison, \citet{Nugis2007} suggest an exponent on the order of unity, while the studies performed by \citet{Crowther2006} resulted in an exponent of $0.8 \pm 0.2$. Our favored exponent of $\alpha = 1.2$ is somewhat higher than the latter value. \citet{Vink2005} theoretically predicted an exponent of 0.86 for cool WN stars in the metallicity range $10^{-3} \leq Z/Z_\odot \leq 1$. 

\begin{figure}
\centering
\includegraphics[width=\hsize]{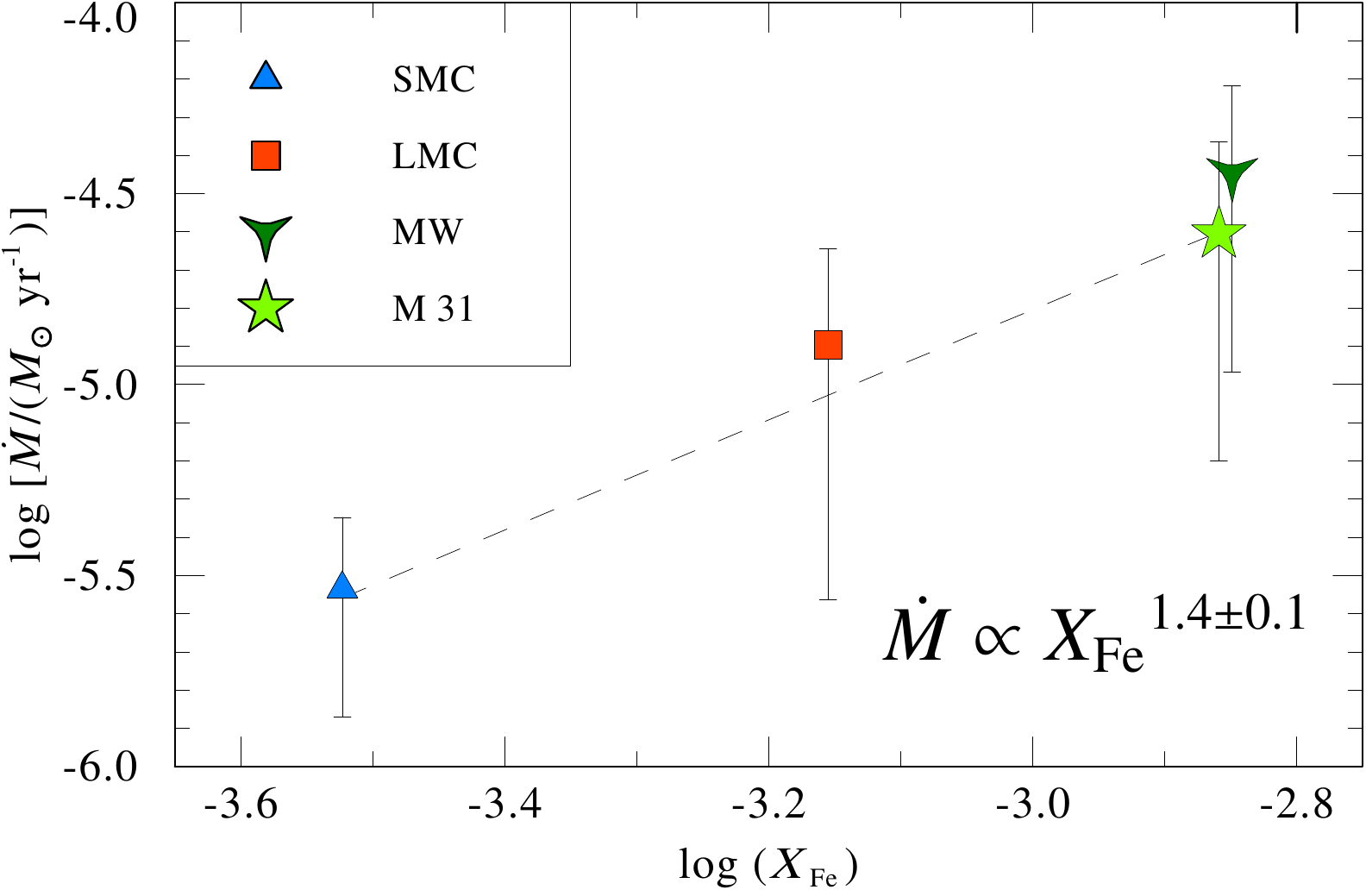}
\caption{Averaged mass-loss rates of the WN stars in the SMC, LMC, MW, and the late type WN stars in M\,31 plotted versus the iron abundance given in Table\,\ref{table:abundances}. The symbols are the same as in Fig.\,\ref{table:abundances}. The dashed line is a linear regression fit that is weighted by the number of objects analyzed in each galaxy.} 
\label{fig:mdot-fe}
\end{figure}

Hydrodynamical stellar atmosphere models \citep{Graefener2005,Graefener2008} and Monte Carlo wind models \citep{Vink2005} argue that the wind of WR stars is mainly driven by radiation pressure through the vast amount of line transitions of the iron group elements. This is valid as long as the metallicity is not too low. For extremely low metallicities, the CNO elements might become more important for the wind driving. In Fig.\,\ref{fig:mdot-fe}, we plot the average mass-loss rates in the three galaxies over the iron abundances given in Table\,\ref{table:abundances}, which is considered to be a proxy for the whole iron group. A linear regression to the data for the full set of analyzed stars in each galaxy results in a power law with an exponent of ${1.4 \pm 0.2}$. This relation is steeper than the $\dot{M}$-$Z$-relation obtained above because the gradient of the iron abundance is shallower than the metallicity gradient. Since the iron group elements are directly connected to the wind driving mechanism of massive stars, the $\dot{M}$-$\element{Fe}$-relation is in principle the more appreciable dependency.

\begin{figure}
\centering
\includegraphics[width=\hsize]{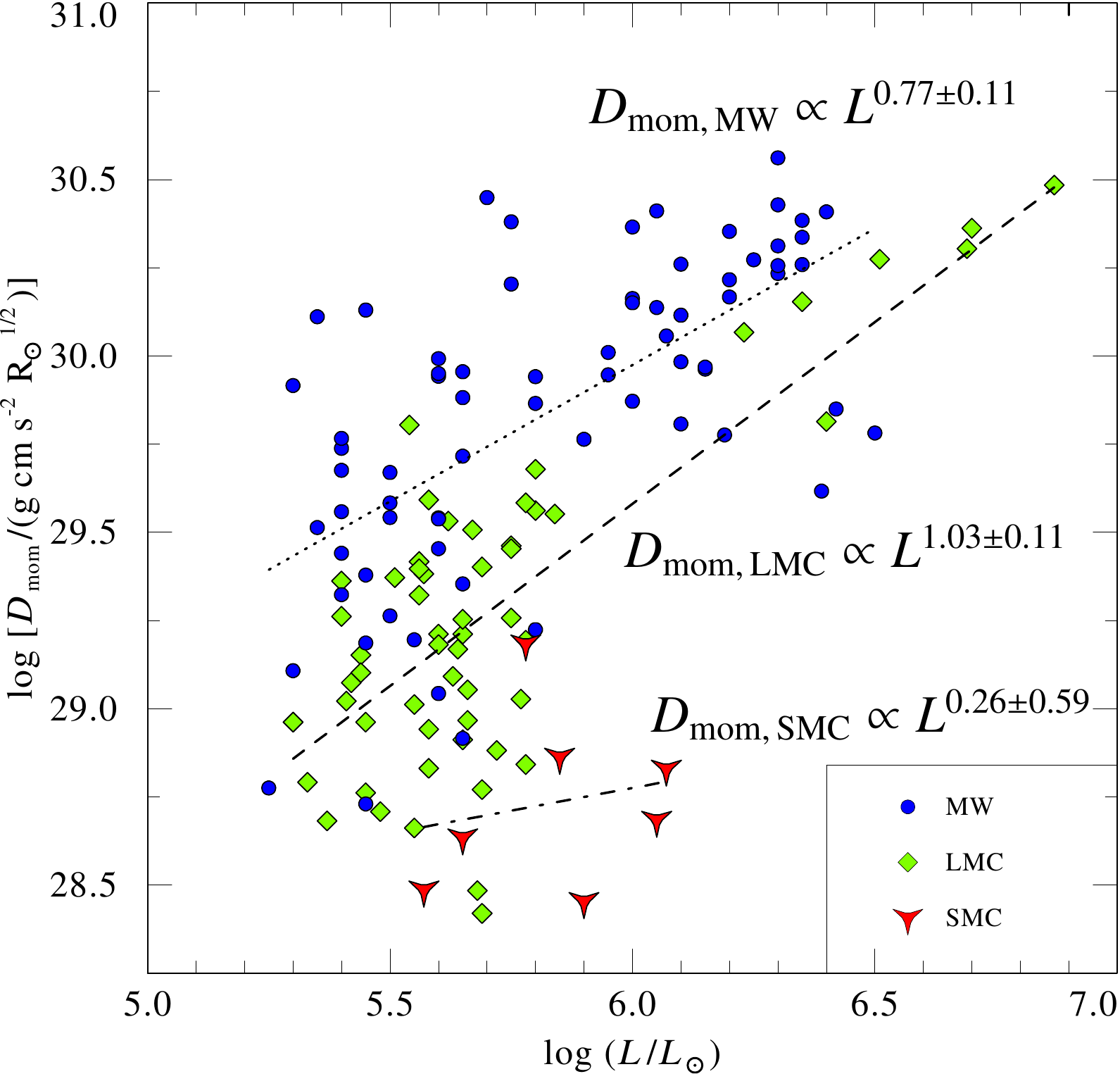}
\caption{
Modified wind momentum $D_\mathrm{mom}$ versus luminosity for the WN stars in the Galaxy, LMC, and SMC. The dashed lines represent linear regressions to the respective samples. 
} 
\label{fig:wlr}
\end{figure}

In addition to the direct method used above, we also derived the dependence of the mass-loss rate on the metallicity by comparing the individual WLRs for the WN stars in the MW, LMC, and SMC. 
In principle, this approach is advantageous because it implicitly corrects for the dependence of the mass-loss rate on the luminosity.
Unfortunately, the WLRs for the WN stars in the MW, LMC, and SMC cannot be determined with high accuracy, primarily because of the large scatter in the derived mass-loss rates \citep{Hamann2006,Hainich2014}. 
As a consequence, the slopes of the different WLRs deviate quite strongly.
Moreover, the exponent of the SMC WLR is very uncertain owing to the low number statistic. 
To the samples of each galaxy, we fitted a relation of the form 
\begin{equation}
\label{eq:wlr_2}
\log {D_\mathrm{mom}} = \log D_{0} + 29.7 + m \log (L/L_\odot - 5.8)~. 
\end{equation}
The corresponding coefficients can be found in Table\,\ref{table:wlr}, while the relations, together with the positions of the individual objects, are plotted in Fig.\,\ref{fig:wlr}.

\begin{table}
\caption{Coefficients for the modified wind momentum luminosity relations}
\label{table:wlr}
\centering  
\begin{tabular}{ccc}
\hline\hline  \rule[0mm]{0mm}{4.0mm} 
    &
   $\log D_{0}$ & 
   $m$
   \\
\hline \rule[0mm]{0mm}{4.0mm} 
   MW   &  $0.12 \pm 0.04$     &  $0.77 \pm 0.11$       \\
   LMC   &  $-0.34 \pm 0.04$     &  $1.03 \pm 0.11$        \\
   SMC  &  $-0.98 \pm 0.10$      &  $0.26 \pm 0.59$       \\
\hline 
\end{tabular}
\end{table}

Since the modified wind momentum is also a function of the terminal wind velocity, the metallicity dependence of $v_\infty$ needs to be considered in the evaluation of the mass-loss rate metallicity relation. 
For WR stars, however, this dependence is weak, if present at all. 
For example, \citet{Niedzielski2004} found only a marginal difference in the terminal velocity between their Galactic and LMC WN sample, studying UV resonance lines such as the \ion{He}{ii}\,$\lambda\,1640$ line. Therefore, we do not consider such a dependence here. 
Thus, combining Eqs.\,(\ref{eq:dmom}) and (\ref{eq:mdotz}) we obtain  
\begin{equation}
\label{eq:dmom_z}
D_\mathrm{mom} \propto Z^{\alpha}~, 
\end{equation}
while, the slope of the mass-loss rate metallicity relation can be written as
\begin{equation}
\label{eq:delta_dmom_z}
\alpha = \frac{\log D_\mathrm{mom}}{\log Z} + \mathrm{const.} = \frac{\log D_{0} + m \log (L/L_\odot)}{\log Z} + \mathrm{const.}~.
\end{equation}

\begin{figure}
\centering
\includegraphics[width=\hsize]{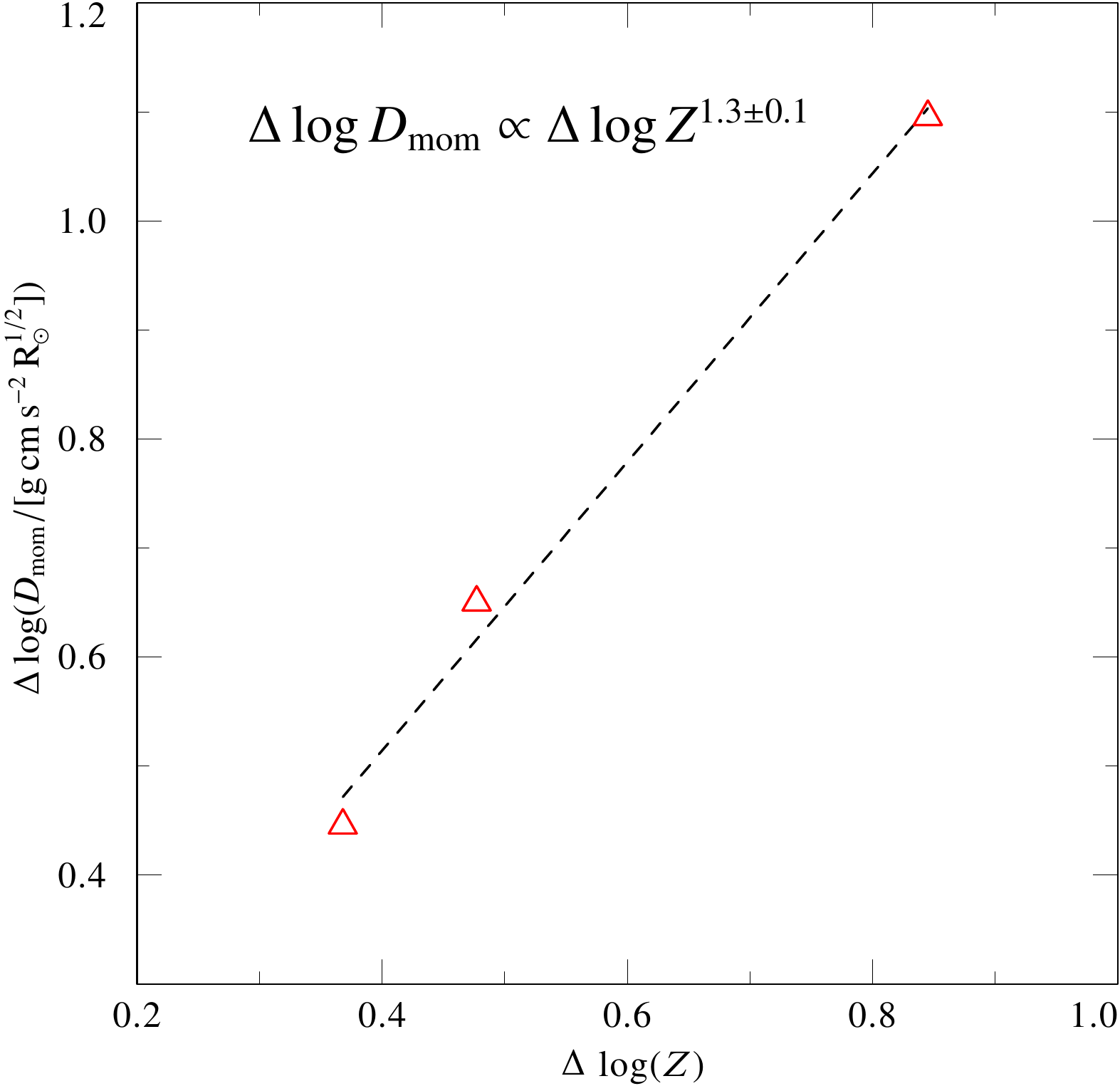}
\caption{
The $\Delta \log D_\mathrm{mom}$ values plotted over the difference of the logarithm of the metallicity. The dashed line is a linear regression. 
} 
\label{fig:DmomZ}
\end{figure}

As the exponents of the WLRs are not identical, one has to evaluate Eq.\,(\ref{eq:delta_dmom_z}) for a specific luminosity. We choose $\log (L/L_\odot) = 5.8$, which is right in the middle of the overlapping range of the different WLRs. 
In Fig.\,\ref{fig:DmomZ}, we plot the differences between the $\log D_\mathrm{mom}$ values for the three WLRs versus the corresponding differences between the $\log Z$ values. 
Since $\Delta \log D_\mathrm{mom}$ vanishes for $\Delta \log Z = 0$, the slope of the resulting line fit gives the exponent of the mass-loss rate metallicity relation. With $\alpha = 1.3 \pm 0.2$, it is close to the value directly derived from the relation shown in Fig.\,\ref{fig:mdot-z}.
We note that the quoted error only characterizes the quality of the fit, while several additional uncertainties would come on top. The largest contribution follows from the uncertainty in the derived slopes of the WLRs, which in turn relate to the adopted luminosity at which the WLRs are evaluated. At a luminosity of $\log (L/L_\odot) = 5.6$ the exponent would be $\alpha = 1.1 \pm 0.3$, while it is $\alpha = 1.5 \pm 0.5$ for a luminosity of $\log (L/L_\odot) = 6.0$.

\subsubsection{Dependence on the Eddington factor}
\label{subsubsect:mdot_eddington}

The strong mass loss and the spectral appearance of WR stars are associated with their proximity to the Eddington limit \citep[][]{Graefener2008,Vink2011}. This proximity can be characterized with the Eddington factor
\begin{equation}
\label{eq:gammae}
\Gamma_\mathrm{e} = \frac{\kappa_\mathrm{e} L}{4 \pi c G M} 
\end{equation}
where $\kappa_\mathrm{e}$ is the Thomson opacity. Assuming a completely ionized plasma, which is a reasonable assumption in the deep layers of the stellar atmosphere, $\Gamma_\mathrm{e}$ can be written in the form 
\begin{equation}
\label{eq:gammae_hydro}
\Gamma_\mathrm{e} = 10^{-4.813}\, (1 + X_\mathrm{H}) \frac{L/L_\odot}{M/M_\odot}~.
\end{equation}
Since $X_\mathrm{H}$ and the luminosity $L$ are known from the normalized line fit and SED fit, respectively, the only unknown quantity is the stellar mass.  Unfortunately, a reliable estimate of the surface gravity and consequently of the stellar mass from the absorption lines in the spectra of the SMC WN stars was unsuccessful because of the low spectral resolution of most optical spectra at hand and the strong impact of the stellar wind on these lines. Following the procedure applied by \citet{Graefener2011} and \citet{Bestenlehner2014}, a rough estimate of the stellar mass of our sample stars (as a function of their luminosity and surface hydrogen abundance) may be obtained by means of the mass-luminosity relation derived by \citet[][Eq.\,11]{Graefener2011}.
In Sect.\,\ref{sect:evolution}, we compare these masses, which in a strict sense are only valid for chemically homogeneous stars with solar metallicities, with masses obtained from stellar evolution models.

\begin{figure}
\centering
\includegraphics[width=\hsize]{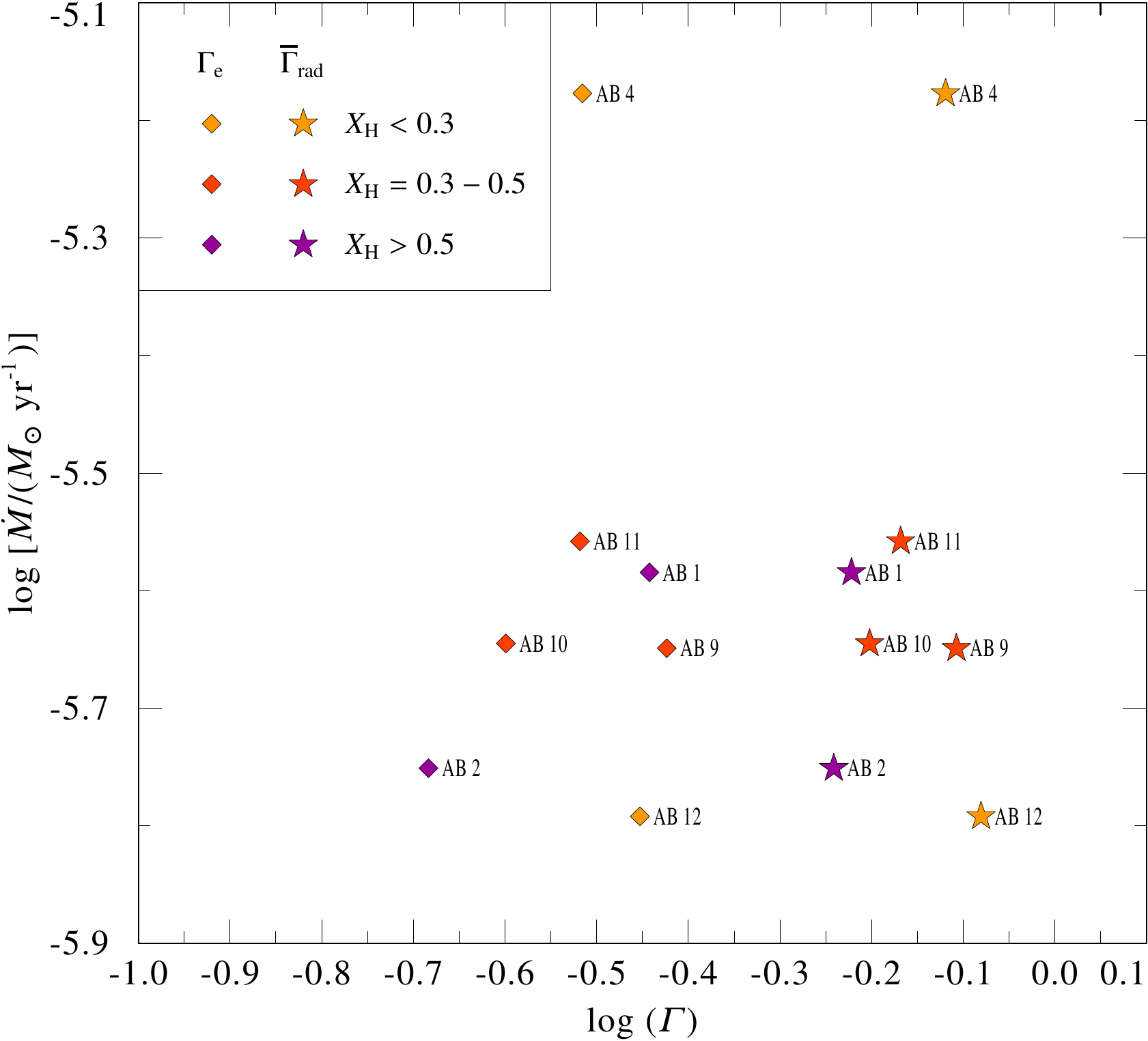}
\caption{Mass loss rates  of the WN stars in the SMC versus Eddington factor. The abscissa has two alternative meanings: the standard Eddington factor $\Gamma_\mathrm{e}$ which only accounts for the radiation pressure on free electrons (stars represented by diamonds), and the Eddington factor $\overline{\Gamma}_\mathrm{rad}$ which includes the total radiation pressure on all absorbers, averaged over the quasi-hydrostatic layers (asterisks).
} 
\label{fig:mdot-gamma_comp}
\end{figure}

It is customary to correlate mass-loss rates of WR stars with their Eddington factors $\Gamma_\mathrm{e}$, which account for electron scattering alone. However, scattering of electrons contributes only a fraction to the total radiative pressure exerted on the atmosphere; in hot stars with dense winds, the contributions of line and continuum transitions are far from being negligible. With PoWR, we are able to accurately calculate the full radiation pressure and the corresponding Eddington factor, which is denoted with $\Gamma_\mathrm{rad}$ and which generally varies with depth more extremely than $\Gamma_\mathrm{e}$. In Fig.\,\ref{fig:mdot-gamma_comp} we plot the mass-loss rates over the Eddington factor $\Gamma_\mathrm{e}$, and alternatively over the complete Eddington factor averaged over the hydrostatic domain, $\overline{\Gamma}_\mathrm{rad}$ \citep{Sander2015}. While $\Gamma_\mathrm{e}$ values are well below unity, the $\overline{\Gamma}_\mathrm{rad}$ values are considerably higher, pushing these stars significantly closer to the Eddington limit already in the quasi-hydrostatic layers of the atmosphere. We note that because of incomplete opacities in the model calculations the true $\overline{\Gamma}_\mathrm{rad}$ values are probably still underestimated.

The largest uncertainty in the evaluation of $\Gamma_\mathrm{e}$ and the $\overline{\Gamma}_\mathrm{rad}$, however, is caused by the uncertain stellar mass. Evolution tracks predict that WR stars with surface hydrogen abundances below approximately 40\,\% (by mass) are already core helium-burning \citep{Meynet2005,Georgy2013}. However, these tracks do not reproduce the HRD positions of our stars (see Sect.\,\ref{sect:evolution}) and can therefore not be used to extract a mass-luminosity relation. As an extreme assumption, one may adopt the $M$-$L$-relation for hydrogen-free stars provided by \citet{Graefener2011}. However, this leads to $\overline{\Gamma}_\mathrm{rad}$ values that exceed unity and do not allow for the existence of quasi-hydrostatic layers.
Alternatively, our program stars might be explained by homogenous evolution (see Sect.\,\ref{sect:evolution}). 
We therefore choose to calculate the stellar masses using the mass-luminosity relation for homogeneous stars provided by \citet{Graefener2011}.

\begin{figure}
\centering
\includegraphics[width=\hsize]{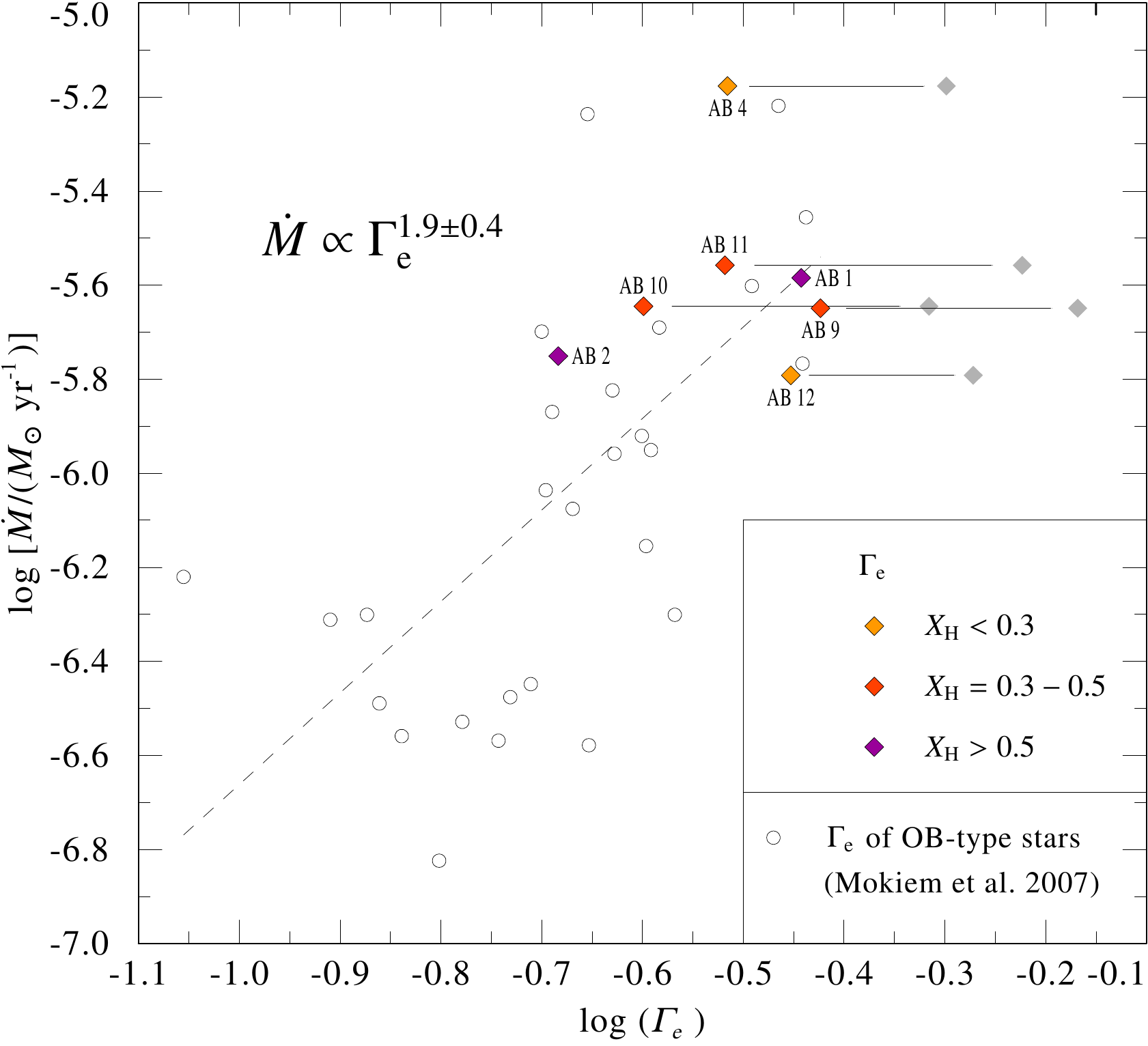}
\caption{The mass-loss rate of the WN stars in the SMC vs. the Eddington factor. The meaning of the colored symbols is the same as in Fig\,\ref{fig:hrd_compare}. The open symbols refer to the OB-type stars studied by \citet{Mokiem2007}. The gray filled symbols depict the position of the SMC WN stars (with $X_{\element{H}} < 0.5$) when the mass is calculated based on a mass-luminosity relation for hydrogen free stars. The dashed line is a linear fit to the OB-type stars and the WN stars (colored symbols).} 
\label{fig:mdot-gamma}
\end{figure}

In Fig.\,\ref{fig:mdot-gamma} we plot the mass-loss rates versus the Eddington factors for our sample stars as well as for the OB-type stars studied by \citet{Mokiem2007}.
For consistency, the masses of the OB-type stars were estimated with the same method as those of the WN stars. 
In comparison to the OB-type sample presented by \citet{Mokiem2007}, the SMC WN stars exhibit about the same mass-loss rates at the equivalent $\Gamma_\mathrm{e}$ values. The fit to the combined sample of OB-type and WN stars results in a power law with an exponent of $1.9 \pm 0.4$. In this context, the WN stars in the SMC seem to constitute an extension to the $\dot{M}$-$\Gamma_\mathrm{e}$-relation observed for the OB-type stars. 
Thus, the strong dependence of the WN mass-loss rate on $\Gamma_\mathrm{e}$ predicted by \citet{Graefener2008} and \citet{Vink2011} is not observed in the SMC. However, because of low number statistics, these statements are somewhat speculative.

According to \citet{Vink2011} a steepening of the $\dot{M}$-$\Gamma_\mathrm{e}$-relation is expected for WR stars in comparison to OB-type stars because the former objects are in the multi-scattering domain. The first observational indications of a steepening of the $\dot{M}$-$\Gamma_\mathrm{e}$-relation were found by \citet{Bestenlehner2014} for stars in 30\,Doradus. However, this trend is not seen in the SMC, which is consistent with the fact all SMC WN stars fall below the single-scattering limit.

For those objects in our sample with hydrogen abundances below $X_{\element{H}} = 0.5$ (mass fraction), we also include in Fig.\,\ref{fig:mdot-gamma} the $\Gamma_\mathrm{e}$ values obtained with the mass-luminosity relation for hydrogen-free stars, indicating the range of possible Eddington factors. For the latter $\Gamma_\mathrm{e}$ values, the observed mass-loss rates of our program stars are even lower than the tentative $\dot{M}$-$\Gamma_\mathrm{e}$-relation (dashed line in Fig.\,\ref{fig:mdot-gamma}).

\subsection{WN mass-loss prescription}
\label{sect:mdot_prescription}

\begin{figure}
\centering
\includegraphics[width=\hsize]{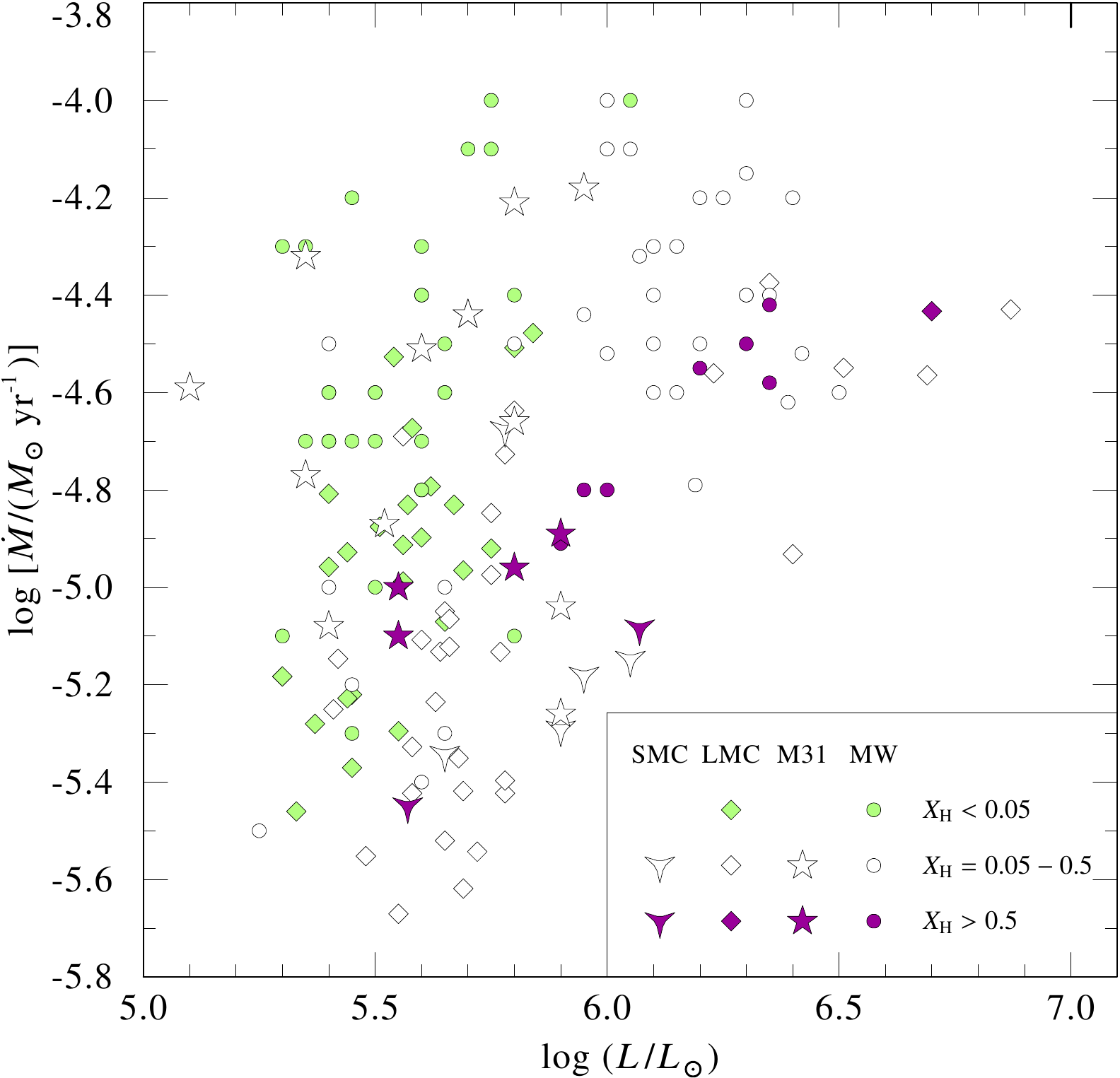}
\caption{The mass-loss rate of the WN stars in the SMC, LMC, M\,31, and MW vs. the luminosity. The meaning of the symbols is the same as in Fig\,\ref{fig:hrd_compare}. Highlighted are the hydrogen free as well as the stars with large surface hydrogen abundances.} 
\label{fig:mdot-l_comp}
\end{figure}

The large number of WN stars analyzed in the MW, M\,31, the LMC, and the SMC should provide the bases for a statistical fit to the mass-loss rate as function of fundamental parameters. 
We adopt a mass-loss prescription that is a function of the stellar luminosity $L$, the stellar temperature $T_*$, the surface helium abundance $X_{\element{He}}$, and the metallicity $Z$ in the form of a power-law
\begin{align}
\label{eq:mdot-fit_obs}
\begin{split}
\log (\dot{M}) = C_1 &+ C_2 \log (L)  + C_3 \log (T_*) \\ &+ C_4 \log (X_{\element{He}}) + C_5 \log (Z)~. 
\end{split}
\end{align}
Based on datasets derived from spectral analyses, the free parameters ($C_1-C_5$) are constrained by means of a $\chi^2$-fit. The best fitting parameters are summarized in Table\,\ref{table:chisquare_obs_new}. When a specific parameter is not considered in the fit, it is depicted by a dash. 

\begin{table}
\caption{Coefficients obtained by a $\chi^2$-fit of Eq.\,(\ref{eq:mdot-fit_obs}) to the stellar parameters of the complete sample of 153 WN stars, and of the sub-samples of the hydrogen-free (56 stars) and the hydrogen-rich objects (14 stars), respectively.}
\label{table:chisquare_obs_new}      
\centering
\small
\begin{tabular}{lSSSSS}
\hline\hline 
\rule[0mm]{0mm}{2.2ex} 
$X_{\element{H}}$ & $C_{1}$  & $C_{2}$ & $C_{3}$ & $C_{4}$ & $C_{5}$ \\
\hline 
$0 - 0.7$   & -5.13 & 0.63                          & -0.23       & 1.3             & 1.02 \rule[0mm]{0mm}{2.5ex} \\
$< 0.05$  & -7.15 & 0.90                            & -0.13       & \multicolumn{1}{c}{-}  & 0.97   \\
$\ge 0.5$ & -6.17 & 0.80                            & -0.44       & \multicolumn{1}{c}{-}  & 0.75 \\
\hline 
\end{tabular}
\end{table}

In Table\,\ref{table:chisquare_obs_new}, the first line gives the coefficients obtained for the whole set of analyzed WN stars, comprising 153 putatively single WN stars from \citet{Hamann2006}, \citet{Martins2008}, \citet{Liermann2010}, \citet{Oskinova2013}, \citet{Sander2014}, and \citet{Hainich2014}.
The mean deviation between the fit and the individual measurements is 0.27\,dex and thus relatively large. 

If we restrict the fit to the hydrogen-free objects ($X_{\element{H}} < 0.05$, second line in Table\,\ref{table:chisquare_obs_new}) the scatter is comparable ($\sigma = 0.26\,\mathrm{dex}$). For the sub-sample of the hydrogen-rich objects (third line), however, the scatter is significantly smaller ($\sigma = 0.07\,\mathrm{dex}$). 
Both these fits exhibit similar $\dot{M}$-$L$ relations, with exponents in the range of 0.8--0.9. 
In comparison to the mass-loss relation derived by \citet{Nugis2000}, \citet{Hamann1995}, and \citet{Hainich2014}, all of the prescriptions given in Table\,\ref{table:chisquare_obs_new} posses a shallower dependence on the luminosity. Moreover, the relation for all objects also exhibits a weaker dependence on the helium abundance ($\dot{M} \propto X_{\element{He}}^{1.3}$ compared to $\dot{M} \propto X_{\element{He}}^{2.22}$ obtained by \citealt{Nugis2000}).
While the temperature dependence is almost negligible in the hydrogen-free case, it is slightly negative for the hydrogen-rich stars. 
Interestingly, the exponent obtained for the $\dot{M}$-$Z$ relation is smaller for the latter objects.
In each case, however, the metallicity dependence obtained from the multidimensional fits is shallower than the one-parameter fit conducted in Sect.\,\ref{subsubsect:mdot-z}. 

Theoretical mass-loss prescriptions for WN stars were derived by \citet{Graefener2008} and \citet{Vink2011}. Both studies encountered a strong dependence of the mass-loss rate on the Eddigton factor. Similar results have been obtained by \citet{Graefener2011} and \citet{Bestenlehner2014} for WN stars in the Arches cluster and 30\,Doradus, respectively. Therefore, we also explored whether this behavior can be observed considering a large number of diverse WN stars. 
As described in Sect.\,\ref{subsubsect:mdot_eddington}, we estimated the mass of the WN stars by means of a mass-luminosity relation predicted by \citet[][their equations\,11 and 13]{Graefener2011} for chemically homogeneous stars and solar metallicity. 
With theses masses, we computed the classical Eddington factor by means of Eq.\,(\ref{eq:gammae_hydro}). Incorporating these values in the determination of the mass-loss prescription results in a relatively weak, or even vanishing, dependence on the Eddigton Gamma when also an explicit hydrogen dependence is considered in the fit.
In each case, the exponent obtained for the power-law dependence of the mass-loss rate on $\Gamma_\mathrm{e}$ is considerably weaker than the values predicted by \citet{Graefener2008} and \citet{Vink2011}.

\subsection{Stellar rotation}
\label{subsect:rotation}

The rotational velocity of OB stars can be directly determined from the profiles of their photospheric absorption lines by means of Fourier analyses or spectral fits. For the latter, synthetic model spectra are convolved with a rotational profile. This method, however, is not applicable to the emission line spectra of WR stars, since the spectral lines of these objects are formed in the stellar wind. Line formation hence requires an integration of the radiative transfer equation in three dimensions. A corresponding extension to the PoWR code has been introduced by \citet{Shenar2014}.

Assuming conservation of angular momentum, the azimuthal component of
the velocity drops rapidly with increasing distance from the star.
Adopting such a kinematic model, rotation might have significant effects on
emission line profiles \citep{Hillier1998,Shenar2014}. More
drastic rotational broadening is obtained when assuming co-rotation
of the wind up to some distance. Such co-rotation could be enforced by
sufficiently strong magnetic fields.

\begin{figure}
\centering
\includegraphics[width=\hsize]{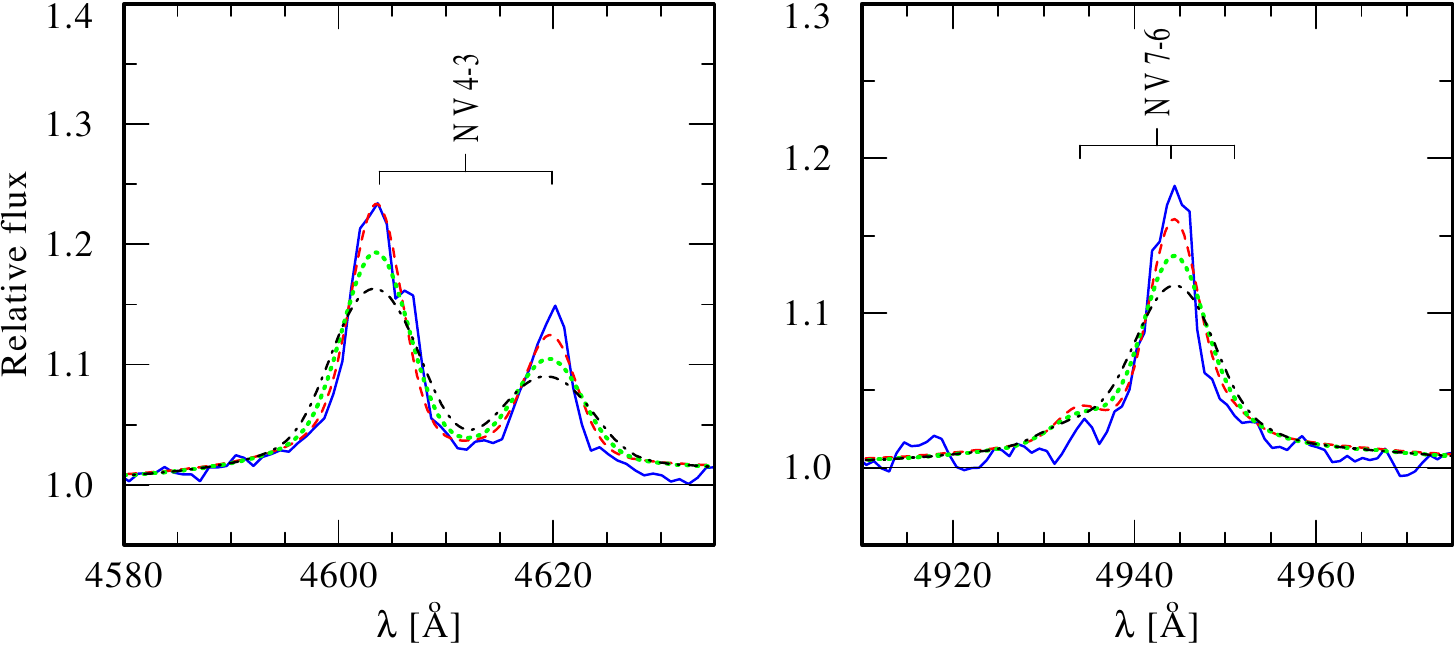}
\caption{Comparison of synthetic spectra for different rotational velocities, accounting for the full 3D line formation in the wind under the assumption of angular momentum conservation. $v \sin \mathrm{i}$ refers to the photosphere. The left panel shows the \ion{N}{v}\,$\lambda\lambda\, 4604, 4620$ lines of SMC\,AB\,12, while the right panel depicts the \ion{N}{v}\,$\lambda\lambda\, 4934, 4944, 4951$ lines of the same object. The blue full line is the observation. The red dashed line, green dotted line, and black dashed-dotted line correspond to a rotational velocity of $v \sin i = 0\,\mathrm{km/s}$, $200\,\mathrm{km/s}$, and $300\,\mathrm{km/s}$, respectively.}
\label{fig:vrot_fit}
\end{figure}

Since our program stars do not show evidence for wind co-rotation, we
adopt angular momentum conservation and perform test calculations with
different rotational velocities $\varv \sin i$ (defined at $R_*$). Figure\,\ref{fig:vrot_fit} depicts the obtained line profiles for two representative \ion{N}{v} lines of SMC\,AB\,12. 
From such comparisons with the observed profiles, rotation velocities
higher than 200 km/s can be excluded for all stars of our sample. Sharper
limits can be drawn for AB1 and AB4 (100 km/s) and for AB2 (50km/s). The
sharper limits rely on characteristic Of lines such as \ion{N}{iv}\,$\lambda\, 4060$ and \ion{N}{iii}\,$\lambda\, 4640$, present in late-type WN spectra. The Of lines are usually formed very close to the photosphere of the star and their line profiles are thus more strongly affected by rotation than lines formed far out in the wind. Therefore, the constrains on the rotational velocity are more restrictive for late spectral types. 
\citet{Martins2009} excluded rotational velocities much larger than $50\,\mathrm{km/s}$ for the SMC WN stars by means of synthetic emission line spectra convolved with rotational profiles. In contrast, our elaborated integration scheme results in a high upper limit of about $200\,\mathrm{km/s}$ for the rotational velocity of the WN\,3 stars.

\section{Discussion of the evolutionary status}
\label{sect:evolution}

The evolution of massive stars, especially in their late phases, is still not completely understood \citep[e.g.,][]{Langer1994,Vanbeveren1998,Langer2012,Sander2014,Meynet2015}. Comparing empirical results with theoretical predictions is the basic instrument to unveil the evolutionary status connections. While stellar evolution models achieved progress in reproducing the observed ratio between the numbers of WN to WC stars \citep{Meynet2003,Eldridge2006,Georgy2012b}, they still partly fail to reproduce the range of luminosities and temperatures observed for the WR population in the LMC \citep{Hainich2014} and the Galaxy \citep{Hamann2006,Sander2012}. Moreover, the current stellar evolution models predict a considerable number of WR stars in a parameter range where no object is actually observed \citep{Hainich2014,Hamann2006}. In the following, we test whether the same holds for the putatively single WN stars in the SMC.

\begin{figure}
\centering
\includegraphics[width=\hsize]{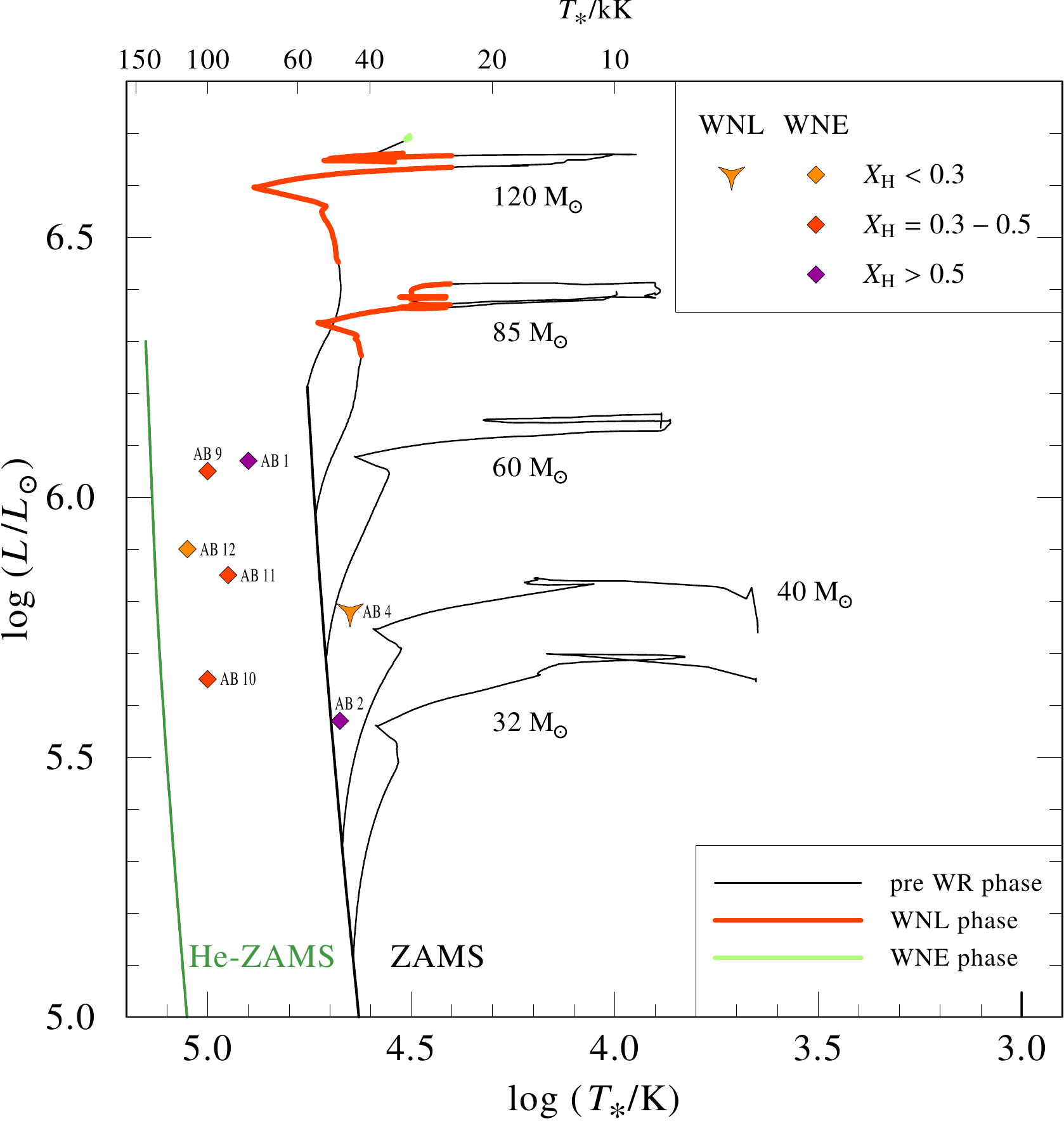}
\caption{The HRD of the single WN stars compared to the recent stellar evolution tracks (with rotation) calculated by the Geneva group \citep{Georgy2013} for a metallicity of $0.14\,Z_\odot$. 
The color coding of the symbols and the tracks during the WN phase reflects the hydrogen abundance at the stellar surface (see inlet). The labels close to the tracks refer to the initial mass.} 
\label{fig:hrd_geneva}
\end{figure}

Figure\,\ref{fig:hrd_geneva} shows the HRD position of the WN stars in the SMC in comparison to single star evolution tracks published recently by the Geneva group \citep{Georgy2013}. These tracks are calculated for an initial metallicity of $Z = 0.14\,Z_\odot$. Based on the stellar temperature and the surface chemical composition, the discrete WR phases are distinguished by different colors and line styles. 
The pre-WR phase is defined by a hydrogen mass-fraction in excess of $X_{\element{H}} = 0.55$ (thin black lines). At hydrogen abundances below $X_{\element{H}} = 0.55$ and stellar temperatures above $T_* = 25\,\mathrm{kK}$, the track is considered to represent the WNL stage (bold red line). Once the hydrogen abundances drops below $X_{\element{H}} = 0.05$, we classify the star as a WNE (thick green line). 

The lowest initial mass for which a track reaches the WR phase is $85\,M_\sun$. Interestingly, the tracks for $85\,M_\sun$ and $120\,M_\sun$ enter the WN stage for the first time already at the end of the main sequence. Subsequently, both tracks evolve toward cooler surface temperatures in the regime of the supergiants before they experience another blueward evolution and enter the WN stage for a second time. While the second WN phase is only short for the $85\,M_\sun$ model, the $120\,M_\sun$ model appears as WNE even during carbon burning, close to the end of the evolution track. 

In contrast to these predictions, the SMC WN stars are found in a region of the HRD ($\log\,(L/L_\odot) = 5.6\,...\,6.1$) that corresponds to considerably lower initial masses in the range from about $30\,M_\sun$ to $60\,M_\sun$. The comparison illustrated in Fig.\,\ref{fig:hrd_geneva} clearly elucidates that the parameter regime occupied by the SMC WN stars is not reproduced by the theoretical prediction of the Geneva evolution models for single stars with an initial metallicity of $Z = 0.14\,Z_\odot$.

\begin{figure}
\centering
\includegraphics[width=\hsize]{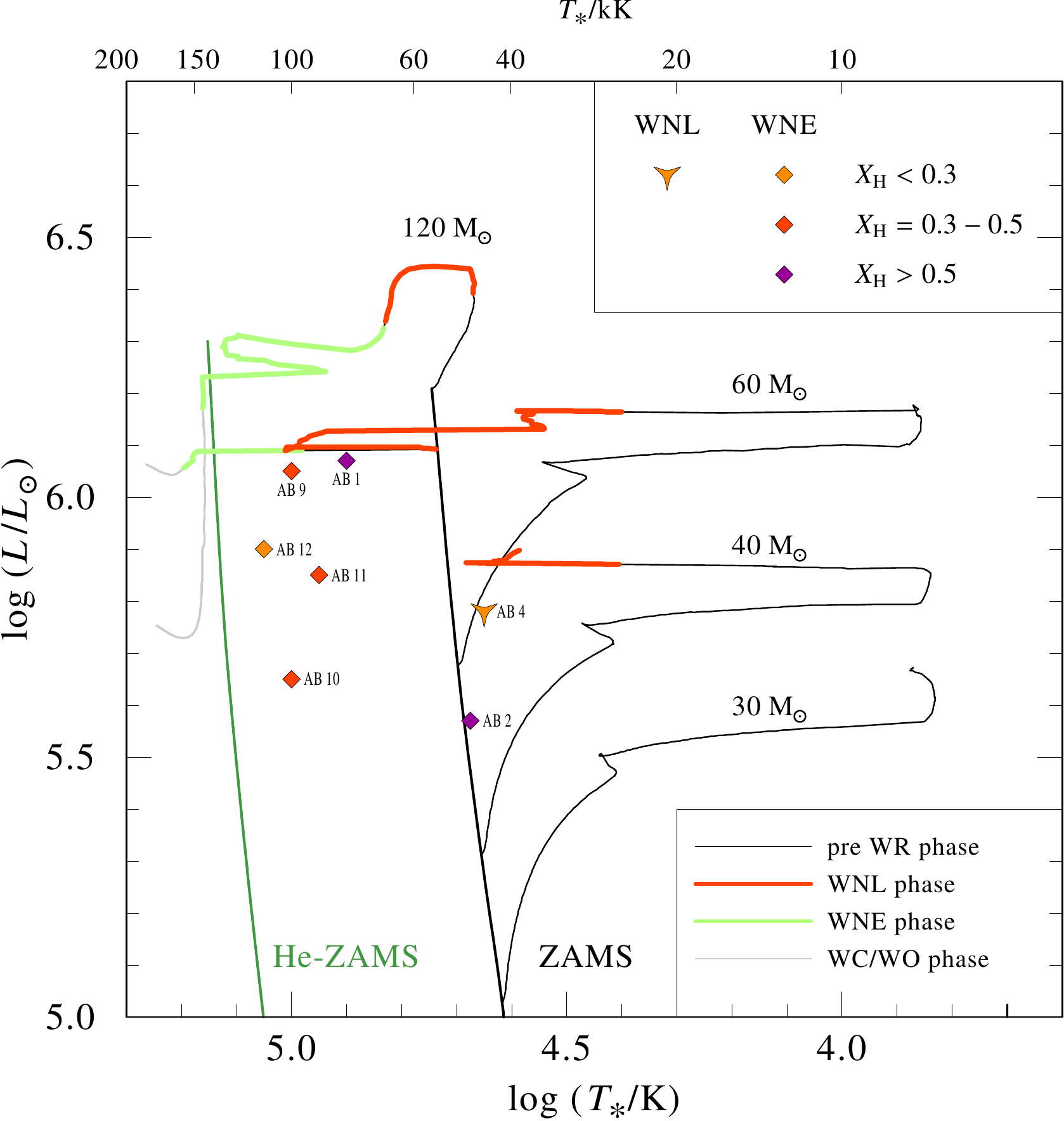}
\caption{Same as Fig.\,\ref{fig:hrd_geneva}, but with evolution tracks for $Z = 0.28\,Z_\odot$ from \citet{Meynet2005} which also account for rotation.\protect\footnotemark
} 
\label{fig:hrd_oldgeneva}
\end{figure}

\footnotetext{The differences between the tracks shown in this figure and the tracks plotted in the left panel of Fig.\,6 in \citet{Martins2009} are attributable to the different kinds of effective temperatures given by \cite{Meynet2005}. \citet{Martins2009} used the temperature that is corrected for the thickness of the WR wind, whereas we plot the temperature that refers to the hydrostatic radius. The latter temperature is comparable to the stellar temperature $T_*$ used in stellar atmosphere codes like PoWR or CMFGEN.}

In Fig.\,\ref{fig:hrd_oldgeneva}, we compare our sample with the older generation of the Geneva evolution models computed by \citet{Meynet2005}, which also account for the effect of rotation but are calculated for a higher initial metallicitiy of $0.28\,Z_\odot$. 
In contrast to the recent Geneva models for $Z = 0.14\,Z_\odot$ \citep{Georgy2013}, the tracks plotted in Fig.\,\ref{fig:hrd_oldgeneva} evolve considerably further to the blue part of the HRD. Some of these tracks even reach the WC phase (gray line), defined when the carbon mass-fraction at the surface exceeds $X_{\element{C}} = 0.2$. Moreover, compared to the evolution models published by \citet{Georgy2013}, the tracks drawn in Fig.\,\ref{fig:hrd_oldgeneva} \citep{Meynet2005} predict a significantly lower initial mass limit for reaching the WR phase of $40\,M_\sun$.
However, the agreement between our empirical HRD positions and the models from \citet{Meynet2005} is still relatively small. 
Only the two most luminous stars in our sample (SMC\,AB\,9 and 1) are reproduced. The relatively low luminosities of SMC\,AB\,2 and 10 suggested that these stars evolved from an initial mass of about $30\,M_\sun$, while the corresponding track never turns into a WR star.

\begin{figure}
\centering
\includegraphics[width=\hsize]{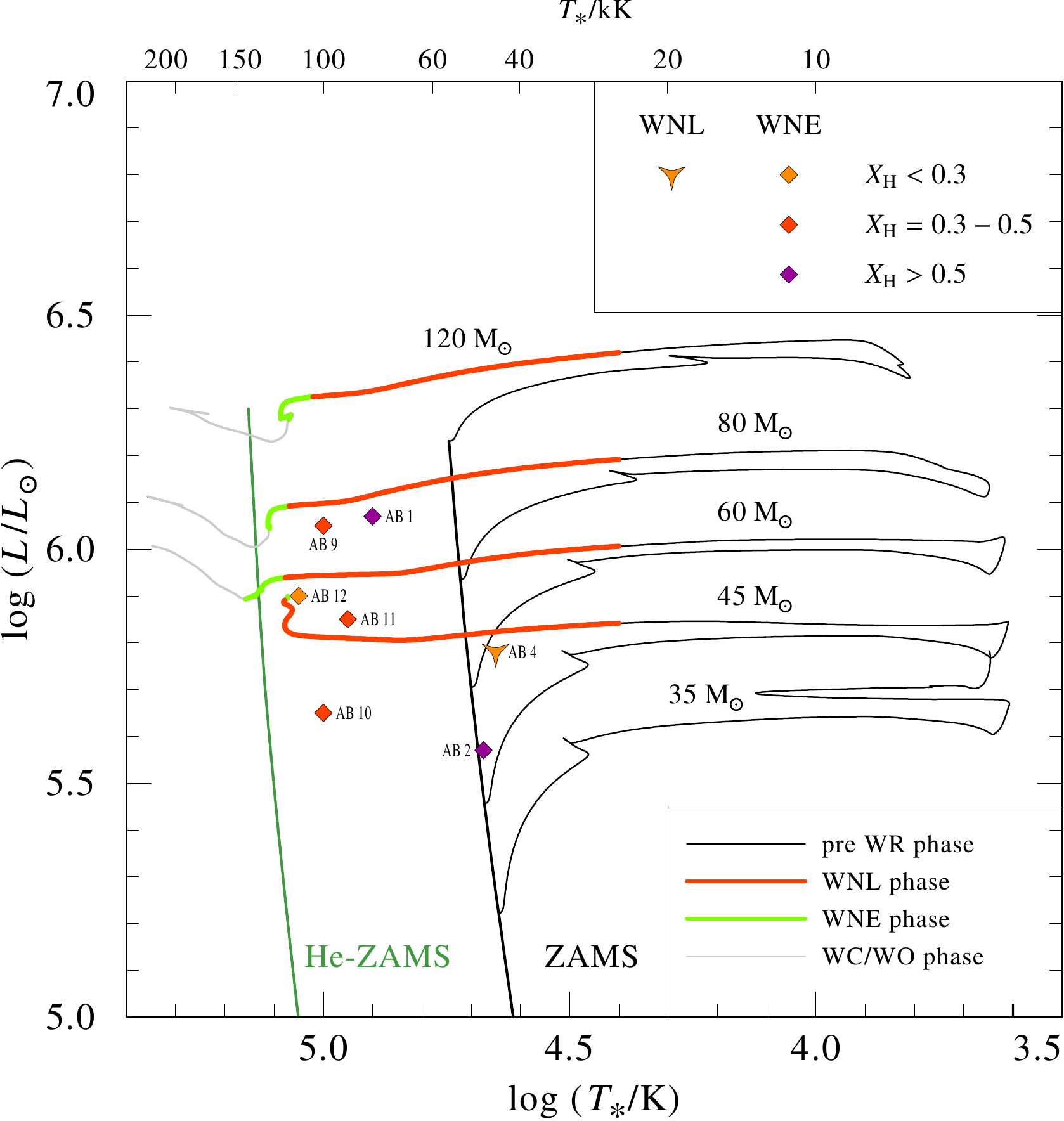}
\caption{Same as Fig.\,\ref{fig:hrd_geneva}, but with non-rotating evolution tracks for $Z = 0.28\,Z_\odot$ calculated by \citet{Eldridge2006}.
} 
\label{fig:hrd_eldridge}
\end{figure}

In the above comparison remains unclear whether the adopted metallicities are causing the different results, or whether other details of the model calculations are the cause. We also compared the HRD positions of our program stars with the non-rotating stellar evolution models from \citet[][see Fig.\,\ref{fig:hrd_eldridge}]{Eldridge2006}, which are calculated for an initial metallicity of $0.28\,Z_\odot$. The mass-loss rates which they adopted for their non-rotating models in the pre-WR phases is comparable to the one used by \cite{Meynet2005} for their rotating models. During the WR phase, the models described in \citet{Eldridge2006} account for a metallicity scaling of the mass-loss rate, while such a dependency was not considered in the older generation of the Geneva models \citep{Meynet2005}. Taking into account that the total WR lifetime is nearly independent from the metallicity scaling of the mass-loss rate during the WR phase \citep{Eldridge2006}, the comparison of these tracks with the ones shown in Figs.\,\ref{fig:hrd_geneva} and \ref{fig:hrd_oldgeneva} might give some indication of the impact of the metallicity. 

For the stars in our sample with a luminosity of $\log\,(L/L_\odot) \geq 5.8$, there is a good agreement with respect to the tracks from \citet{Eldridge2006}, whereas the two faintest objects (SMC\,AB\,2 and 10) cannot be reproduced at all. Although the tracks from \citet{Eldridge2006}, compared to the Geneva tracks, coincide better with our results, these models also require a minimum initial mass of $45\,M_\sun$ to reach the WR phase and therefore do not reach the WN phase for luminosities as low as those of SMC\,AB\,2 and 10.

The higher the metallicity of the stellar evolution models, the higher the degree of agreement that can be achieved with the HRD positions of the WN stars in the SMC. Our sample is best reproduced with tracks for $Z = 0.56\,Z_\odot$ from \citet{Meynet2005}, which account for the implications of rotation. The model atmospheres used in our analysis have a metallicity of $Z = 0.13\,Z_\odot$. Our relatively young WN stars might have been born with a slightly higher metallicity than this average value for the SMC.
However, we do not see evidence for a considerably higher iron abundance than the value given in Table\,\ref{table:abundances}. An exception is SMC\,AB\,4, which might be reproduced with stellar atmosphere models of higher metallicities as well.
The enhanced nitrogen abundance observed (see Table\,\ref{table:derived_abundances}) can be well understood in terms of self-enrichment with products from central helium burning. In conclusion, we suggest that it is not the metallicity itself, but some other effect which is not properly accounted for in the evolution models that are causing the difference between the tracks and the empirical HRD positions. 

Apart from the discrepancies described above, another disagreement between our analysis and the stellar evolution models is the surface hydrogen abundances. When the evolution tracks cross the empirical HRD positions of our SMC WN stars, they predict a considerably lower hydrogen abundance at the surface than observed. This coincides with the fact that the mass-loss rates derived in our analysis (see Sect.\,\ref{subsect:mdot}) are on average lower than the values given by the mass-loss relation from \citet{Nugis2000}, which is often applied in stellar evolution models.\footnote{In some cases the more recent description derived by \citet{Graefener2008} is also used \citep[e.g.,][]{Ekstroem2012,Georgy2013}.} Thus, the discrepancy in the hydrogen abundance might be caused by too high WR mass-loss rates in the evolution models, which remove too much of the hydrogen-rich envelope. Therefore, higher mass-loss rates in the WR phase would not bring the stellar evolution tracks into accordance with the empirical HRD positions of the SMC WN stars. We conclude that either the mass-loss in a pre-WR stage is significantly higher (via either continous mass-loss or episodic bursts), as suggested by e.g., \citet{Sander2012} to explain the disagreement between WC stars and evolution models at Galactic metallicity \citep[see also][]{Vanbeveren1998,Vanbeveren2007, Georgy2012}, or that standard models for single star evolution still miss essential physical ingredients. 

Such a missing ingredient could be very fast rotation, which lead to full internal mixing and quasi-homogeneous evolution. This type of evolution proceeds roughly speaking directly from the hydrogen zero age main sequence to the helium main sequence \citep[e.g.,][]{Maeder1987, Langer1992, Yoon2005, Woosley2006}. Homogeneous evolution is expected to be more common at low metallicities, since rotational mixing is more efficient \citep{Brott2011} and angular momentum loss is lower because of the reduced wind mass-loss (see Sect.\,\ref{subsubsect:mdot-z}). Thus, rapid initial rotation will be conserved for a longer period, facilitating the evolution toward the blue part of the HRD even at relatively low initial masses.

\begin{figure}
\centering
\includegraphics[width=\hsize]{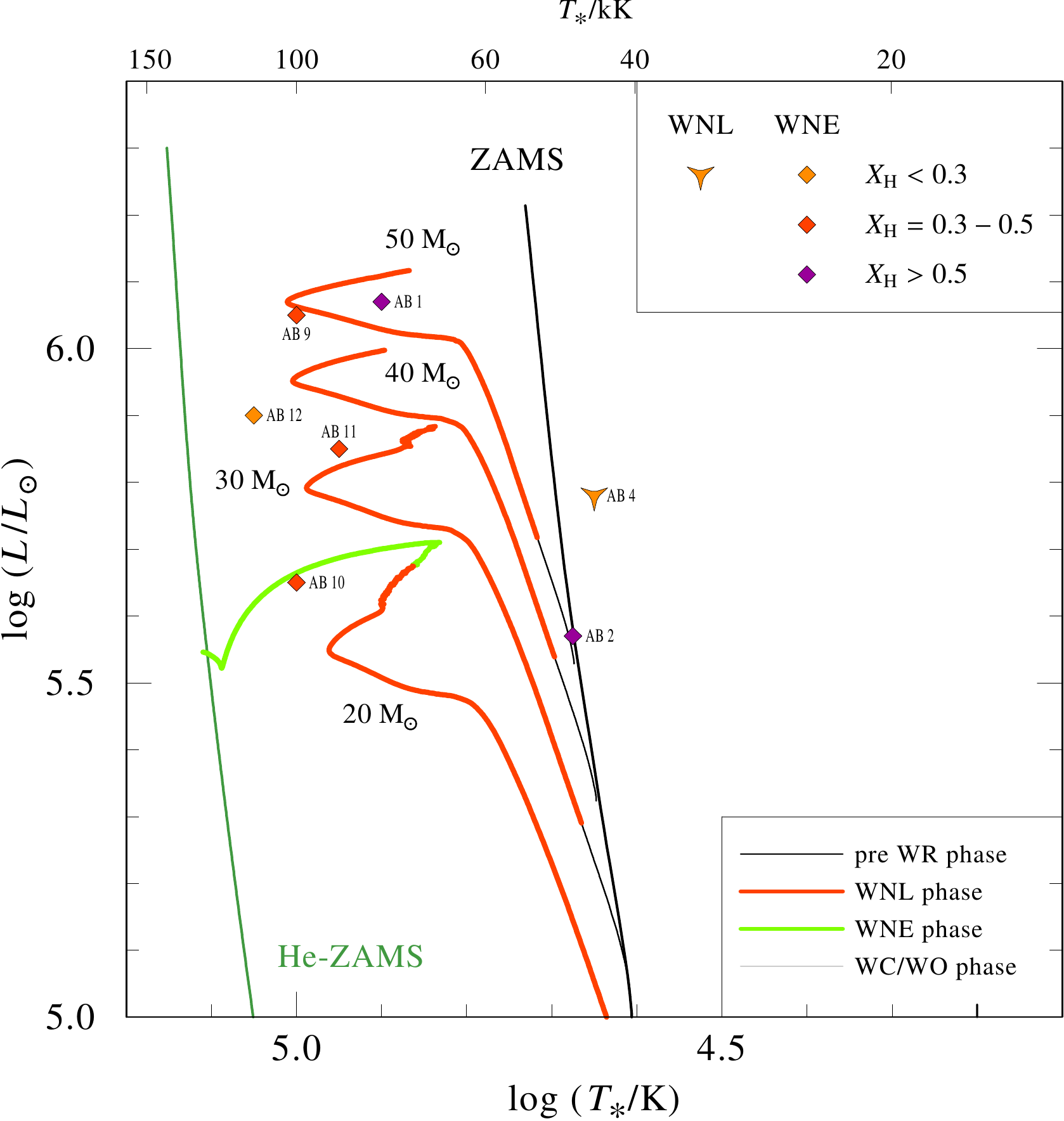}
\caption{The HRD of the single WN stars compared to the evolution tracks that account for very rapid initial rotation \citep[$v_\mathrm{rot} > 500\,\mathrm{km/s}$, ][]{Brott2011}. These evolution models assume an initial metallicity of $0.14\,Z_\odot$ and were calculated roughly until the central hydrogen exhaustion. The labels close to the tracks refer to the initial mass. The color coding in the same as in Fig.\,\ref{fig:hrd_geneva} and Fig.\,\ref{fig:hrd_oldgeneva} (see also the inlet).} 
\label{fig:hrd_brott}
\end{figure}

For SMC\,AB\,1 and 2, this scenario was already proposed by \citet{Martins2009}, mainly because of the high hydrogen abundance and the location of these two objects in the HRD. As evident from Fig.\,\ref{fig:hrd_brott}, stellar evolution tracks \citep{Brott2011}, when accounting for vary rapid initial rotation ($v_\mathrm{rot} > 500\,\mathrm{km/s}$) and a metallicity of $0.14\,Z_\odot$, can at first glance explain the HRD positions of most SMC WN stars. Taking a closer look at the evolution of chemically homogeneous stars, however, reveals that the hydrogen abundance of these models in the corresponding parameter range is considerably lower than derived from the spectral analysis. The $50\,M_\sun$ track close to the position of SMC\,AB\,1, e.g., predicts a hydrogen mass-fraction below $9\,\%$, while our analysis (in agreement with the findings of \citealt{Martins2009}) results in a value of $X_{\element{H}} = 50\,\%$. For comparison, the $80\,M_\sun$ track from \citet{Eldridge2006} exhibit a hydrogen abundance of about $15\,\%$ close to the position of SMC\,AB\,1. 
The disagreement between our results and the evolution tracks published by \citet{Brott2011} is confirmed by the BONNSAI\footnote{The BONNSAI web-service is available at www.astro.uni-bonn.de/stars/bonnsai.} Bayesian tool \citep{Schneider2014}, which allows us to qualitatively match the Bonn stellar evolution models \citep[][]{Brott2011,Koehler2015} to the results obtained from our stellar atmosphere analyses, accurately accounting for observational uncertainties. With the exception of SMC\,AB\,4, the parameters derived in our analysis could not be consistently reproduced by this tool.  
Thus, the discrepancy between the hydrogen abundance predicted by the evolution models and the values derived by means of the spectral analysis does not support the scenario of homogeneous evolution. 

\begin{figure}
\centering
\includegraphics[width=\hsize]{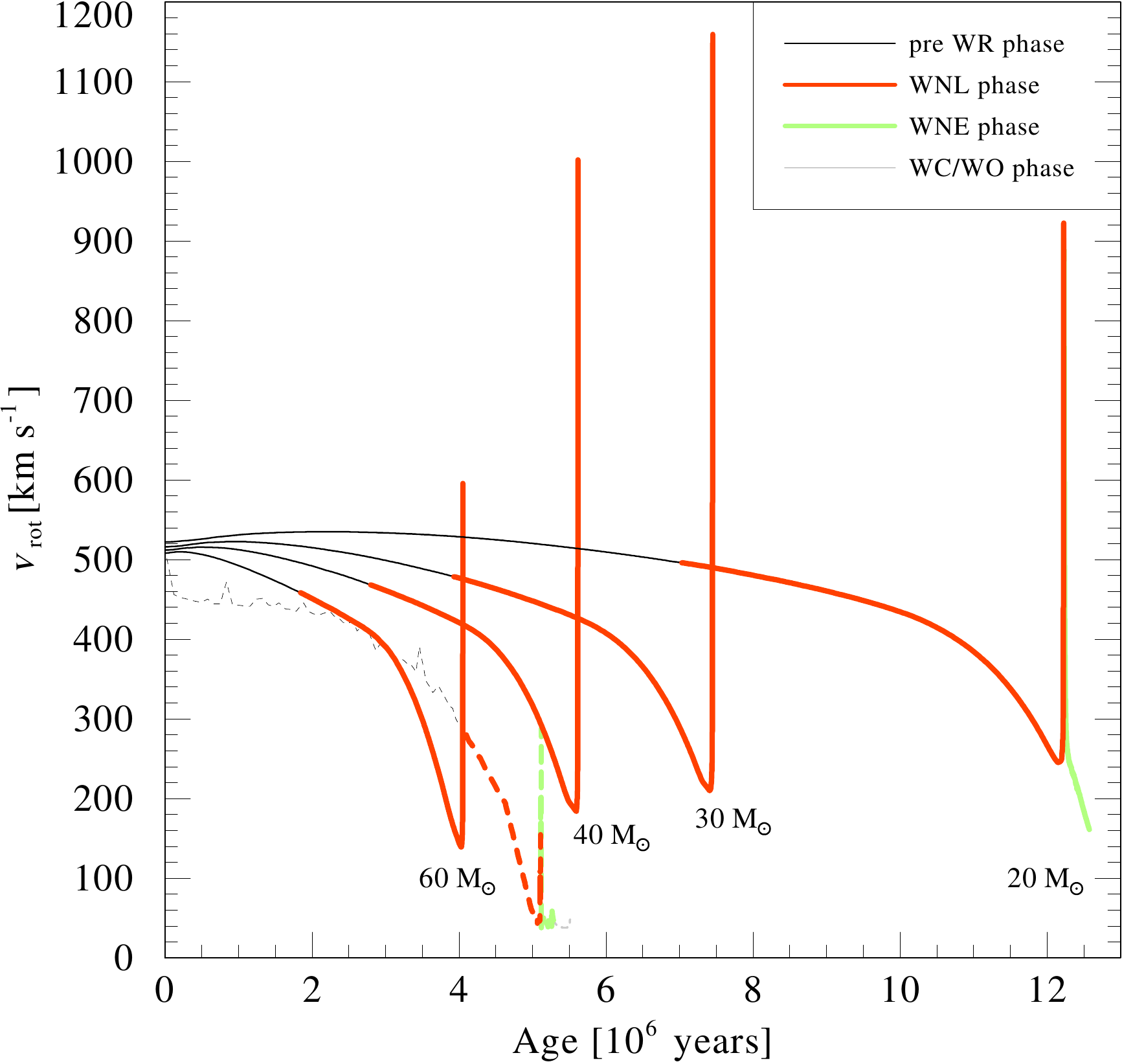}
\caption{The evolution of the equatorial rotational velocity as a function of time for the stellar evolution models calculated by \citet[][solid lines]{Brott2011} and a $60\,M_\odot$ track (dashed line) published by \citet{Meynet2005}. Both studies account for very rapid rotation ($v_\mathrm{rot, ini} > 500\,\mathrm{km/s}$) and an initial metallicity of $0.14\,Z_\odot$. The labels close to the tracks refer to the initial mass. The color coding is the same as in Figs.\,\ref{fig:hrd_geneva} and \ref{fig:hrd_oldgeneva} (see also the inlet).
} 
\label{fig:vrot_brott}
\end{figure}

From homogeneous models, it is expected that the rotational velocity is substantial ($v_\mathrm{rot} > 100\,\mathrm{km/s}$) during a considerable part of the stellar evolution. 
For the quasi-homogeneous models shown in Fig.\,\ref{fig:hrd_brott} \citep[][$v_\mathrm{rot, ini} > 500\,\mathrm{km/s}$]{Brott2011}, the evolution of the equatorial velocity as a function of time is illustrated in Fig.\,\ref{fig:vrot_brott}. The rotational velocity in these models never falls below $100\,\mathrm{km/s}$. 
For comparison we include in this figure a $60\,M_\odot$ track published by \citet{Meynet2005} that also accounts for fast rotation ($v_\mathrm{rot, ini} = 500\,\mathrm{km/s}$) and quasi homogeneous evolution. In contrast to the tracks calculated by \citet{Brott2011}, the Geneva track exhibits considerably lower rotational velocities during the WR phase. 
As discussed in Sect.\,\ref{subsect:rotation}, rotational velocities as high as $200\,\mathrm{km/s}$ at the stellar photosphere could not be excluded for the analyzed WN\,3 stars, as long as the winds do not co-rotate with the photosphere. Thus, homogeneous evolution remains a promising possibility to explain the origin of the WN stars in the SMC.

The influence of binary evolution on the rotational velocities of massive stars was recently studied by \citet{deMink2013} by means of a rapid binary evolution code. Assuming a uniform initial distribution of the rotational velocity with $v_\mathrm{rot} < 200\,\mathrm{km/s}$, these authors show that the rapid rotators observed in populations of main-sequence stars might consist exclusively of stars that already underwent Roche-lobe overflow. Hence, initially fast rotation could be very rare among massive stars. However, \citet{Cantiello2007} demonstrate that binary interaction can provide high rotation rates and quasi-homogeneous evolution very similar to the single star evolution models. 

The effects of binary evolution on WR populations were studied by various authors \citep{Paczynski1967,Vanbeveren2007,Eldridge2008,Eldridge2013}. The initial mass required to enter the WR phase is considerably reduced, if angular momentum and mass transfer by Roche lobe overflow occurs in the course of the evolution. Therefore, binary evolution can in principle account for the HRD positions observed for the SMC WN stars. The current binary status of the WR stars in the SMC was carefully investigated by \citet[][see also Sect.\,\ref{sect:sample}]{Foellmi2003a}, inferring that the stars in our sample are not short periodic binaries (with a lower certainty for SMC\,AB\,9). However, according to \cite{deMink2011,deMink2014}, a significant fraction of the stars that are now single might be products of previous binary interaction. Thus, our sample of apparently single stars could be polluted or even completely consist of objects affected by binary evolution. 
A study of the properties of the SMC WR binaries is currently underway, and should shed light on the systematic differences between the single and binary WR population (Shenar et al., in prep.).

\section{Summary and Conclusions}
\label{sect:conclusions}

We performed a comprehensive study on the putatively single WN stars in the SMC. Optical and, if available, UV spectra were analyzed by means of PoWR model atmospheres. The sample stars prove to be significantly different from their counterparts in the LMC and the Galaxy. In contrast to these galaxies, all SMC WN stars show a significant amount of hydrogen at their surface ($X_{\element{H}} > 0.2$). With the exception of SMC\,AB\,4, all objects exhibit weaker winds. The mass-loss rates are on average lower by about a factor of four and twelve compared to the WN stars in the LMC and MW, respectively. This can be interpreted as a correlation between the metallicity $Z$ and the mass-loss rates in the form $\dot{M} \propto Z^{1.2}$. This finding substantiates that the winds of WN stars are driven by radiation pressure that is intercepted by metal lines. We also derived multidimensional fits to express the mass-loss rate as a function of the stellar parameters. 

Standard stellar evolution models only partly prove to reproduce the empirical HRD positions of the WN stars in the SMC and their observed surface hydrogen abundance. In general, these evolution tracks require an initial mass to reach the WR phase that is relatively high, so that they cannot reproduce the low luminosities obtained for some of the analyzed stars. 
Our results actually agree better with evolution models calculated for higher metallicity. 
This could indicate enhanced or episodic mass-loss in a pre WR stage. Higher mass-loss rates in the WN phase do not agree with our results, since the mass-loss rates applied in stellar evolution models are already higher than we find empirically. Moreover, the high hydrogen abundances derived in our analysis are in general not reproduced by the stellar evolution models. Enhanced mass loss in an earlier phase of the stellar evolution, e.g., in a luminous blue variable (LBV) phase, might explain the blueward evolution of our sample stars, while it could preserve enough of the hydrogen-rich atmosphere to explain the observed surface abundances.

An alternative explanation for the evolutionary origin of the SMC WN stars is by fast rotation which entails a quasi-homogeneous evolution. Corresponding tracks can reproduce the observed HDR positions. However, lower mass-loss rates in the quasi homogeneous evolution models are necessary to bring the predicted hydrogen abundance in accordance with the observations.

We also cannot completely exclude that our sample stars have been affected by binary evolution. 
With respect to the high hydrogen abundances of the SMC WN stars two scenarios seem possible:
Either the objects that we currently observe as single WN stars were mass gainers in interacting binary systems, which were already disrupted, or these objects were formed via merger events.

Summarizing, the Wolf-Rayet population in the Small Magellanic Cloud is found to be strikingly different from corresponding stars in environments with higher metallicity (our Galaxy, the LMC and M\,31). This reveals fundamental differences in how massive stars evolve under metal-poor conditions. 
With a metallicity that is about six times lower than the Sun's, the Small Magellanic Cloud comes close to conditions that are typical for the early Universe.

\begin{acknowledgements}

This research made use of the SIMBAD database, operated at CDS,
Strasbourg, France, and of data products from the Two Micron All Sky
Survey, which is a joint project of the University of Massachusetts and
the Infrared Processing and Analysis Center/California Institute of
Technology, funded by the National Aeronautics and Space Administration
and the National Science Foundation (NASA). 
This work is partly based on INES data from the IUE satellite, and on observations
with the Spitzer Space Telescope, which is operated by the Jet
Propulsion Laboratory, California Institute of Technology under a
contract with NASA. 
This research made also use of NASA's Astrophysics Data System
and of the VizieR catalog access tool, CDS, Strasbourg, France.
Some of the data presented in this paper were retrieved from the
Mikulski Archive for Space Telescopes (MAST). STScI is operated by the
Association of Universities for Research in Astronomy, Inc., under NASA
contract NAS5-26555. Support for MAST for non-HST data is provided by
the NASA Office of Space Science via grant NNX09AF08G. 
The work also based on data 
made available through Wikimbad, hosted in the LAOG, France (http://wikimbad.org). 
Based on data obtained from the ESO Science Archive Facility under request number 95173.
Based on observations made with ESO Telescopes at the La Silla Paranal Observatory under programme ID 077.D-0029.
This publication makes use of data products from the Wide-field Infrared Survey Explorer, which is a joint project of the University of California, Los Angeles, and the Jet Propulsion Laboratory/California Institute of Technology, funded by the National Aeronautics and Space Administration. 
A.S. is supported by the Deutsche Forschungsgemeinschaft (DFG) under grant HA 1455/22. T.S. is grateful for financial support from the Leibniz Graduate School for Quantitative Spectroscopy in Astrophysics, a joint project of the Leibniz Institute for Astrophysics Potsdam (AIP) and the Institute of Physics and Astronomy of the University of Potsdam.

\end{acknowledgements}

\bibliographystyle{aa}
\bibliography{paper}


\Online
\label{onlinematerial}

\begin{appendix} 
\section{Additional tables}
\label{sec:addtables}

\begin{table}[htbp]
\caption{Atomic model used in the stellar atmosphere calculations} 
\label{table:model_atoms}
\centering  
\begin{tabular}{lSc}
\hline\hline 
   Ion &
   \multicolumn{1}{c}{Number of levels} &
   Number of lines \rule[0mm]{0mm}{3.5mm}
   \\
\hline  
   \ion{H}{i}                      &  22   &  231 \rule[0mm]{0mm}{4.0mm}\\
   \ion{H}{ii}                      &  1    & 0   \\
   \ion{He}{i}                     &  35    & 595 \\
   \ion{He}{ii}                    &  26   &  325 \\
   \ion{He}{iii}                   &  1    &  0   \\
   \ion{N}{i}                      &  1    &  45  \\
   \ion{N}{ii}                     &  38   &  703 \\
   \ion{N}{iii}                    &  36   &  630 \\
   \ion{N}{iv}                    &  38    &  703 \\
   \ion{N}{v}                     &  20    &  190 \\
   \ion{N}{vi}                    &  1    &   91  \\
   \ion{C}{i}                      &  1    &  105 \\
   \ion{C}{ii}                     &  32   &  496 \\
   \ion{C}{iii}                    &  40   &  780 \\
   \ion{C}{iv}                    &  21    &  300 \\
   \ion{C}{v}                     &  1    &   10  \\
   \ion{O}{ii}                     &  37    & 666 \\
   \ion{O}{iii}                    &  33    & 528 \\
   \ion{O}{iv}                    &  25    &  300 \\
   \ion{O}{v}                     &  36    &  630 \\
   \ion{O}{vi}                     &  16   &  120 \\
   \ion{O}{vii}                    &  15   &  105 \\
   \ion{P}{iv}                    &  12    &  66  \\
   \ion{P}{v}                     &  11    &  55  \\
   \ion{P}{vi}                     &  1    &  0   \\
   \ion{G}{ii}\tablefootmark{a}    &  1   &   0   \\
   \ion{G}{iii}\tablefootmark{a}   &   8  &   25  \\
   \ion{G}{iv}\tablefootmark{a}   &  11  &    49  \\
   \ion{G}{v}\tablefootmark{a}    &  13  &    69  \\
   \ion{G}{vi}\tablefootmark{a}   &  17  &    121 \\
   \ion{G}{vii}\tablefootmark{a}  &  11  &    52  \\
   \ion{G}{viii}\tablefootmark{a}  &  9  &    34  \\
   \ion{G}{ix}\tablefootmark{a}   &  9  &     35  \\
   \ion{G}{x}\tablefootmark{a}    &  1  &     0   \\
\hline 
\end{tabular}
\tablefoot{
\tablefoottext{a}{G denotes a generic atom which incorporates the following iron group elements \element{Fe}, \element{Sc}, \element{Ti}, \element{Cr}, \element{Mn}, \element{Co}, and \element{Ni}. The  corresponding ions are treated by means of a superlevel approach \citep[for details see][]{Graefener2002}.}
}
\end{table}

\begin{table}
\caption{The spectra used in the analysis of the putatively single WN stars in the SMC}
\label{table:spectra}
\centering  
\begin{tabular}{cccccc}
\hline\hline  \rule[0mm]{0mm}{4.0mm} 
   SMC\,AB &
   FUSE & 
   IUE &
   UVES &
   FMG\tablefootmark{a}  &
   TM\tablefootmark{b} 
   \\
\hline \rule[0mm]{0mm}{4.0mm} 
   1   &               & $\times$ & $\times$ & $\times$ & $\times$ \\
   2   & $\times$ & $\times$ & $\times$ & $\times$ & $\times$ \\
   4   & $\times$ & $\times$ & $\times$ & $\times$ & $\times$ \\
   9   & $\times$ &              &               & $\times$ & \\
   10 & $\times$ &               &              & $\times$ & \\
   11 &               &              &               & $\times$ & \\
   12 &               &              &               & $\times$ & \\
\hline 
\end{tabular}
\tablefoot{A $\times$ indicates observation used in our analysis. 
\tablefoottext{a}{observations conducted by \citet{Foellmi2003a} and \citet{Foellmi2004}}
\tablefoottext{b}{flux calibrated observations conducted by \citet{Torres-Dodgen1988}}
}
\end{table}

\begin{table}
\caption{Abundances derived in the analysis of the putatively single WN stars in the SMC (mass fraction in percent)} 
\label{table:derived_abundances}
\centering  
\begin{tabular}{cScSc}
\hline\hline  \rule[0mm]{0mm}{4.0mm} 
   SMC\,AB &
   \multicolumn{1}{c}{$X_{\element{N}}$} & 
   $X_{\element{C}}$ &
   \multicolumn{1}{c}{$X_{\element{O}}$} &
   $X_{\element{P}}$ 
   \\
\hline \rule[0mm]{0mm}{4.0mm} 
   1   & 0.225      &  -           & \multicolumn{1}{c}{-} & -             \\
   2   & 0.45       &  0.005       & \multicolumn{1}{c}{-} & 0.00006  \\
   4   & 0.675      &  -           & \multicolumn{1}{c}{-} & 0.00006   \\
   9   & 0.45       &  -           & 0.00075               & -             \\
   10  & 0.45       &  -           & 0.0025                & -             \\
   11  & 0.225      &  -           & \multicolumn{1}{c}{-} & -             \\
   12  & 0.9        &  -           & \multicolumn{1}{c}{-} & -             \\
\hline 
\end{tabular}
\tablefoot{
When a specific abundance is not determined in the fit, it is depicted by a dash.
}
\end{table}

\begin{table}
\caption{Number of ionizing photons and Zanstra temperatures for the putatively single WN stars in the SMC} 
\label{table:zansT}
\centering  
\begin{tabular}{ccSccS}
\hline\hline  \rule[0mm]{0mm}{4.0mm} 
   SMC\,AB &
   \multicolumn{2}{c}{\ion{H}{i}}  & 
   \ion{He}{i} &
   \multicolumn{2}{c}{\ion{He}{ii}}
   \\
    &  
    $\log Q$ &
    \multicolumn{1}{c}{$T_\mathrm{Zanstra}$} &
    $\log Q$ &
    $\log Q$ &
    \multicolumn{1}{c}{$T_\mathrm{Zanstra}$}
    \\
    &  
    $[\mathrm{s^{-1}}]$ &
    \multicolumn{1}{c}{$[\mathrm{K}]$} &
    $[\mathrm{s^{-1}}]$ &
    $[\mathrm{s^{-1}}]$ &
    \multicolumn{1}{c}{$[\mathrm{K}]$}
    \\
\hline \rule[0mm]{0mm}{4.0mm} 
\input{table-zansT.tex}
\hline 
\end{tabular}
\end{table}

\begin{table}
\caption{Mass-loss rates and luminosities derived by \citet{Nugis2007}}
\label{table:comp_mdot}
\centering  
\begin{tabular}{c|SS|SS}
\hline\hline  \rule[0mm]{0mm}{4.0mm} 
   SMC\,AB &
   \multicolumn{1}{c}{$\log \dot{M}_\mathrm{Nugis}$} &
   \multicolumn{1}{c}{$\log \dot{M}~$\tablefootmark{a}} & 
   \multicolumn{1}{c}{$\log {L}_\mathrm{Nugis}$} & 
   \multicolumn{1}{c}{$\log {L}~$\tablefootmark{a}} 
   \\
    &
    $[M_{\odot}/\mathrm{yr}]$ &
    $[M_{\odot}/\mathrm{yr}]$ &
    $[L_{\odot}]$  &
    $[L_{\odot}]$
   \\
\hline \rule[0mm]{0mm}{4.0mm} 
   1   & -5.41    &   -5.58      &  5.88      &  6.07         \\
   2   & -5.58    &   -5.75      &  6.24      &  5.57         \\
   9   & -5.62    &   -5.65      &  5.62      &  6.05         \\
   10  & -5.7     &   -5.64      &  5.38      &  5.65         \\
   11  & -5.69    &   -5.56      &  5.59      &  5.85         \\
   12  & -5.85    &   -5.79      &  5.23      &  5.90          \\
\hline 
\end{tabular}
\tablefoot{
\tablefoottext{a}{this work}
}
\end{table}

\clearpage

\section{Comments on individual stars}
\label{sect:comments}

\paragraph{SMC\,AB\,1} (alias: AzV\,2a, SMC-WR1): We achieved the best fit with a stellar temperature of $79.4\,\mathrm{kK}$, while \citet{Martins2009} derived a value of $65.4\,\mathrm{kK}$. At this low temperature, however, all observed \ion{N}{v} lines are considerably underpredicted by our models. In addition, the synthetic spectra of these models posses a distinct \ion{N}{iv}\,$\lambda\,4060$ emission line that is not observed for this object. The \ion{N}{iv} emission in the model spectra already vanishes at approximately $70\,\mathrm{kK}$. 
Because of a better fit of the \ion{N}{v} lines we favor, however, the model with a slightly higher temperature. 
Another indication of the higher temperature is the \ion{C}{iv}\,$\lambda\,1551$ line, which also vanishes at approximately $70\,\mathrm{kK}$ and which is indeed not observed for this object.
The higher temperature in comparison to the findings by \citet{Martins2009} also results in a stellar luminosity which is $0.32\,\mathrm{dex}$ higher. 
Although both studies are based on the same ESO VLT UVES observations, the reduced data have some minor differences. Furthermore, the differences in the stellar parameters are partly due to the fact that the models do not reproduce all lines in the observed spectrum
consistently. Hence the derived stellar parameters depend on which lines are considered to be most significant for the line
fit.
We note that the mass-loss rate is almost identical in both studies.

\paragraph{SMC\,AB\,2} (alias: AzV\,39a, SMC-WR2): 
\citet{Testor2001} found this star to be embedded in the H\,II region N\,28.
The stellar parameters of SMC\,AB\,2 were already studied by \citet{Martins2009}. Within the expected error margins, the stellar temperature and the luminosity obtained in this study are compatible with the findings of the previous authors. \citet{Martins2009} assume a clumping factor of $D = 10$, while our fit points to a lower value of about four. Taking this into account, the mass-loss rates obtained in both studies deviate by only $0.1\,\mathrm{dex}$, which is within the expected error range as well.

\paragraph{SMC\,AB\,4} (alias: AzV\,81, Sk\,41, SMC-WR4): It is the only WN6h star in the SMC which appears to be single \citep{Foellmi2003a}. It was analyzed before by \citet{Crowther2000} and \citet{Martins2009}. Compared to the findings of these two studies, our analysis results in a slightly higher stellar temperature (by approximately $2-3\,\mathrm{kK}$) and a mass-loss rate that is lower by about 0.1\,dex when the different clumping factors are taken into account. The luminosity deduced from the SED fit lies within the range determined by the previous analyses \citep{Crowther2000,Martins2009}, while the deviation amounts to approximately 0.1\,dex in both cases. We also note that the hydrogen abundance deduced in our analysis is lower by 5\,\% and 10\,\% in relation to the results of \citet{Crowther2000} and \citet{Martins2009}, respectively.

\paragraph{SMC\,AB\,9} (alias: MVD\,1, SMC-WR9): This object belongs to the hottest stars in the sample. The temperature determination is base primarily on the antagonistic behavior of the nitrogen lines and the helium lines (especially \ion{He}{ii}\,$\lambda\,5412$). While the nitrogen lines tend to favor lower temperatures, a superior fit of the \ion{He}{ii}\,$\lambda\,5412$ line requires higher model temperatures. We note that a weak \ion{O}{iii}\,$\lambda\,5007$ nebular line is visible in the optical spectrum, which could indicate an associated nebula.  

Unfortunately, the quality of the fit in the FUSE range is poor, since the continuum shape of the FUSE spectra cannot be reproduced by the models. This seems to be an issue of the observation rather than a model deficiency. The observed flux shortward of 1000\,\AA\ is approximately a factor of two lower than predicted by the model. Since no significant ionization edge is expected in this spectral region, we assume that the observation at hand is affected by inaccurate flux in at least one of the two SiC channels (SiC refers to the silicon carbide coating of the corresponding mirror) of the FUSE satellite. This is consistent with a bump in the spectrum at about 1085\,\AA, where a gap between the segments of the LiF channels (LiF refers to the lithium fluoride coating of the corresponding mirror) is covered by the SiC channels.

\paragraph{SMC\,AB\,10} (alias: SMC-WR10): The fit of the FUSE spectral range suffers from the same caveats as described for SMC\,AB\,9 (see corresponding comment). 
In contrast to the other objects in our sample, the best fitting model assumes a velocity field in the form of a double-$\beta$ law \citep{Hillier1999} as defined by \citet{Todt2015} with $\beta_{1} = 1$ and $\beta_{2} = 4$.  
For SMC\,AB\,10, the choice of a double-$\beta$ law is mainly motivated the exceptional line profile of the \ion{He}{ii}\,$\lambda\,4686$ line that has a broad base and a very narrow line peak. 
With respect to a model using a single-$\beta$ law with $\beta = 1$, the mass-loss rate is increased by about 0.1\,dex and the terminal wind velocity is higher by about 600\,km/s.

\begin{figure}
\centering
\includegraphics[width=\hsize]{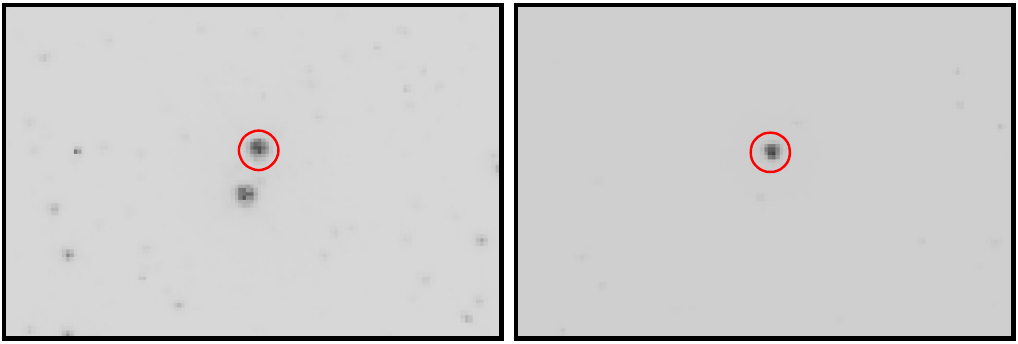}
\caption{HST-WFPC2 images of the region within about $8\,\arcsec$ distance of SMC\,AB\,11. Left panel: Image taken in the F606W filter. Right panel: Image taken in the F300W filter. The red circle marks the position of SMC\,AB\,11. The images were retrieved from the HST archive. North is up and east is left.} 
\label{fig:hst_ab11}
\end{figure}

\paragraph{SMC\,AB\,11} (alias: SMC-WR11): A bright red object is located about $1.2\,\arcsec$ from SMC\,AB\,11 \citep[see also\ ][]{Crowther2006b}. Since the spectra at hand \citep{Foellmi2003a} have a spatial resolution on the order of $1\,\arcsec$, we note the possibility of a contamination. Figure\,\ref{fig:hst_ab11} shows two images taken with the Wide-Field Planetary Camera 2 (WFPC2) aboard the {\em Hubble Space Telescope} (HST) in the F300W and F606W filter, respectively. The position of SMC\,AB\,11 is marked with a red circle. 
The HST observations clearly show that the WR star dominates the flux in the UV, while both objects exhibit approximately the same flux level in the F606W filter. 
In the IR on the other hand, the flux is dominated by the non-WR object. 

\begin{figure*}
  \centering
  \includegraphics[trim=0cm 6.0cm 0cm 0cm,width=0.92\textwidth]{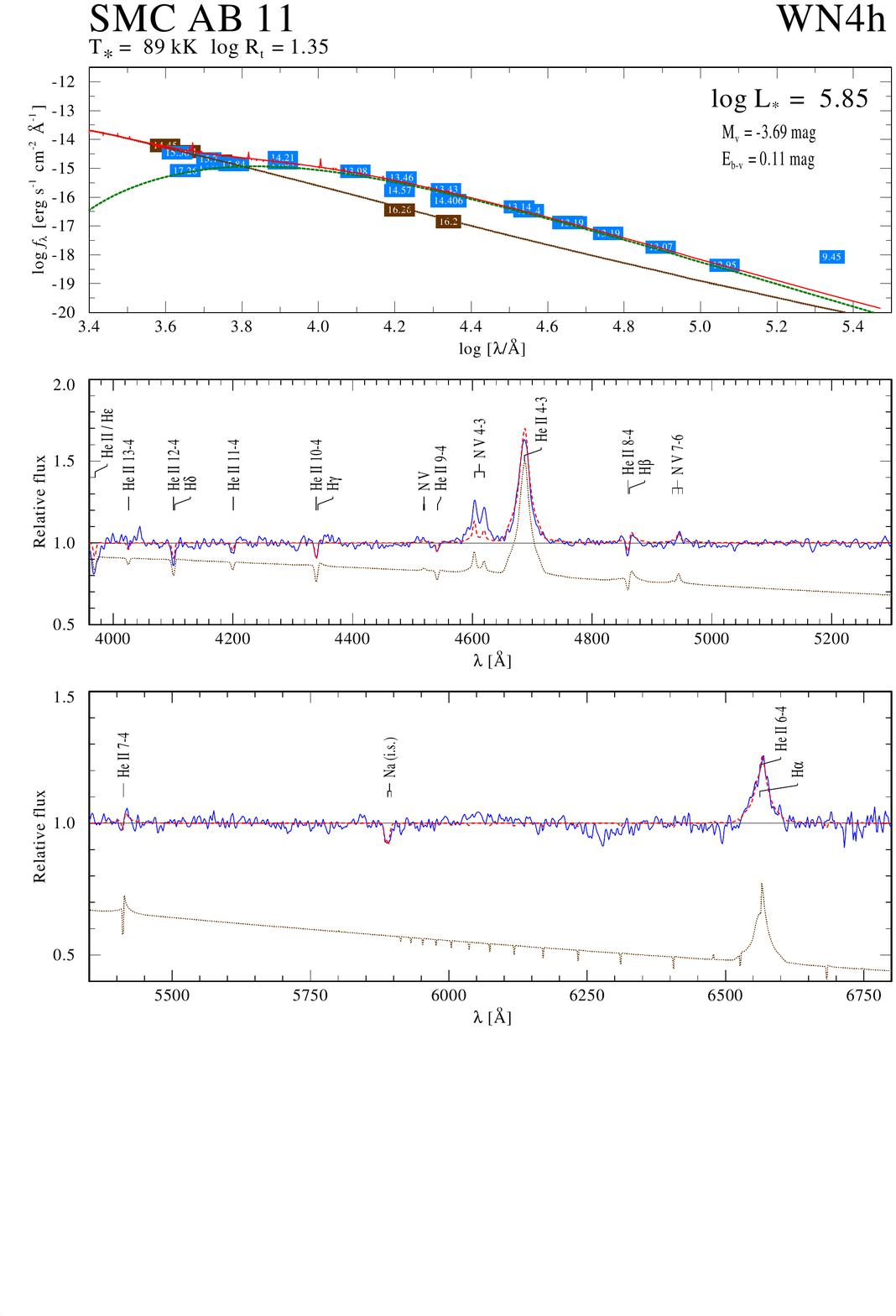}
  \caption{The spectral fit for SMC\,AB\,11. The brown boxes in the top panel are the photometry attributed to the WR star alone, while the blue boxes refer to photometry that comprises the flux from a companion separated by approximately $1.2\,\arcsec$ (see Appendix\,\ref{sect:comments} for details). The brown dotted line represents the best fitting PoWR model for the WN star, while the green dashed line shows a black body spectrum at $4.6\,\mathrm{kK}$ as a rough approximation to the nearby red star. The photometry with large apertures (blue boxes) are obviously determined by the red star for $\lambda \gtrsim 8000$\,\AA. The straight red line is the combined model SED that incorporates the flux from the WN star and the line-of-sight companion.}
  \label{fig:ab11}
\end{figure*}

The first panel of Figure\,\ref{fig:ab11} shows the SED fit derived in our analysis. Since no flux-calibrated spectra are available for this object, the luminosity are derived solely from photometry. A variety of photometric measurements is available within a radius of $2\,\arcsec$ from the coordinates given for SMC\,AB\,11 in the SIMBAD database. In comparison to the line-of-sight companion, SMC\,AB\,11 is located farther to the north (see Fig.\,\ref{fig:hst_ab11}). Since the measurements performed by \citet{Zaritsky2002} and \citet{Kato2007} were obtained at a similar position, we mainly used these values as well as the spectrophotometry obtained by \citet{Crowther2006b} to constrain the luminosity of SMC\,AB\,11. In Fig.\,\ref{fig:ab11}, the corresponding photometry marks and the model SED for SMC\,AB\,11 are shown by brown boxes and a brown dotted line, respectively. Other photometry \citep{Zaritsky2002,Monet2003,Oey2004,Gordon2011,Cutri2012a,Cutri2012b}, which presumably incorporates contributions from both objects, were used to construct the observed SED that combines both objects (blue boxes in Fig.\,\ref{fig:ab11}). In this figure, the red straight line represents the model that incorporates both the WN star as well as flux contribution from the non-WR object, which is assumed to be a black body (green dashed line). Because of the flux contribution of the line-of-sight companion, the uncertainty in the derived luminosity of SMC\,AB\,11 is higher than for the other stars in our sample. In the lower panels of Fig.\,\ref{fig:ab11}, the relative contribution of the WN star to the combined spectrum (red dashed line) is shown by the brown dotted line. 

\citet{Foellmi2003a} mentions the presence of \ion{N}{iv}\,$\lambda\,4060$ emission in an ESO-NTT spectrum of SMC\,AB\,11, although neither the spectrum at hand nor the spectrum published by \citet{Massey2001} exhibits this line. 
We note that stellar atmosphere models with temperatures of about $65\,\mathrm{kK}$ would be necessary to reproduce this line simultaneously with the \ion{N}{v} emission lines. This would be considerably lower than the $89\,\mathrm{kK}$ derived in this analysis. 

\paragraph{SMC\,AB\,12} (alias: SMC-WR12): It is the object with the highest surface temperature in our sample. In the same way as the other WN3 stars only one ionization stage per element is visible in our spectra of SMC\,AB\,12, so that the temperature determination has to rely on the \ion{N}{V} lines and the \ion{He}{ii} lines. At least $112\,\mathrm{kK}$ are necessary to reproduce the \ion{He}{ii}\,$\lambda\,5412$ line. Moreover at these temperatures the model spectra do not show \ion{He}{ii} lines originating from the fifth quantum level, which is consistent with the observations.

\section{Spectral fits}
\label{sect:specfits}

The following figures show the final spectral fit of each object analyzed in this paper. In all plots the upper panel shows the fit of the SED, while the lower panels exhibit the fit of the normalized optical and UV spectrum, if available. The best-fitting model is plotted in red, while all observations are shown in blue.

\clearpage

\begin{figure*}
  \centering
  \includegraphics[width=0.92\textwidth]{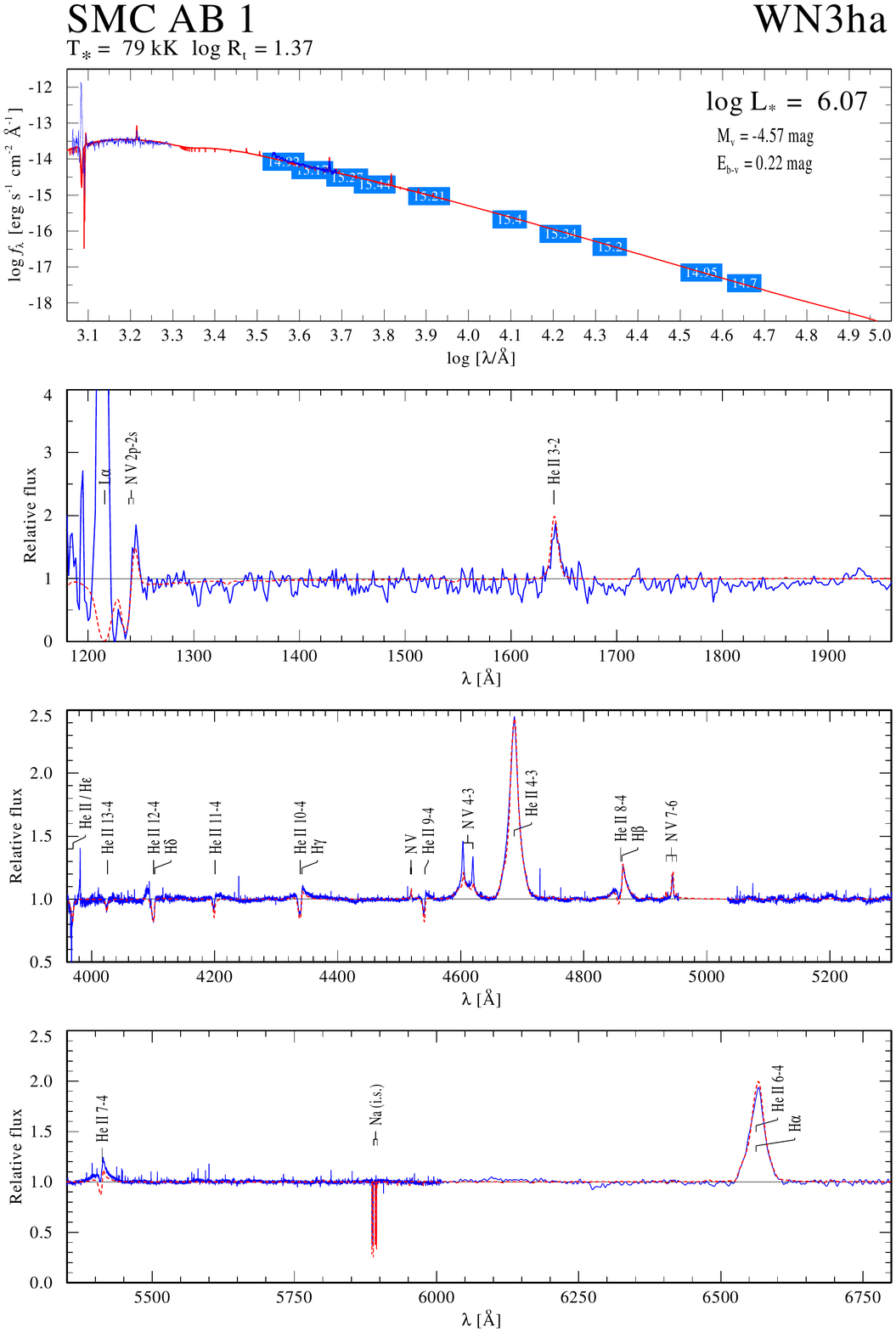}
  \caption{Spectral fit for SMC\,AB\,1}
  \label{fig:ab1}
\end{figure*}

\clearpage

\begin{figure*}
  \centering
  \includegraphics[page=1,width=0.92\textwidth]{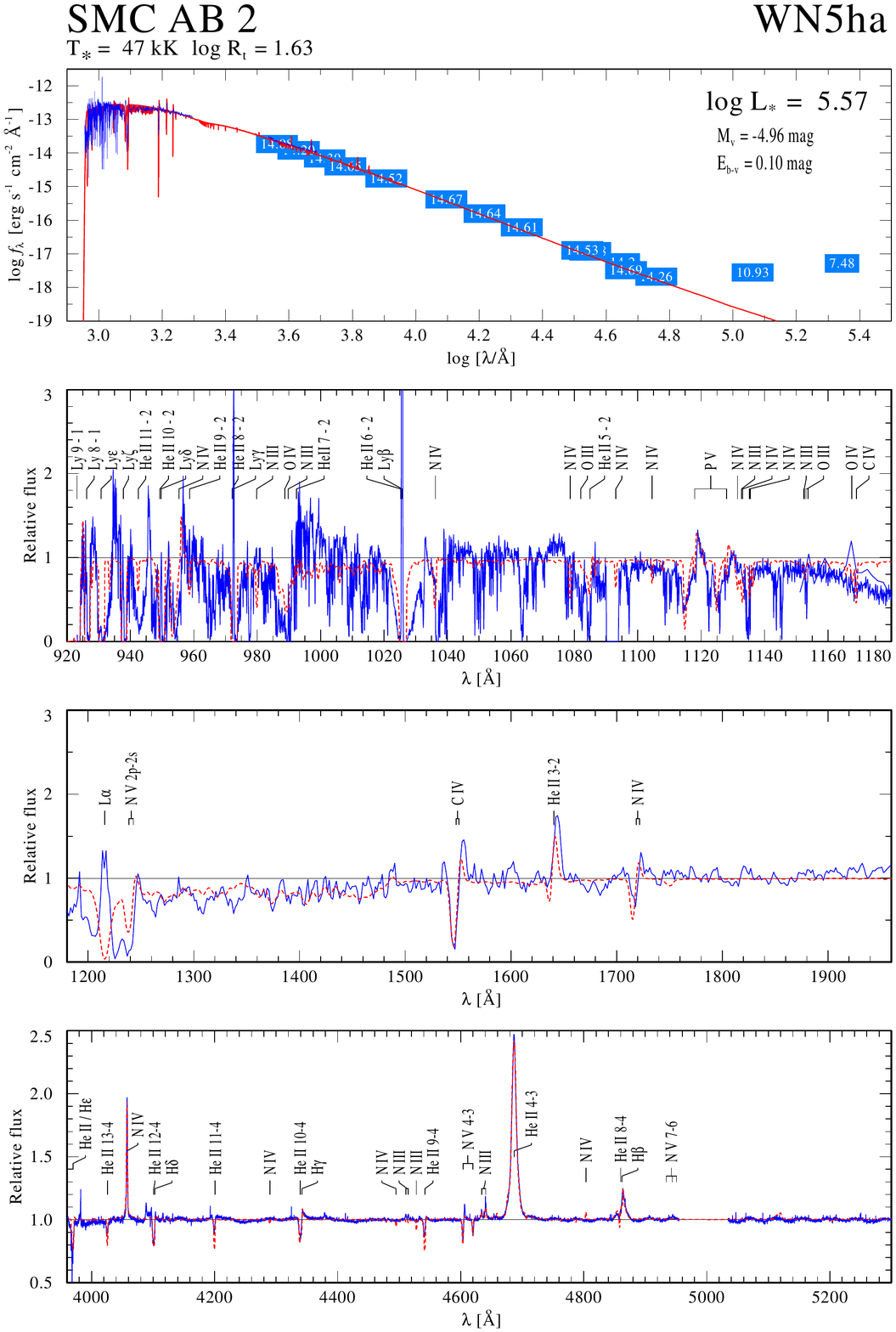}
  \caption{Spectral fit for SMC\,AB\,2}
  \label{fig:ab2}
\end{figure*}

\begin{figure*}
  \centering
  \includegraphics[page=2,trim=0cm 21cm 0cm 0cm,width=0.92\textwidth]{online/ab2.pdf}
  \caption{Spectral fit for SMC\,AB\,2 (continued)}
\end{figure*}

\clearpage

\begin{figure*}
  \centering
  \includegraphics[page=1,width=0.92\textwidth]{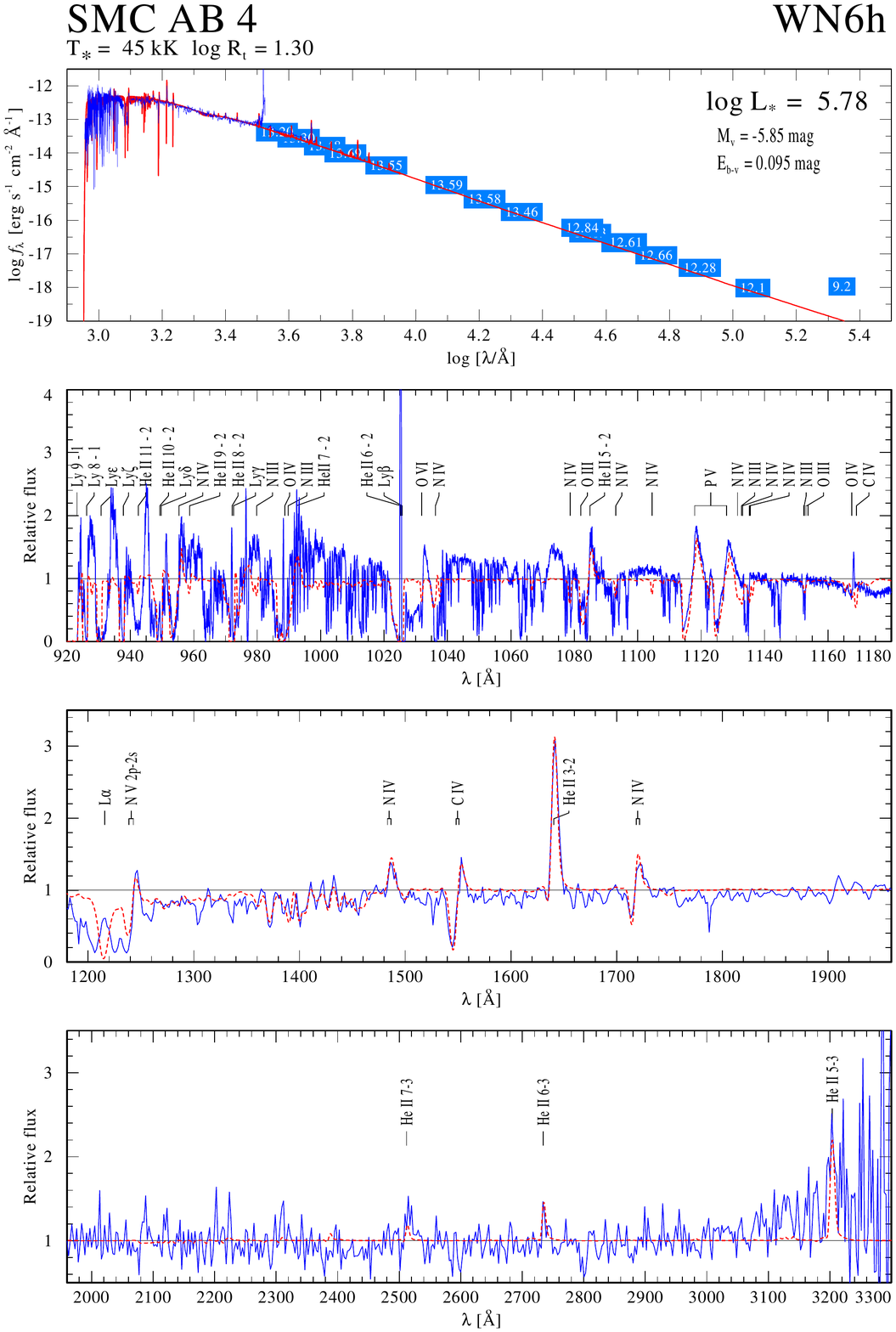}
  \caption{Spectral fit for SMC\,AB\,4}
  \label{fig:ab4}
\end{figure*}

\begin{figure*}
  \centering
  \includegraphics[page=2,trim=0cm 14.5cm 0cm 0cm,width=0.92\textwidth]{online/ab4.pdf}
  \caption{Spectral fit for SMC\,AB\,4 (continued)}
\end{figure*}

\clearpage

\begin{figure*}
  \centering
  \includegraphics[width=0.92\textwidth]{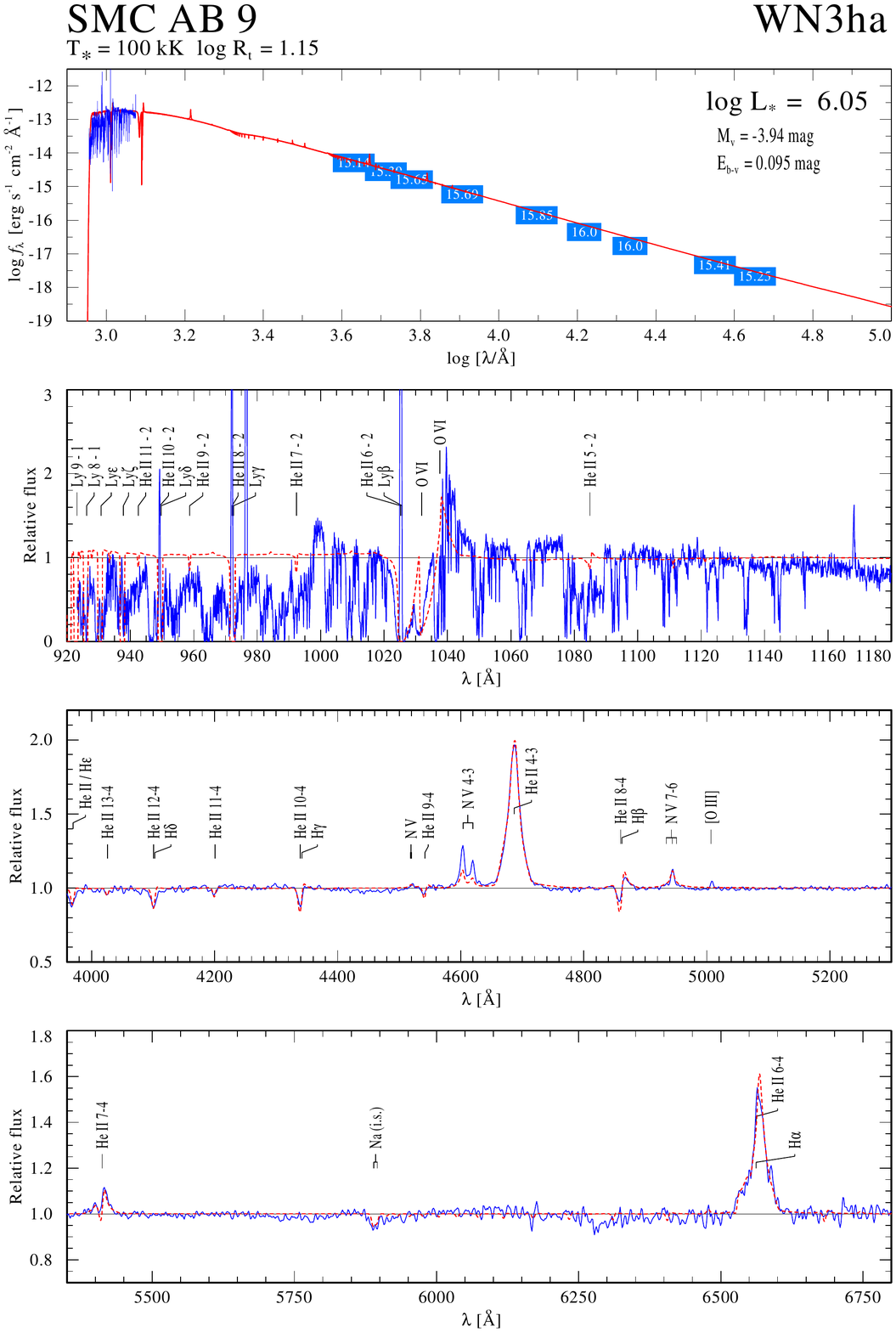}
  \caption{Spectral fit for SMC\,AB\,9}
  \label{fig:ab9}
\end{figure*}

\clearpage

\begin{figure*}
  \centering
  \includegraphics[width=0.92\textwidth]{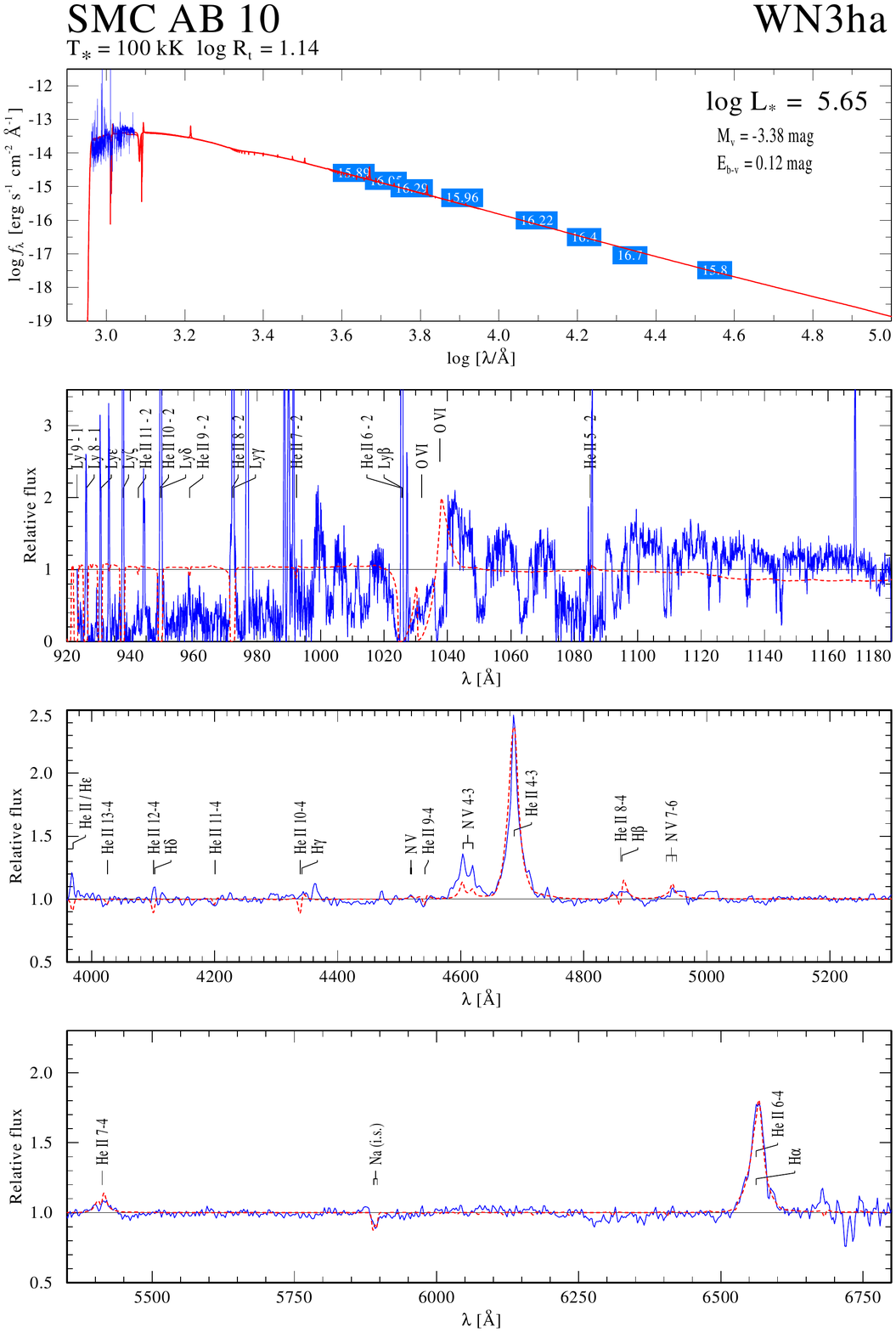}
  \caption{Spectral fit for SMC\,AB\,10}
  \label{fig:ab10}
\end{figure*}
\clearpage

\clearpage

\begin{figure*}
  \centering
  \includegraphics[trim=0cm 7.0cm 0cm 0cm,width=0.92\textwidth]{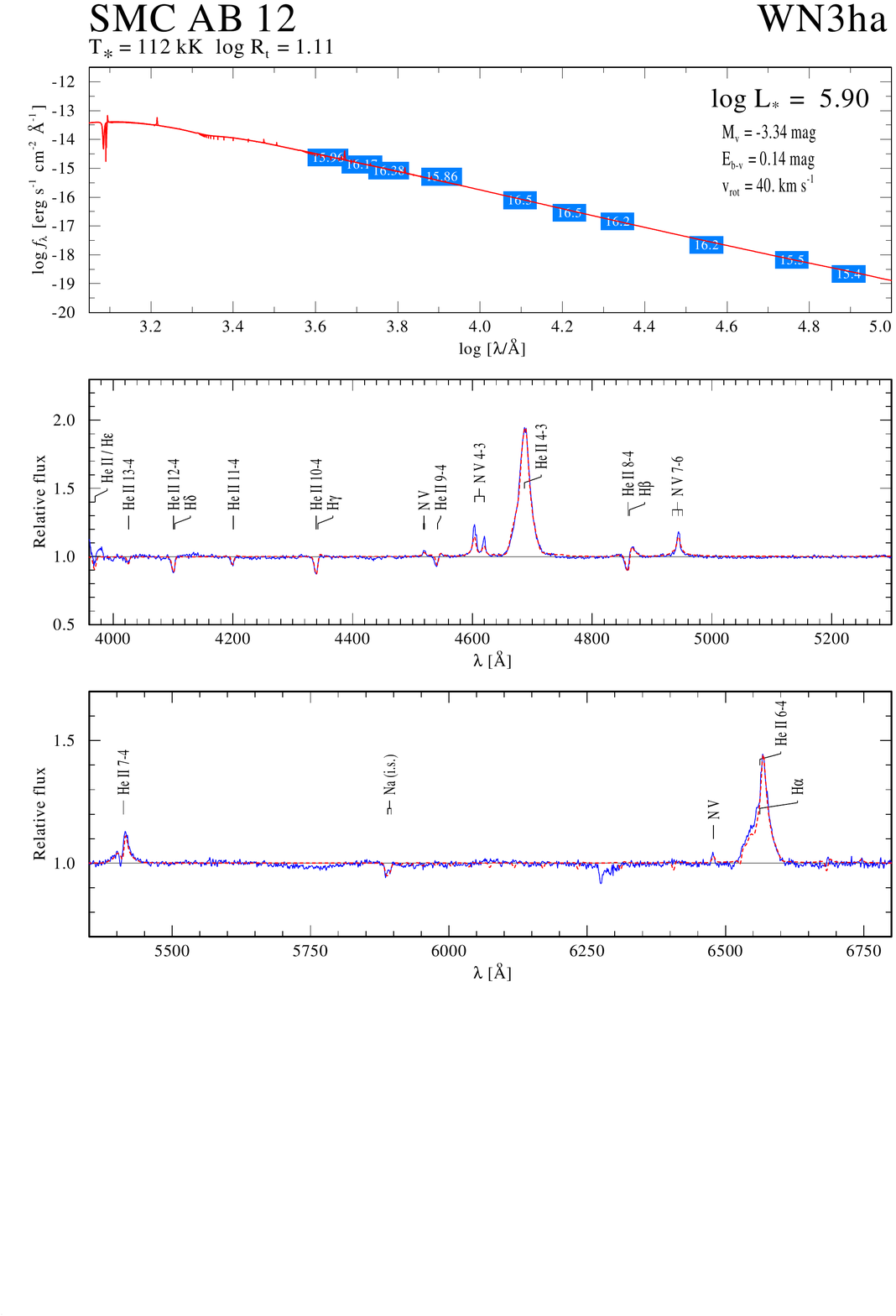}
  \caption{Spectral fit for SMC\,AB\,12}
  \label{fig:ab12}
\end{figure*}

\end{appendix} 

\end{document}

%% file: table-parameters.tex
1  & WN3ha 	& 79 	& 1.37 & 1700 	& 0.22 & -4.57 & 5.7 	& -5.58 & 10 	& 6.07 & 0.2 & 75 & 0.5 &  \\ 
2  & WN5ha 	& 47 	& 1.63 & 900 	& 0.10 & -4.96 & 9.1 	& -5.75 & 4 	& 5.57 & 0.2 & 43 & 0.55 &  \\ 
4  & WN6h 	& 45 	& 1.30 & 1000 	& 0.09 & -5.85 & 13.0 	& -5.18 & 10 	& 5.78 & 0.5 & 38 & 0.25 &  \\ 
9  & WN3ha 	& 100 	& 1.22 & 1800 	& 0.09 & -3.94 & 3.5 	& -5.65 & 10 	& 6.05 & 0.2 & 62 & 0.35 &  \\ 
10  & WN3ha 	& 100 	& 1.18 & 2000 	& 0.12 & -3.38 & 2.2 	& -5.64 & 4 	& 5.65 & 0.5 & 37 & 0.35 &  \\ 
11  & WN4h:a 	& 89 	& 1.35 & 2200 	& 0.11 & -3.69 & 3.5 	& -5.56 & 4 	& 5.85 & 0.4 & 50 & 0.4 & composite \\ 
12  & WN3ha 	& 112 	& 1.14 & 1800 	& 0.14 & -3.34 & 2.4 	& -5.79 & 10 	& 5.90 & 0.2 & 42 & 0.2 &  \\ 

%% file: table-zansT.tex
1  &  49.94  &  87.7 &  49.68 &  47.68 &  70.2 \\ 
2  &  49.40  &  51.5 &  48.78 &  38.40 &  20.3 \\ 
4  &  49.61  &  47.4 &  48.94 &  37.51 &  18.6 \\ 
9  &  49.90  &  106.6 &  49.70 &  48.49 &  96.3 \\ 
10  &  49.50  &  104.9 &  49.30 &  48.10 &  95.8 \\ 
11  &  49.72  &  99.3 &  49.49 &  47.88 &  82.0 \\ 
12  &  49.73  &  118.0 &  49.55 &  48.57 &  109.3 \\ 

%% file: paper.bbl
\begin{thebibliography}{122}
\expandafter\ifx\csname natexlab\endcsname\relax\def\natexlab#1{#1}\fi

\bibitem[{{Abbott}(1982)}]{Abbott1982}
{Abbott}, D.~C. 1982, \apj, 259, 282

\bibitem[{{Asplund} {et~al.}(2009){Asplund}, {Grevesse}, {Sauval}, \&
  {Scott}}]{Asplund2009}
{Asplund}, M., {Grevesse}, N., {Sauval}, A.~J., \& {Scott}, P. 2009, \araa, 47,
  481

\bibitem[{{Azzopardi} \& {Breysacher}(1979)}]{Azzopardi1979}
{Azzopardi}, M. \& {Breysacher}, J. 1979, \aap, 75, 120

\bibitem[{{Balser} {et~al.}(2011){Balser}, {Rood}, {Bania}, \&
  {Anderson}}]{Balser2011}
{Balser}, D.~S., {Rood}, R.~T., {Bania}, T.~M., \& {Anderson}, L.~D. 2011,
  \apj, 738, 27

\bibitem[{{Barniske} {et~al.}(2008){Barniske}, {Oskinova}, \&
  {Hamann}}]{Barniske2008}
{Barniske}, A., {Oskinova}, L.~M., \& {Hamann}, W.-R. 2008, \aap, 486, 971

\bibitem[{{Bessell}(1991)}]{Bessell1991}
{Bessell}, M.~S. 1991, \aap, 242, L17

\bibitem[{{Bestenlehner} {et~al.}(2014){Bestenlehner}, {Gr{\"a}fener}, {Vink},
  {Najarro}, {de Koter}, {Sana}, {Evans}, {Crowther}, {H{\'e}nault-Brunet},
  {Herrero}, {Langer}, {Schneider}, {Sim{\'o}n-D{\'{\i}}az}, {Taylor}, \&
  {Walborn}}]{Bestenlehner2014}
{Bestenlehner}, J.~M., {Gr{\"a}fener}, G., {Vink}, J.~S., {et~al.} 2014, \aap,
  570, A38

\bibitem[{{Bonanos} {et~al.}(2010){Bonanos}, {Lennon}, {K{\"o}hlinger}, {van
  Loon}, {Massa}, {Sewilo}, {Evans}, {Panagia}, {Babler}, {Block}, {Bracker},
  {Engelbracht}, {Gordon}, {Hora}, {Indebetouw}, {Meade}, {Meixner}, {Misselt},
  {Robitaille}, {Shiao}, \& {Whitney}}]{Bonanos2010}
{Bonanos}, A.~Z., {Lennon}, D.~J., {K{\"o}hlinger}, F., {et~al.} 2010, \aj,
  140, 416

\bibitem[{{Bouret} {et~al.}(2003){Bouret}, {Lanz}, {Hillier}, {Heap}, {Hubeny},
  {Lennon}, {Smith}, \& {Evans}}]{Bouret2003}
{Bouret}, J.-C., {Lanz}, T., {Hillier}, D.~J., {et~al.} 2003, \apj, 595, 1182

\bibitem[{{Bovy} {et~al.}(2014){Bovy}, {Nidever}, {Rix}, {Girardi}, {Zasowski},
  {Chojnowski}, {Holtzman}, {Epstein}, {Frinchaboy}, {Hayden}, {Rodrigues},
  {Majewski}, {Johnson}, {Pinsonneault}, {Stello}, {Allende Prieto}, {Andrews},
  {Basu}, {Beers}, {Bizyaev}, {Burton}, {Chaplin}, {Cunha}, {Elsworth},
  {Garc{\'{\i}}a}, {Garc{\'{\i}}a-Her{\'n}andez}, {Garc{\'{\i}}a P{\'e}rez},
  {Hearty}, {Hekker}, {Kallinger}, {Kinemuchi}, {Koesterke},
  {M{\'e}sz{\'a}ros}, {Mosser}, {O'Connell}, {Oravetz}, {Pan}, {Robin},
  {Schiavon}, {Schneider}, {Schultheis}, {Serenelli}, {Shetrone}, {Silva
  Aguirre}, {Simmons}, {Skrutskie}, {Smith}, {Stassun}, {Weinberg}, {Wilson},
  \& {Zamora}}]{Bovy2014}
{Bovy}, J., {Nidever}, D.~L., {Rix}, H.-W., {et~al.} 2014, \apj, 790, 127

\bibitem[{{Brott} {et~al.}(2011){Brott}, {de Mink}, {Cantiello}, {Langer}, {de
  Koter}, {Evans}, {Hunter}, {Trundle}, \& {Vink}}]{Brott2011}
{Brott}, I., {de Mink}, S.~E., {Cantiello}, M., {et~al.} 2011, \aap, 530, A115

\bibitem[{{Cantiello} {et~al.}(2007){Cantiello}, {Yoon}, {Langer}, \&
  {Livio}}]{Cantiello2007}
{Cantiello}, M., {Yoon}, S.-C., {Langer}, N., \& {Livio}, M. 2007, \aap, 465,
  L29

\bibitem[{{Castor} {et~al.}(1975){Castor}, {Abbott}, \& {Klein}}]{Castor1975}
{Castor}, J.~I., {Abbott}, D.~C., \& {Klein}, R.~I. 1975, \apj, 195, 157

\bibitem[{{Castor} \& {Lamers}(1979)}]{Castor1979}
{Castor}, J.~I. \& {Lamers}, H.~J.~G.~L.~M. 1979, \apjs, 39, 481

\bibitem[{{Conti} {et~al.}(1989){Conti}, {Garmany}, \& {Massey}}]{Conti1989}
{Conti}, P.~S., {Garmany}, C.~D., \& {Massey}, P. 1989, \apj, 341, 113

\bibitem[{{Crowther}(2000)}]{Crowther2000}
{Crowther}, P.~A. 2000, \aap, 356, 191

\bibitem[{{Crowther}(2006)}]{Crowther2006}
{Crowther}, P.~A. 2006, in Astronomical Society of the Pacific Conference
  Series, Vol. 353, Stellar Evolution at Low Metallicity: Mass Loss,
  Explosions, Cosmology, ed. H.~J.~G.~L.~M. {Lamers}, N.~{Langer}, T.~{Nugis},
  \& K.~{Annuk}, 157

\bibitem[{{Crowther} \& {Hadfield}(2006)}]{Crowther2006b}
{Crowther}, P.~A. \& {Hadfield}, L.~J. 2006, \aap, 449, 711

\bibitem[{{Cutri} \& {et al.}(2012)}]{Cutri2012a}
{Cutri}, R.~M. \& {et al.} 2012, VizieR Online Data Catalog, 2311, 0

\bibitem[{{Cutri} {et~al.}(2012){Cutri}, {Skrutskie}, {van Dyk}, {Beichman},
  {Carpenter}, {Chester}, {Cambresy}, {Evans}, {Fowler}, {Gizis}, {Howard},
  {Huchra}, {Jarrett}, {Kopan}, {Kirkpatrick}, {Light}, {Marsh}, {McCallon},
  {Schneider}, {Stiening}, {Sykes}, {Weinberg}, {Wheaton}, {Wheelock}, \&
  {Zacharias}}]{Cutri2012b}
{Cutri}, R.~M., {Skrutskie}, M.~F., {van Dyk}, S., {et~al.} 2012, VizieR Online
  Data Catalog, 2281, 0

\bibitem[{{de Koter}(2006)}]{deKoter2006}
{de Koter}, A. 2006, in Astronomical Society of the Pacific Conference Series,
  Vol. 353, Stellar Evolution at Low Metallicity: Mass Loss, Explosions,
  Cosmology, ed. H.~J.~G.~L.~M. {Lamers}, N.~{Langer}, T.~{Nugis}, \&
  K.~{Annuk}, 99

\bibitem[{{de Mink} {et~al.}(2011){de Mink}, {Langer}, \&
  {Izzard}}]{deMink2011}
{de Mink}, S.~E., {Langer}, N., \& {Izzard}, R.~G. 2011, Bulletin de la Societe
  Royale des Sciences de Liege, 80, 543

\bibitem[{{de Mink} {et~al.}(2013){de Mink}, {Langer}, {Izzard}, {Sana}, \& {de
  Koter}}]{deMink2013}
{de Mink}, S.~E., {Langer}, N., {Izzard}, R.~G., {Sana}, H., \& {de Koter}, A.
  2013, \apj, 764, 166

\bibitem[{{de Mink} {et~al.}(2014){de Mink}, {Sana}, {Langer}, {Izzard}, \&
  {Schneider}}]{deMink2014}
{de Mink}, S.~E., {Sana}, H., {Langer}, N., {Izzard}, R.~G., \& {Schneider},
  F.~R.~N. 2014, \apj, 782, 7

\bibitem[{{Ekstr{\"o}m} {et~al.}(2012){Ekstr{\"o}m}, {Georgy}, {Eggenberger},
  {Meynet}, {Mowlavi}, {Wyttenbach}, {Granada}, {Decressin}, {Hirschi},
  {Frischknecht}, {Charbonnel}, \& {Maeder}}]{Ekstroem2012}
{Ekstr{\"o}m}, S., {Georgy}, C., {Eggenberger}, P., {et~al.} 2012, \aap, 537,
  A146

\bibitem[{{Eldridge} {et~al.}(2013){Eldridge}, {Fraser}, {Smartt}, {Maund}, \&
  {Crockett}}]{Eldridge2013}
{Eldridge}, J.~J., {Fraser}, M., {Smartt}, S.~J., {Maund}, J.~R., \&
  {Crockett}, R.~M. 2013, \mnras, 436, 774

\bibitem[{{Eldridge} {et~al.}(2008){Eldridge}, {Izzard}, \&
  {Tout}}]{Eldridge2008}
{Eldridge}, J.~J., {Izzard}, R.~G., \& {Tout}, C.~A. 2008, \mnras, 384, 1109

\bibitem[{{Eldridge} \& {Vink}(2006)}]{Eldridge2006}
{Eldridge}, J.~J. \& {Vink}, J.~S. 2006, \aap, 452, 295

\bibitem[{{Foellmi}(2004)}]{Foellmi2004}
{Foellmi}, C. 2004, \aap, 416, 291

\bibitem[{{Foellmi} {et~al.}(2003){Foellmi}, {Moffat}, \&
  {Guerrero}}]{Foellmi2003a}
{Foellmi}, C., {Moffat}, A.~F.~J., \& {Guerrero}, M.~A. 2003, \mnras, 338, 360

\bibitem[{{Georgy}(2012)}]{Georgy2012}
{Georgy}, C. 2012, \aap, 538, L8

\bibitem[{{Georgy} {et~al.}(2013){Georgy}, {Ekstr{\"o}m}, {Eggenberger},
  {Meynet}, {Haemmerl{\'e}}, {Maeder}, {Granada}, {Groh}, {Hirschi}, {Mowlavi},
  {Yusof}, {Charbonnel}, {Decressin}, \& {Barblan}}]{Georgy2013}
{Georgy}, C., {Ekstr{\"o}m}, S., {Eggenberger}, P., {et~al.} 2013, \aap, 558,
  A103

\bibitem[{{Georgy} {et~al.}(2012){Georgy}, {Ekstr{\"o}m}, {Meynet}, {Massey},
  {Levesque}, {Hirschi}, {Eggenberger}, \& {Maeder}}]{Georgy2012b}
{Georgy}, C., {Ekstr{\"o}m}, S., {Meynet}, G., {et~al.} 2012, \aap, 542, A29

\bibitem[{{Gordon} {et~al.}(2003){Gordon}, {Clayton}, {Misselt}, {Landolt}, \&
  {Wolff}}]{Gordon2003}
{Gordon}, K.~D., {Clayton}, G.~C., {Misselt}, K.~A., {Landolt}, A.~U., \&
  {Wolff}, M.~J. 2003, \apj, 594, 279

\bibitem[{{Gordon} {et~al.}(2011){Gordon}, {Meixner}, {Meade}, {Whitney},
  {Engelbracht}, {Bot}, {Boyer}, {Lawton}, {Sewi{\l}o}, {Babler}, {Bernard},
  {Bracker}, {Block}, {Blum}, {Bolatto}, {Bonanos}, {Harris}, {Hora},
  {Indebetouw}, {Misselt}, {Reach}, {Shiao}, {Tielens}, {Carlson},
  {Churchwell}, {Clayton}, {Chen}, {Cohen}, {Fukui}, {Gorjian}, {Hony},
  {Israel}, {Kawamura}, {Kemper}, {Leroy}, {Li}, {Madden}, {Marble},
  {McDonald}, {Mizuno}, {Mizuno}, {Muller}, {Oliveira}, {Olsen}, {Onishi},
  {Paladini}, {Paradis}, {Points}, {Robitaille}, {Rubin}, {Sandstrom}, {Sato},
  {Shibai}, {Simon}, {Smith}, {Srinivasan}, {Vijh}, {Van Dyk}, {van Loon}, \&
  {Zaritsky}}]{Gordon2011}
{Gordon}, K.~D., {Meixner}, M., {Meade}, M.~R., {et~al.} 2011, \aj, 142, 102

\bibitem[{{Gr{\"a}fener} \& {Hamann}(2005)}]{Graefener2005}
{Gr{\"a}fener}, G. \& {Hamann}, W.-R. 2005, \aap, 432, 633

\bibitem[{{Gr{\"a}fener} \& {Hamann}(2008)}]{Graefener2008}
{Gr{\"a}fener}, G. \& {Hamann}, W.-R. 2008, \aap, 482, 945

\bibitem[{{Gr{\"a}fener} {et~al.}(2002){Gr{\"a}fener}, {Koesterke}, \&
  {Hamann}}]{Graefener2002}
{Gr{\"a}fener}, G., {Koesterke}, L., \& {Hamann}, W.-R. 2002, \aap, 387, 244

\bibitem[{{Gr{\"a}fener} {et~al.}(2011){Gr{\"a}fener}, {Vink}, {de Koter}, \&
  {Langer}}]{Graefener2011}
{Gr{\"a}fener}, G., {Vink}, J.~S., {de Koter}, A., \& {Langer}, N. 2011, \aap,
  535, A56

\bibitem[{{Grieve} \& {Madore}(1986)}]{Grieve1986}
{Grieve}, G.~R. \& {Madore}, B.~F. 1986, \apjs, 62, 427

\bibitem[{{Hainich} {et~al.}(2014){Hainich}, {R{\"u}hling}, {Todt}, {Oskinova},
  {Liermann}, {Gr{\"a}fener}, {Foellmi}, {Schnurr}, \& {Hamann}}]{Hainich2014}
{Hainich}, R., {R{\"u}hling}, U., {Todt}, H., {et~al.} 2014, \aap, 565, A27

\bibitem[{{Hamann} \& {Gr{\"a}fener}(2003)}]{Hamann2003}
{Hamann}, W.-R. \& {Gr{\"a}fener}, G. 2003, \aap, 410, 993

\bibitem[{{Hamann} \& {Gr{\"a}fener}(2004)}]{Hamann2004}
{Hamann}, W.-R. \& {Gr{\"a}fener}, G. 2004, \aap, 427, 697

\bibitem[{{Hamann} {et~al.}(2006){Hamann}, {Gr{\"a}fener}, \&
  {Liermann}}]{Hamann2006}
{Hamann}, W.-R., {Gr{\"a}fener}, G., \& {Liermann}, A. 2006, \aap, 457, 1015

\bibitem[{{Hamann} \& {Koesterke}(1998)}]{Hamann1998}
{Hamann}, W.-R. \& {Koesterke}, L. 1998, \aap, 335, 1003

\bibitem[{{Hamann} {et~al.}(1995){Hamann}, {Koesterke}, \&
  {Wessolowski}}]{Hamann1995}
{Hamann}, W.-R., {Koesterke}, L., \& {Wessolowski}, U. 1995, \aap, 299, 151

\bibitem[{{Harries} {et~al.}(2003){Harries}, {Hilditch}, \&
  {Howarth}}]{Harries2003}
{Harries}, T.~J., {Hilditch}, R.~W., \& {Howarth}, I.~D. 2003, \mnras, 339, 157

\bibitem[{{Haschke} {et~al.}(2011){Haschke}, {Grebel}, \&
  {Duffau}}]{Haschke2011}
{Haschke}, R., {Grebel}, E.~K., \& {Duffau}, S. 2011, \aj, 141, 158

\bibitem[{{Hayden} {et~al.}(2014){Hayden}, {Holtzman}, {Bovy}, {Majewski},
  {Johnson}, {Allende Prieto}, {Beers}, {Cunha}, {Frinchaboy}, {Garc{\'{\i}}a
  P{\'e}rez}, {Girardi}, {Hearty}, {Lee}, {Nidever}, {Schiavon}, {Schlesinger},
  {Schneider}, {Schultheis}, {Shetrone}, {Smith}, {Zasowski}, {Bizyaev},
  {Feuillet}, {Hasselquist}, {Kinemuchi}, {Malanushenko}, {Malanushenko},
  {O'Connell}, {Pan}, \& {Stassun}}]{Hayden2014}
{Hayden}, M.~R., {Holtzman}, J.~A., {Bovy}, J., {et~al.} 2014, \aj, 147, 116

\bibitem[{{Hilditch} {et~al.}(2005){Hilditch}, {Howarth}, \&
  {Harries}}]{Hilditch2005}
{Hilditch}, R.~W., {Howarth}, I.~D., \& {Harries}, T.~J. 2005, \mnras, 357, 304

\bibitem[{{Hillier}(1991)}]{Hillier1991}
{Hillier}, D.~J. 1991, \aap, 247, 455

\bibitem[{{Hillier} \& {Miller}(1998)}]{Hillier1998}
{Hillier}, D.~J. \& {Miller}, D.~L. 1998, \apj, 496, 407

\bibitem[{{Hillier} \& {Miller}(1999)}]{Hillier1999}
{Hillier}, D.~J. \& {Miller}, D.~L. 1999, \apj, 519, 354

\bibitem[{{Hunter} {et~al.}(2007){Hunter}, {Dufton}, {Smartt}, {Ryans},
  {Evans}, {Lennon}, {Trundle}, {Hubeny}, \& {Lanz}}]{Hunter2007}
{Hunter}, I., {Dufton}, P.~L., {Smartt}, S.~J., {et~al.} 2007, \aap, 466, 277

\bibitem[{{Kato} {et~al.}(2007){Kato}, {Nagashima}, {Nagayama}, {Kurita},
  {Koerwer}, {Kawai}, {Yamamuro}, {Zenno}, {Nishiyama}, {Baba}, {Kadowaki},
  {Haba}, {Hatano}, {Shimizu}, {Nishimura}, {Nagata}, {Sato}, {Murai},
  {Kawazu}, {Nakajima}, {Nakaya}, {Kandori}, {Kusakabe}, {Ishihara},
  {Kaneyasu}, {Hashimoto}, {Tamura}, {Tanab{\'e}}, {Ita}, {Matsunaga},
  {Nakada}, {Sugitani}, {Wakamatsu}, {Glass}, {Feast}, {Menzies}, {Whitelock},
  {Fourie}, {Stoffels}, {Evans}, \& {Hasegawa}}]{Kato2007}
{Kato}, D., {Nagashima}, C., {Nagayama}, T., {et~al.} 2007, \pasj, 59, 615

\bibitem[{{Keller} \& {Wood}(2006)}]{Keller2006}
{Keller}, S.~C. \& {Wood}, P.~R. 2006, \apj, 642, 834

\bibitem[{{K{\"o}hler} {et~al.}(2015){K{\"o}hler}, {Langer}, {de Koter}, {de
  Mink}, {Crowther}, {Evans}, {Gr{\"a}fener}, {Sana}, {Sanyal}, {Schneider}, \&
  {Vink}}]{Koehler2015}
{K{\"o}hler}, K., {Langer}, N., {de Koter}, A., {et~al.} 2015, \aap, 573, A71

\bibitem[{{Korn} {et~al.}(2000){Korn}, {Becker}, {Gummersbach}, \&
  {Wolf}}]{Korn2000}
{Korn}, A.~J., {Becker}, S.~R., {Gummersbach}, C.~A., \& {Wolf}, B. 2000, \aap,
  353, 655

\bibitem[{{Kudritzki} {et~al.}(1995){Kudritzki}, {Lennon}, \&
  {Puls}}]{Kudritzki1995}
{Kudritzki}, R.-P., {Lennon}, D.~J., \& {Puls}, J. 1995, in Science with the
  VLT, ed. J.~R. {Walsh} \& I.~J. {Danziger}, 246

\bibitem[{{Kudritzki} {et~al.}(1987){Kudritzki}, {Pauldrach}, \&
  {Puls}}]{Kudritzki1987}
{Kudritzki}, R.~P., {Pauldrach}, A., \& {Puls}, J. 1987, \aap, 173, 293

\bibitem[{{Kudritzki} {et~al.}(1999){Kudritzki}, {Puls}, {Lennon}, {Venn},
  {Reetz}, {Najarro}, {McCarthy}, \& {Herrero}}]{Kudritzki1999}
{Kudritzki}, R.~P., {Puls}, J., {Lennon}, D.~J., {et~al.} 1999, \aap, 350, 970

\bibitem[{{Kurt} \& {Dufour}(1998)}]{Kurt1998}
{Kurt}, C.~M. \& {Dufour}, R.~J. 1998, in Revista Mexicana de Astronomia y
  Astrofisica Conference Series, Vol.~7, Revista Mexicana de Astronomia y
  Astrofisica Conference Series, ed. R.~J. {Dufour} \& S.~{Torres-Peimbert},
  202

\bibitem[{{Langer}(1992)}]{Langer1992}
{Langer}, N. 1992, \aap, 265, L17

\bibitem[{{Langer}(2012)}]{Langer2012}
{Langer}, N. 2012, \araa, 50, 107

\bibitem[{{Langer} {et~al.}(1994){Langer}, {Hamann}, {Lennon}, {Najarro},
  {Pauldrach}, \& {Puls}}]{Langer1994}
{Langer}, N., {Hamann}, W.-R., {Lennon}, M., {et~al.} 1994, \aap, 290, 819

\bibitem[{{Larsen} {et~al.}(2000){Larsen}, {Clausen}, \& {Storm}}]{Larsen2000}
{Larsen}, S.~S., {Clausen}, J.~V., \& {Storm}, J. 2000, \aap, 364, 455

\bibitem[{{Liermann} {et~al.}(2010){Liermann}, {Hamann}, {Oskinova}, {Todt}, \&
  {Butler}}]{Liermann2010}
{Liermann}, A., {Hamann}, W.-R., {Oskinova}, L.~M., {Todt}, H., \& {Butler}, K.
  2010, \aap, 524, A82

\bibitem[{{L{\'o}pez-S{\'a}nchez} \& {Esteban}(2010)}]{Lopez-Sanchez2010}
{L{\'o}pez-S{\'a}nchez}, {\'A}.~R. \& {Esteban}, C. 2010, \aap, 516, A104

\bibitem[{{Maeder}(1987)}]{Maeder1987}
{Maeder}, A. 1987, \aap, 178, 159

\bibitem[{{Martins} {et~al.}(2009){Martins}, {Hillier}, {Bouret}, {Depagne},
  {Foellmi}, {Marchenko}, \& {Moffat}}]{Martins2009}
{Martins}, F., {Hillier}, D.~J., {Bouret}, J.~C., {et~al.} 2009, \aap, 495, 257

\bibitem[{{Martins} {et~al.}(2008){Martins}, {Hillier}, {Paumard},
  {Eisenhauer}, {Ott}, \& {Genzel}}]{Martins2008}
{Martins}, F., {Hillier}, D.~J., {Paumard}, T., {et~al.} 2008, \aap, 478, 219

\bibitem[{{Massey} \& {Duffy}(2001)}]{Massey2001}
{Massey}, P. \& {Duffy}, A.~S. 2001, \apj, 550, 713

\bibitem[{{Massey} {et~al.}(2003){Massey}, {Olsen}, \& {Parker}}]{Massey2003}
{Massey}, P., {Olsen}, K.~A.~G., \& {Parker}, J.~W. 2003, \pasp, 115, 1265

\bibitem[{{Mayor}(1976)}]{Mayor1976}
{Mayor}, M. 1976, \aap, 48, 301

\bibitem[{{Meynet} {et~al.}(2015){Meynet}, {Chomienne}, {Ekstr{\"o}m},
  {Georgy}, {Granada}, {Groh}, {Maeder}, {Eggenberger}, {Levesque}, \&
  {Massey}}]{Meynet2015}
{Meynet}, G., {Chomienne}, V., {Ekstr{\"o}m}, S., {et~al.} 2015, \aap, 575, A60

\bibitem[{{Meynet} \& {Maeder}(2003)}]{Meynet2003}
{Meynet}, G. \& {Maeder}, A. 2003, \aap, 404, 975

\bibitem[{{Meynet} \& {Maeder}(2005)}]{Meynet2005}
{Meynet}, G. \& {Maeder}, A. 2005, \aap, 429, 581

\bibitem[{{Mokiem} {et~al.}(2007){Mokiem}, {de Koter}, {Vink}, {Puls}, {Evans},
  {Smartt}, {Crowther}, {Herrero}, {Langer}, {Lennon}, {Najarro}, \&
  {Villamariz}}]{Mokiem2007}
{Mokiem}, M.~R., {de Koter}, A., {Vink}, J.~S., {et~al.} 2007, \aap, 473, 603

\bibitem[{{Monet} {et~al.}(2003){Monet}, {Levine}, {Canzian}, {Ables}, {Bird},
  {Dahn}, {Guetter}, {Harris}, {Henden}, {Leggett}, {Levison}, {Luginbuhl},
  {Martini}, {Monet}, {Munn}, {Pier}, {Rhodes}, {Riepe}, {Sell}, {Stone},
  {Vrba}, {Walker}, {Westerhout}, {Brucato}, {Reid}, {Schoening}, {Hartley},
  {Read}, \& {Tritton}}]{Monet2003}
{Monet}, D.~G., {Levine}, S.~E., {Canzian}, B., {et~al.} 2003, \aj, 125, 984

\bibitem[{{Morgan} {et~al.}(1991){Morgan}, {Vassiliadis}, \&
  {Dopita}}]{Morgan1991}
{Morgan}, D.~H., {Vassiliadis}, E., \& {Dopita}, M.~A. 1991, \mnras, 251, 51P

\bibitem[{{Niedzielski} {et~al.}(2004){Niedzielski}, {Nugis}, \&
  {Skorzynski}}]{Niedzielski2004}
{Niedzielski}, A., {Nugis}, T., \& {Skorzynski}, W. 2004, Acta Astronomica, 54,
  405

\bibitem[{{Nordstr{\"o}m} {et~al.}(2004){Nordstr{\"o}m}, {Mayor}, {Andersen},
  {Holmberg}, {Pont}, {J{\o}rgensen}, {Olsen}, {Udry}, \&
  {Mowlavi}}]{Nordstroem2004}
{Nordstr{\"o}m}, B., {Mayor}, M., {Andersen}, J., {et~al.} 2004, \aap, 418, 989

\bibitem[{{Nugis} {et~al.}(2007){Nugis}, {Annuk}, \& {Hirv}}]{Nugis2007}
{Nugis}, T., {Annuk}, K., \& {Hirv}, A. 2007, Baltic Astronomy, 16, 227

\bibitem[{{Nugis} \& {Lamers}(2000)}]{Nugis2000}
{Nugis}, T. \& {Lamers}, H.~J.~G.~L.~M. 2000, \aap, 360, 227

\bibitem[{{Oey} {et~al.}(2004){Oey}, {King}, \& {Parker}}]{Oey2004}
{Oey}, M.~S., {King}, N.~L., \& {Parker}, J.~W. 2004, \aj, 127, 1632

\bibitem[{{Oskinova} {et~al.}(2013){Oskinova}, {Steinke}, {Hamann}, {Sander},
  {Todt}, \& {Liermann}}]{Oskinova2013}
{Oskinova}, L.~M., {Steinke}, M., {Hamann}, W.-R., {et~al.} 2013, \mnras, 436,
  3357

\bibitem[{{Paczy{\'n}ski}(1967)}]{Paczynski1967}
{Paczy{\'n}ski}, B. 1967, \actaa, 17, 355

\bibitem[{{Pauldrach} {et~al.}(1986){Pauldrach}, {Puls}, \&
  {Kudritzki}}]{Pauldrach1986}
{Pauldrach}, A., {Puls}, J., \& {Kudritzki}, R.~P. 1986, \aap, 164, 86

\bibitem[{{Piatti}(2011)}]{Piatti2011}
{Piatti}, A.~E. 2011, \mnras, 418, L69

\bibitem[{{Piatti}(2012)}]{Piatti2012}
{Piatti}, A.~E. 2012, \mnras, 422, 1109

\bibitem[{{Piatti} \& {Geisler}(2013)}]{Piatti2013}
{Piatti}, A.~E. \& {Geisler}, D. 2013, \aj, 145, 17

\bibitem[{{Przybilla} {et~al.}(2008){Przybilla}, {Nieva}, \&
  {Butler}}]{Przybilla2008}
{Przybilla}, N., {Nieva}, M.-F., \& {Butler}, K. 2008, \apjl, 688, L103

\bibitem[{{Puls} {et~al.}(1996){Puls}, {Kudritzki}, {Herrero}, {Pauldrach},
  {Haser}, {Lennon}, {Gabler}, {Voels}, {Vilchez}, {Wachter}, \&
  {Feldmeier}}]{Puls1996}
{Puls}, J., {Kudritzki}, R.-P., {Herrero}, A., {et~al.} 1996, \aap, 305, 171

\bibitem[{{Puls} {et~al.}(2000){Puls}, {Springmann}, \& {Lennon}}]{Puls2000}
{Puls}, J., {Springmann}, U., \& {Lennon}, M. 2000, \aaps, 141, 23

\bibitem[{{Puls} {et~al.}(2008){Puls}, {Vink}, \& {Najarro}}]{Puls2008}
{Puls}, J., {Vink}, J.~S., \& {Najarro}, F. 2008, \aapr, 16, 209

\bibitem[{{Sander} {et~al.}(2012){Sander}, {Hamann}, \& {Todt}}]{Sander2012}
{Sander}, A., {Hamann}, W.-R., \& {Todt}, H. 2012, \aap, 540, A144

\bibitem[{{Sander} {et~al.}(2015){Sander}, {Shenar}, {Hainich},
  {G{\'{\i}}menez-Garc{\'{\i}}a}, {Todt}, \& {Hamann}}]{Sander2015}
{Sander}, A., {Shenar}, T., {Hainich}, R., {et~al.} 2015, \aap, 577, A13

\bibitem[{{Sander} {et~al.}(2014){Sander}, {Todt}, {Hainich}, \&
  {Hamann}}]{Sander2014}
{Sander}, A., {Todt}, H., {Hainich}, R., \& {Hamann}, W.-R. 2014, \aap, 563,
  A89

\bibitem[{{Schaerer} {et~al.}(1993){Schaerer}, {Meynet}, {Maeder}, \&
  {Schaller}}]{Schaerer1993}
{Schaerer}, D., {Meynet}, G., {Maeder}, A., \& {Schaller}, G. 1993, \aaps, 98,
  523

\bibitem[{{Schlegel} {et~al.}(1998){Schlegel}, {Finkbeiner}, \&
  {Davis}}]{Schlegel1998}
{Schlegel}, D.~J., {Finkbeiner}, D.~P., \& {Davis}, M. 1998, \apj, 500, 525

\bibitem[{{Schmutz} {et~al.}(1989){Schmutz}, {Hamann}, \&
  {Wessolowski}}]{Schmutz1989}
{Schmutz}, W., {Hamann}, W., \& {Wessolowski}, U. 1989, \aap, 210, 236

\bibitem[{{Schneider} {et~al.}(2014){Schneider}, {Langer}, {de Koter}, {Brott},
  {Izzard}, \& {Lau}}]{Schneider2014}
{Schneider}, F.~R.~N., {Langer}, N., {de Koter}, A., {et~al.} 2014, \aap, 570,
  A66

\bibitem[{{Schwering} \& {Israel}(1991)}]{Schwering1991}
{Schwering}, P.~B.~W. \& {Israel}, F.~P. 1991, \aap, 246, 231

\bibitem[{{Seaton}(1979)}]{Seaton1979}
{Seaton}, M.~J. 1979, \mnras, 187, 73P

\bibitem[{{Shenar} {et~al.}(2014){Shenar}, {Hamann}, \& {Todt}}]{Shenar2014}
{Shenar}, T., {Hamann}, W.-R., \& {Todt}, H. 2014, \aap, 562, A118

\bibitem[{{Smith}(1968)}]{Smith1968}
{Smith}, L.~F. 1968, \mnras, 140, 409

\bibitem[{{Smith} {et~al.}(2005){Smith}, {Points}, {Chu}, {Winkler},
  {Aguilera}, {Leiton}, \& {MCELS Team}}]{Smith2005}
{Smith}, R.~C., {Points}, S.~D., {Chu}, Y.-H., {et~al.} 2005, in Bulletin of
  the American Astronomical Society, Vol.~37, American Astronomical Society
  Meeting Abstracts, 1200

\bibitem[{{Testor}(2001)}]{Testor2001}
{Testor}, G. 2001, \aap, 372, 667

\bibitem[{{Todt} {et~al.}(2013){Todt}, {Kniazev}, {Gvaramadze}, {Hamann},
  {Buckley}, {Crause}, {Crawford}, {Gulbis}, {Hettlage}, {Hooper}, {Husser},
  {Kotze}, {Loaring}, {Nordsieck}, {O'Donoghue}, {Pickering}, {Potter},
  {Romero-Colmenero}, {Vaisanen}, {Williams}, \& {Wolf}}]{Todt2013}
{Todt}, H., {Kniazev}, A.~Y., {Gvaramadze}, V.~V., {et~al.} 2013, \mnras, 430,
  2302

\bibitem[{{Todt} {et~al.}(2015){Todt}, {Sander}, {Hainich}, {Hamann}, {Quade},
  \& {Shenar}}]{Todt2015}
{Todt}, H., {Sander}, A., {Hainich}, R., {et~al.} 2015, \aap, {accepted}

\bibitem[{{Torres-Dodgen} \& {Massey}(1988)}]{Torres-Dodgen1988}
{Torres-Dodgen}, A.~V. \& {Massey}, P. 1988, \aj, 96, 1076

\bibitem[{{Trundle} {et~al.}(2007){Trundle}, {Dufton}, {Hunter}, {Evans},
  {Lennon}, {Smartt}, \& {Ryans}}]{Trundle2007}
{Trundle}, C., {Dufton}, P.~L., {Hunter}, I., {et~al.} 2007, \aap, 471, 625

\bibitem[{{Vanbeveren} {et~al.}(1998){Vanbeveren}, {De Donder}, {van Bever},
  {van Rensbergen}, \& {De Loore}}]{Vanbeveren1998}
{Vanbeveren}, D., {De Donder}, E., {van Bever}, J., {van Rensbergen}, W., \&
  {De Loore}, C. 1998, \na, 3, 443

\bibitem[{{Vanbeveren} {et~al.}(2007){Vanbeveren}, {Van Bever}, \&
  {Belkus}}]{Vanbeveren2007}
{Vanbeveren}, D., {Van Bever}, J., \& {Belkus}, H. 2007, \apjl, 662, L107

\bibitem[{{Vink} \& {de Koter}(2005)}]{Vink2005}
{Vink}, J.~S. \& {de Koter}, A. 2005, \aap, 442, 587

\bibitem[{{Vink} {et~al.}(2000){Vink}, {de Koter}, \& {Lamers}}]{Vink2000}
{Vink}, J.~S., {de Koter}, A., \& {Lamers}, H.~J.~G.~L.~M. 2000, \aap, 362, 295

\bibitem[{{Vink} {et~al.}(2001){Vink}, {de Koter}, \& {Lamers}}]{Vink2001}
{Vink}, J.~S., {de Koter}, A., \& {Lamers}, H.~J.~G.~L.~M. 2001, \aap, 369, 574

\bibitem[{{Vink} {et~al.}(2011){Vink}, {Muijres}, {Anthonisse}, {de Koter},
  {Gr{\"a}fener}, \& {Langer}}]{Vink2011}
{Vink}, J.~S., {Muijres}, L.~E., {Anthonisse}, B., {et~al.} 2011, \aap, 531,
  A132

\bibitem[{{Westerlund}(1997)}]{Westerlund1997}
{Westerlund}, B.~E. 1997, Cambridge Astrophysics Series, 29

\bibitem[{{Woosley} \& {Heger}(2006)}]{Woosley2006}
{Woosley}, S.~E. \& {Heger}, A. 2006, \apj, 637, 914

\bibitem[{{Yoon} \& {Langer}(2005)}]{Yoon2005}
{Yoon}, S.-C. \& {Langer}, N. 2005, \aap, 443, 643

\bibitem[{{Zaritsky} {et~al.}(2002){Zaritsky}, {Harris}, {Thompson}, {Grebel},
  \& {Massey}}]{Zaritsky2002}
{Zaritsky}, D., {Harris}, J., {Thompson}, I.~B., {Grebel}, E.~K., \& {Massey},
  P. 2002, \aj, 123, 855

\end{thebibliography}
